\documentclass[3p,12pt]{elsarticle}
\usepackage{threeparttable,booktabs,tabularx}
\usepackage[fleqn]{amsmath}
\usepackage{cases}
\usepackage{multirow}
\usepackage{xcolor}
\usepackage[normalem]{ulem}
\usepackage{tabularx}
\usepackage{siunitx}
\usepackage{comment}
\usepackage{algorithm}
\usepackage{algpseudocode}
\usepackage{algorithmicx}
\usepackage{grffile}
\usepackage{rotating}
\usepackage{array}
\usepackage{lineno}
\usepackage{hyperref}
\usepackage{makecell}
\usepackage{graphicx,enumerate}
\usepackage{amssymb}
\usepackage{bm,amsmath}
\allowdisplaybreaks  
\usepackage{subfigure}
\usepackage{caption,color}
\usepackage{threeparttable}
\usepackage{subeqnarray}
\usepackage{multirow}
\usepackage{lscape}
\usepackage{rotating}
\usepackage{graphics}
\floatname{algorithm}{Algorithm}
\usepackage{epstopdf}
\journal{}

\biboptions{numbers,sort&compress} 

\usepackage{natbib}
\AtBeginEnvironment{thebibliography}{%
  \footnotesize
  \interlinepenalty=10000
}

\begin{document}
	
\begin{frontmatter}

\title{Efficient simulation of chemical reaction in DSMC}

\author[]{Hong Deng\corref{equal}}
\author[]{Liyan Luo\corref{equal}}
\author[]{Lei Wu\corref{Boss}}

\cortext[equal]{Both authors contribute equally.}
\cortext[Boss]{Corresponding author: wul@sustech.edu.cn}

\address{Department of Mechanics and Aerospace Engineering, Southern University of Science and Technology, 518055 Shenzhen, China}

\begin{abstract}
A macroscopic‑mesoscopic, deterministic‑stochastic coupling strategy is proposed to accelerate the direct simulation Monte Carlo (DSMC) method for chemical reaction. 
First, a macroscopic synthetic equation is formulated by integrating continuum constitutive relations for diffusion, stress, and heat flux, along with higher‑order constitutive relations that capture nonequilibrium transport effects. Second,  higher‑order constitutive relations and chemical reaction source terms are sampled from DSMC and embedded into the macroscopic synthetic equation. Third, the macroscopic system is solved to the steady state, whose solution is then employed to correct particle distributions in DSMC intermittently. 
This coupling features asymptotic‑preserving, fast‑converging and noise‑reduction properties, supporting efficient, accurate simulations with coarse spatiotemporal grids and reduced evolution/sampling steps. Accordingly, it mitigates DSMC’s major computational bottlenecks for near‑continuum flows by several orders of magnitude.

\end{abstract}


\end{frontmatter}


\section{Introduction}\label{sec:1}

Driven by the demands of space exploration, accurate numerical simulation of chemical reactions within hypersonic flows has become a critical research subject. Hypersonic flows are featured by pronounced thermochemical non-equilibrium, in which molecular collisions, internal energy exchange, and chemical reactions take place simultaneously. For such flow problems, the direct simulation Monte Carlo (DSMC) method is widely used \cite{bird1994molecular}, owing to its moderate memory overhead and inherent ability to resolve intricate molecular‑scale physical and chemical processes \cite{Borgnakke1975JCP, Hass1993PoF, Bird2011PoF}.

However, DSMC is inefficient in the near-continuum flow regime. The decoupled streaming–collision algorithm requires the computational cell size and time step to be smaller than the molecular mean free path and collision time, respectively. As the Knudsen number decreases, these constraints lead to extremely small spatiotemporal resolutions and prohibitive computational cost. 
To balance computational efficiency and physical fidelity across different flow regimes, hybrid approaches coupling macroscopic Navier–Stokes (NS) solvers in continuum regions with mesoscopic DSMC in non-equilibrium regions are proposed \cite{sun2004hybrid, Schwartzentruber2007JCP, Burt2009JCP}. These methods improve efficiency by relaxing the strict DSMC resolution requirements in continuum regions; however, determining appropriate domain decomposition criteria remains challenging, and DSMC simulations near the continuum–rarefied interface can still be computationally expensive.

In recent years, great progress has been made in the development of efficient stochastic approaches for multiscale gas flows. For examples, a stochastic particle method has been developed to solve the Fokker-Planck equation \cite{gorji2011fokkerplanck}, where binary collisions are replaced by a Langevin process, thus enabling the use of larger time steps in continuum regimes;
The unified stochastic particle method \cite{Fei2020Zhang,fei-2023} mitigates numerical dissipation arising from particle transport via compensation within the collision term of the BGK kinetic equation;
The unified gas‑kinetic wave–particle method \cite{ZHU2019UGKWP, liu2020unified}, also developed for BGK‑type kinetic equations, samples only collisionless particles while treating the remaining contribution deterministically, thereby significantly reducing the computational cost in the near‑continuum regime. 
However, extending such simplified collision models to reactive flows involving multiple chemical species and detailed reaction mechanisms remains highly challenging \cite{wei2025unified}, mainly due to difficulty in modeling reactive collision processes.

Since chemically reactive processes can be naturally captured in DSMC~\cite{bird1994molecular, Bird2011PoF}, improving the computational efficiency of solving the full Boltzmann equation under near‑continuum conditions constitutes a critical step toward high‑performance multiscale simulations. Along this line, the time‑relaxed Monte Carlo method~\cite{pareschi-2001}, as well as its revised variants~\cite{dimarco-2011,ren-2014}, have been devised to mitigate the stiffness of the Boltzmann collision operator by splitting it into a stiff linear term and a weakly stiff nonlinear term. The linear part is resolved via a relaxation‑time approximation, permitting larger time‑step sizes. Nevertheless, these schemes only recover the Euler limit. More recently, novel strategies for both single‑species~\cite{fei-2023} and multi‑species~\cite{FEI2025114196} gases have been proposed to reproduce the NS limit on coarse spatial grids, thereby substantially boosting simulation efficiency.

Despite these advances, developing full‑collision stochastic methods to chemically reactive multiscale flows remains an unaddressed challenge. A promising candidate is the recently proposed direct intermittent GSIS–DSMC coupling (DIG) method~\cite{Luo2024AiA}.
Following the core concept of the general synthetic iterative scheme (GSIS), which deterministically couples mesoscopic and macroscopic equations via iterative solution~\cite{Zhu2021JCP,Su2021CMAME,Hu2026JCP}, the DIG exhibits fast‑converging, asymptotic‑preserving, and noise‑reduction characteristics. It recovers standard DSMC behavior in rarefied regimes and NS solutions in continuum regimes, respectively.
By permitting spatiotemporal discretizations coarser than kinetic scales and suppressing DSMC statistical noise through particle guidance from deterministic macroscopic synthetic‑equation solutions, the DIG substantially improves computational efficiency for near‑continuum flows.
For instance, for hypersonic argon flow passing over a cylinder, DIG requires only 40,000 computational cells compared to 2 million cells in the original DSMC, while the computational time is reduced by nearly two orders of magnitude during the transit evolution. 

This work aims to develop the DIG method to efficiently simulate nonequilibrium chemically reactive flows. This is non‑trivial, as it retains the quantum‑kinetic (QK) model for molecular‑scale reactive collisions while ensuring recovery of the chemically reactive NS equations in the continuum regime. Notably, only a single synthetic equation is required even for chemically reacting systems involving numerous species, which substantially mitigates numerical complexity and improves computational stability. 

The remainder of this paper is organized as follows. Section~\ref{sec:2} reviews the standard DSMC method for chemically reacting flows. Section~\ref{sec:3} formulates the single-velocity and single-temperature macroscopic synthetic equations, and presents high-order closures for the constitutive relations and chemical source terms. Section~\ref{sec:4} outlines the general algorithm and implementation of the DIG. Section~\ref{sec:Num} validates the proposed DIG method through simulations of a two-dimensional dissociating nitrogen flow passing over a cylinder under different Knudsen number conditions. Finally, Section~\ref{sec:conclusion} provides conclusions and outlooks for future work.


\section{Mesoscopic description of gas mixture with chemical reactions}\label{sec:2}

In this section, we first review the kinetic equation for chemically reacting gas mixtures based on the Wang–Chang \& Uhlenbeck (WCU) formulation~\cite{WangCS}. We then briefly introduce the DSMC method. Finally, we present the QK reaction model that characterizes reactive molecular collisions.

\subsection{The WCU-type equation for multi-species gas flow}

For a rarefied gas flow involving $N$ species, the molecular state of species $s$ is described by the velocity distribution function $f_{s,i}(t,\bm{x},\bm{v},I_\text{r})$, where $t$ denotes time, $\bm{x}$ the spatial coordinates, $\bm{v}$ the molecular velocity, $I_r$ the continuous rotational energy. Assuming that $N_l$ vibrational energy levels are considered, with $i=0,1,\ldots,N_l-1$ being the discrete vibrational energy level, and $I_{\text{v},i}$ the corresponding vibrational energy. The evolution of distribution function without external force term can be written as,
\begin{equation}
\label{eq:wcu-type-reaction}
	\begin{aligned}
		\frac{\partial f_{s,i}}{\partial t} +\bm{v} \cdot \frac{\partial f_{s,i}}{\partial \bm{x}} 
        = \underbrace{\sum_{r=1}^{N} (Q_{sr,i}+Q_{sr,i}^{\text{chem}})}_{Q_s}, \quad 
        s=1, 2, ..., N,  
	\end{aligned}
\end{equation}
where $Q_{sr,i}$ denotes the binary gas-gas collisions modeled within the WCU framework, and $Q_{sr,i}^{\text{chem}}$ represents chemically reactive collisions, which will be described in the later section. $Q_s$ is the overall collision operator. When the rotational mode is treated by classical mechanics, the collision operator can be written as
\begin{equation}
Q_{sr,i}=\sum_{i',j'}\sum_{j}\int_{\mathbb{R}^{3}}\int_{\mathbb{S}^{2}}\left(\frac{g_i g_j}{g_{i'} g_{j'}}
f_{s,i'} f_{r,j'}-f_{s,i} f_{r,j}\right)c_{sr}\sigma_{sr,ij}^{\,i'j'}\,\mathrm{d}\bm{\Omega}\,\mathrm{d}\bm{v}_r.
\end{equation}
Here, $i$ and $j$ denote the pre-collision vibrational energy levels of species $s$ and $r$, respectively, while $i'$ and $j'$ denote the corresponding post-collision vibrational energy levels. 
$g_i$ and $g_j$ are the degeneracies of vibrational levels $i$ and $j$, respectively. $c_{sr}=|\bm{v}_s-\bm{v}_r|$ is the relative translational speed between the two colliding molecules. $\sigma_{sr,ij}^{\,i'j'}$ is the state-to-state differential scattering cross section, and $\bm{\Omega}$ is the solid angle. Since the total energy is conserved during the collision process, such a transition occurs only when
\begin{equation}
\label{eq:csr2}
|\bm{c}_{sr}'|^2=|\bm{c}_{sr}|^2+\frac{2(m_s+m_r)}{m_s m_r}\left(I_{s,i}+I_{r,j}-I_{s,i'}-I_{r,j'}\right)>0,
\end{equation}
where $I_{s,i}$ denotes the total internal energy associated with vibrational level $i$ of species $s$. 
When the collision is admissible, the post-collision velocities are given by
\begin{equation}
\label{eq:velocitytwopart}
\begin{aligned}
\bm{v}_s'=\bm{v}_*+\frac{m_r}{m_s+m_r}|\bm{c}_{sr}'|\bm{\Omega},\\
\bm{v}_r'=\bm{v}_*-\frac{m_s}{m_s+m_r}|\bm{c}_{sr}'|\bm{\Omega},
\end{aligned}
\end{equation}
where $\bm{v}_*=({m_s\bm{v}_s+m_r\bm{v}_r})/({m_s+m_r})$ is the center-of-mass velocity of two colliding molecules.

By taking moments of the velocity distribution function, the macroscopic properties of species $s$, including the number density $n_s$, mass density $\rho_s$, flow velocity $\bm{u}_s$, translational, rotational and vibrational temperatures $T_{\mathrm{tra},s}$, $T_{\mathrm{rot},s}$ and $T_{\mathrm{vib},s}$, as well as the corresponding energies per unit mass $E_{\text{tra},s}$, $E_{\text{rot},s}$ and $E_{\text{vib},s}$, deviatoric stress tensor $\bm{\sigma}_s$, and translational, rotational and vibrational heat fluxes $\bm{q}_{\mathrm{tra},s}$, $\bm{q}_{\mathrm{rot},s}$ and $\bm{q}_{\mathrm{vib},s}$, can be obtained as
\begin{equation}
\label{eq:fstomacro}
\begin{aligned}
n_s&=\sum_i\left<1,f_{s,i}\right>,\quad
\rho_s=\sum_i\left<m_s,f_{s,i}
\right>,\quad\rho_s\bm{u}_s=\sum_i\left<m_s\bm{v}_s,f_{s,i}\right>,\\
\bm{\sigma}_s&=\sum_i\left<m_s\left(\bm{c}_s\bm{c}_s-\frac{c_s^2}{3}\bm{I}\right),f_{s,i}\right>,
\\
\rho_sE_{\text{tra},s}&=\sum_i\left<\frac{1}{2}m_s v_s^2,f_{s,i}\right>=\frac{3}{2}n_sk_BT_{\text{tra},s}+\frac{1}{2}\rho_s|\bm{u}_s|^2,\,
\\
\rho_sE_{\text{rot},s}&=\sum_i\left<I_{\text{r}},f_{s,i}\right>=\frac{d_{\mathrm{rot},s}}{2}n_sk_B T_{\mathrm{rot},s},\\
\rho_sE_{\text{vib},s}&=\sum_i\left<I_{\text{v},i},f_{s,i}\right>=\frac{d_{\mathrm{vib},s}}{2}n_sk_BT_{\mathrm{vib},s},
\\
\bm{q}_{\mathrm{tra},s}&=\sum_i\left<\frac{1}{2}m_s\bm{c}_s c_s^2,f_{s,i}\right>,\quad
\bm{q}_{\mathrm{rot},s}=\sum_i\left<\bm{c}_s I_{\text{r}},f_{s,i}\right>,\quad
\bm{q}_{\mathrm{vib},s}=\sum_i\left<\bm{c}_s I_{\text{v},i},f_{s,i}\right>.
\end{aligned}
\end{equation}
Here, the operator $\left<h,\psi\right>$ is defined as the integral of $h\psi$ over the velocity and rotational energy spaces.
$k_B$ is the Boltzmann constant.
$\bm{I}$ is the $3\times3$ identity matrix, and $\bm{c}_s=\bm{v}_s-\bm{u}_s$ is the peculiar velocity. $d_{\text{rot,s}}$ is the rotational degrees of freedom for species $s$, which is taken to be constant.
According to the harmonic oscillator model, the effective vibrational degrees of freedom $d_{\text{vib,s}}$ can be written as
\begin{equation}
    d_{\text{vib},s}=\frac{2\Theta_{\text{vib},s}/T_{\text{vib},s}}{\exp\left(\Theta_{\text{vib},s}/T_{\text{vib},s}\right)-1}, 
\end{equation}
where $\Theta_{\text{vib},s}$ is the characteristic vibrational temperature for species $s$.

Macroscopic parameters of the mixture, such as the total mass density $\rho$, flow velocity $\bm{u}$, pressure tensor $\bm{P}$, stress tensor $\bm{\sigma}$, heat flux $\bm{q}$, can be computed by
\begin{equation} \label{equ:sampleMar}
	\begin{aligned}[c] 
        n = &\sum_{s} n_{s},\quad
        \rho = \sum_{s} \rho_{s}, \quad
        \rho \bm{u}= \sum_{s} \rho_{s} \bm{u}_{s}, \\
        \bm{\sigma} =& \underbrace{\sum_s\bm{\sigma}_s}_{\bm{\sigma_{\text{mix}}}} + \bm{\Pi_\text{sp}},\quad 
        \bm{P}=\bm{\sigma}+ \sum_{s} p_{\text{t},s}\bm{I}, \quad 
        \bm{\Pi}_{\text{sp}}=\sum_s\rho_s\bm{V}_s\bm{V}_s,
        \\
        \bm{q}= &  \underbrace{\sum_{s}\left( \bm{q}_{\text{tra},s}+ \bm{q}_{\text{rot},s} + \bm{q}_{\text{vib},s}\right)}_{\bm{q}_\text{mix}}+\underbrace{\sum_s \frac{1}{2}  \rho_{s}\left|\bm{V}_{s}\right|^{2}{\bm{V}_{s}}}_{\bm{K}_\text{diff}}+\sum_s\left(\frac{3}{2}p_{\text{t},s}{\bm{V}_{s}} 
        +\bm{P}_s \cdot{\bm{V}_s} 
        \right),
	\end{aligned}
\end{equation}
where $\bm{V}_{s}=\bm{u}_{s}-\bm{u}$ is the diffusion velocity, $p_{\text{t},s}=n_sk_BT_{\text{tra},s}$ is the translational pressure of the species $s$, $\bm{\Pi}_{\text{sp}}$ is the diffusive shear stress, and $\bm{K}_\text{diff}$ is the diffusion kinetic energy. Note that $\bm{\sigma}_\text{mix}$ and $\bm{q}_\text{mix}$ represent the mixture viscous stress tensor and conductive heat flux, respectively.

The overall temperature $T_{\text{total}}$ constructed from the species temperatures $T_s$ can be written as
\begin{equation}
    T_{\text{total}}=\frac{\sum_{s}(3+d_{\text{rot},s}+d_{\text{vib},s})X_sT_s}{\sum_{s}(3+d_{\text{rot},s}+d_{\text{vib},s})X_s},\quad
    T_s=\frac{3T_{\text{tra},s}+d_{\text{rot},s}T_{\text{rot},s}+d_{\text{vib},s} T_{\text{vib},s}}{{3+d_{\text{rot},s}+d_{\text{vib},s}}},
\end{equation}
where $X_s=n_s/n$ is the molar fraction of species $s$.



\subsection{The DSMC method with Larsen-Borgnakke model}

The deterministic solution of the WCU equation remains computationally prohibitive due to the inherently high-dimensional phase space associated with molecular velocity and internal energy variables~\cite{WCUtransport}. To overcome this challenge, the DSMC method employs a finite set of simulation particles (each representative of a large ensemble of real gas molecules) to bypass the high dimensionality inherent to deterministic velocity‑space discretization and facilitate efficient numerical simulation of rarefied gas flows~\cite{bird1994molecular}. The distribution function $f_{s,i}$ in a single cell can be represented by these simulation particles:
\begin{equation}
    f_{s,i}(\bm{x},\bm{v},I_{\text{r}})=\frac{N_\text{eff}}{V_\text{cell}}\sum_{p=1}^{N_s} \delta(\bm{x}-\bm{x}^{(p)})\delta(\bm{v}-\bm{v}^{(p)})\delta(I_{\text{r}}-I_{\text{r}}^{(p)})\delta_{i,i^{(p)}},
\end{equation}
where $\delta(\cdot)$ and $\delta_{ij}$ denote the Dirac and Kronecker delta functions, respectively. Here, $N_{\mathrm{eff}}$ is the number of real molecules represented by each simulated particle, $V_{\mathrm{cell}}$ is the cell volume, and $N_s$ is the total number of simulated particles of species $s$ within the cell. According to Eq.~\eqref{eq:fstomacro}, macroscopic variables in DSMC can be obtained by taking moments of the distribution function $f_{s,i}$:
\begin{equation}
\label{equ_2}
	\begin{aligned}
        n_s&=\frac{N_{\text{eff}}}{V_{\text{cell}}}N_s, \quad \rho_s=m_sn_s,\quad \bm{u}_s=\frac{1}{N_s}\sum_{p=1}^{N_s}\bm{v}_s, \quad\bm{\sigma}_{s}=\frac{m_sN_{\text{eff}}}{V_{\text{cell}}}\sum_{p=1}^{N_s}
       \left(\bm{c}_s\bm{c}_s-\frac{c^2_s}{3}\mathbf{I}\right), \\
        \rho_sE_{\text{tra},s}&=\frac{N_{\text{eff}}}{V_{\text{cell}}}\sum_{p=1}^{N_s}\frac{1}{2}m_s|\bm{v}_s|^2,
        \quad 
        \rho_sE_{\text{rot},s}=\frac{N_{\text{eff}}}{V_{\text{cell}}}\sum_{p=1}^{N_s} I_{\text{r},s}
        ,\quad 
        \rho_sE_{\text{vib},s}=\frac{N_{\text{eff}}}{V_{\text{cell}}}\sum_{p=1}^{N_s} I_{\text{v},i,s}
        ,\\
        \bm{q}_{\text{tra},s}&=\frac{m_sN_{\text{eff}}}{2V_{\text{cell}}}\sum_{p=1}^{N_s} \bm{c}_s c^2_s, \quad 
        \bm{q}_{\text{rot},s}=\frac{N_{\text{eff}}}{V_{\text{cell}}}\sum_{p=1}^{N_s} \bm{c}_s I_{\text{r},s}, \quad
        \bm{q}_{\text{vib},s}=\frac{N_{\text{eff}}}{V_{\text{cell}}}\sum_{p=1}^{N_s}{\bm{c}_{s}}I_{\text{v},i,s}.
	\end{aligned}
\end{equation}
It should be noted that, according to the simple harmonic model, the vibrational energy of each particle is determined as $I_{\text{v},i}=ik_\text{B}\Theta_\text{vib}$.

In DSMC, the kinetic equation is numerically split into two distinct physical processes~\cite{bird1994molecular}: the molecular advection and intermolecular collisions, i.e.,
\begin{equation}
    \begin{aligned}
        &\text{Advection:}\quad \frac{\partial f_{s,i}}{\partial t} +\bm{v} \cdot \frac{\partial f_{s,i}}{\partial \bm{x}} = 0,\\
        & \text{Collision:}\quad\quad  \left[\frac{\partial f_{s,i}}{\partial t}\right]_{\text{coll}}=Q_s.
    \end{aligned}
\end{equation}
For the advection part, the velocities of simulation particles remain unchanged, while their positions are modified according to $\bm{x}_s^{(p)}(t+\Delta t)=\bm{x}_s^{(p)}(t)+\bm{v}_s^{(p)}\Delta t$. 
After the advection step, binary collisions are performed between particles in each computational cell, including both inter-species and intra-species interactions. 

For polyatomic gases, collisions involve not only translational energy exchange but also redistribution between translational and internal energy modes. In DSMC, such internal energy relaxation is commonly modeled using the Larsen--Borgnakke scheme~\cite{Borgnakke1975JCP}. When a collision is regarded as inelastic, the total collision energy, including the relative translational energy and internal energy, is redistributed among translational, rotational and vibrational modes according to equilibrium energy distributions. Only a fraction of collisions are treated as inelastic collisions, and the corresponding probability is related to the relaxation collision number $Z$. For the variable-soft-sphere (VSS) collision model, the fraction of inelastic collisions can be written as~\cite{bird1994molecular}
\begin{equation}
    P_{\text{inelastic}}=\frac{\alpha(5-2\omega)(7-2\omega)}{5(\alpha+1)(\alpha+2)Z},
\end{equation}
where $\omega$ is the viscosity temperature exponent and $\alpha$ is the angular scattering parameter in the VSS collision model.
Here, $Z$ may correspond to either the rotational ($Z_\text{r}$) or vibrational ($Z_\text{v}$) relaxation collision number.
During vibrational relaxation, the maximum accessible vibrational energy level is given by $\left \lfloor \frac{E_\text{c}}{k_\text{B}\Theta_\text{vib}} \right \rfloor$, where $E_\text{c}$ is the sum of relative translational energy and vibrational energy. The post-collision vibrational level $i^*$ must satisfy $i^* = \left \lfloor \frac{-\ln(R_\text{f})E_\text{c}}{k_\text{B}\Theta_\text{vib}} \right \rfloor $, with $R_f$ being a uniformly distributed random number between 0 and 1. Subsequently, $i^*$ can be selected using an acceptance-rejection algorithm, with the selection probability given by,
\begin{equation}
	\begin{aligned}
		P_{\text{vib}}={\left(1-i^*\frac{k_\text{B}\Theta_\text{vib}}{E_\text{c}} \right)}^{\frac{3}{2}-\omega},
	\end{aligned}
\end{equation}
It should be noted that, different from $P_{\text{inelastic}}$, which determines whether a collision is treated as an inelastic collision, $P_{\text{vib}}$ determines whether the sampled post-collision vibrational energy level is accepted in the acceptance--rejection procedure. After the redistribution of internal energy, the relative translational energy is updated according to Eq.~\eqref{eq:csr2}, and the post-collision velocities of the two particles are then determined from Eq.~\eqref{eq:velocitytwopart}.

\subsection{QK chemical reaction model in DSMC}

In Eq.~\eqref{eq:wcu-type-reaction}, the chemical collision operator $Q_{sr,i}^{\mathrm{chem}}$ represents chemically reactive collisions that alter both the species composition and internal energy states. In DSMC, this operator is realized through stochastic binary collision sampling, in which a reaction model determines whether a chemical reaction occurs and specifies the corresponding post-collision products~\cite{bird1994molecular,Bird2011PoF}.
Over the past decades, a variety of chemical reaction model have been developed, reflecting different balances between physical fidelity and computational efficiency. 
The total collision energy method is one of the most widely used, in which the probability of reaction is determined on the basis of Arrhenius-type rate coefficients. Later, Bird subsequently developed a QK model to characterize chemically reactive processes~\cite{Bird2011PoF}, which can accurately recover target equilibrium reaction rates for variable hard‑sphere (VHS) gases. The performance of this QK model was thoroughly investigated by Gallis \textit{et al.} using a zero‑dimensional DSMC solver~\cite{gallis2009jchemialp,gallis2010jtht}. Most recently, Civrais \textit{et al.} extended the original framework by proposing a vibronic QK model that yields improved predictions of thermochemical equilibrium characteristics~\cite{Civrais2025pof}.

In this study, the QK model is adopted~\cite{gallis2009jchemialp,gallis2010jtht}. 
Generally, a chemical reaction form $\text{AB+C}\to \text{A+B+C}$ is regarded as a dissociation reaction. In the QK model, particle AB is considered dissociable only when the maximum vibrational energy level 
$i_\text{max}$ exceeds the dissociation threshold, i.e.,
\begin{equation}
    i_\text{max}=\left \lfloor \frac{E_c}{k_B \Theta_{vib}} \right \rfloor >\frac{\Theta_\text{dis}}{\Theta_\text{vib}},
\end{equation}
where $\Theta_\text{dis}$ is the characteristic dissociation temperature corresponding to the dissociation energy. Upon occurrence of a dissociation reaction, the residual collision energy,defined as the pre‑reaction collision energy minus the chemical activation energy, is first redistributed among the internal energy modes of the resulting products, provided such products possess rotational or vibrational degrees of freedom. Following the completion of this internal energy redistribution, the post‑collision velocities of individual particles are then determined. Although a dissociation reaction formally generates three distinct products, A, B, and C, the velocity update process during the collision is performed in a two-body framework by temporarily treating particles A and B as one composite particle, denoted by $\text{AB}'$. Under this treatment, the post-collision velocities of $\text{AB}'$ and C are first evaluated in the same manner as elastic collision model, while the dissociating products A and B inherit the translational velocity of the composite particle $\text{AB}'$. The verification of the present in-house DSMC code with chemical reaction is provided in~\ref{Appendix_QK}.

\section{Macroscopic synthetic equation for reacting gas flows}\label{sec:3}

In the prior DIG formulation for gas mixtures without chemical reaction, macroscopic synthetic equations were constructed from multi‑fluid governing equations, wherein each species is assigned its own density, velocity, and temperature~\cite{Luo2026_APDSMCMixtures}. While this framework explicitly resolves inter‑species nonequilibrium effects, its performance is  sensitive to the statistical quality of DSMC sampling. This issue becomes especially pronounced in chemically reacting flows, where low‑density reactive species can trigger substantial statistical fluctuations and degrade the robustness of the stochastic‑deterministic coupling scheme.
For continuum reacting flows, especially when the mass ratios among species are not significantly different, the gas mixture can often be reasonably described using a common mixture-averaged velocity and temperature~\cite{shuen1990inviscid,shuen1993coupled}. Therefore, instead of the multi-fluid formulation, the present study adopts a single-velocity single-temperature framework to describe the chemically reacting gas mixture. Non‑equilibrium contributions will be encapsulated within the higher‑order constitutive relations derived from DSMC simulations. Such a formulation significantly reduces the number of macroscopic variables and improves the robustness of the stochastic-deterministic coupling procedure.



\subsection{Macroscopic synthetic equation for chemical reacting flows}

By taking moments of Eq.~\eqref{eq:wcu-type-reaction} over molecular velocity and internal‑energy spaces, alongside summation over discrete vibrational levels, the species‑resolved multi‑fluid governing equations can be derived~\cite{Luo2026_APDSMCMixtures}. These equations are further summed over all constituent species to yield a mixture‑averaged single‑fluid formulation, see the derivation in~\ref{Appendix_meq_derivation}. Thus, the macroscopic equations for the $N$-component mixture with chemically reacting flows can be written as:
\begin{equation}
\begin{aligned}
\frac{\partial\rho}{\partial t}+\nabla\cdot(\rho\bm{u})&=0,\\
\frac{\partial}{\partial t}\left(\rho\boldsymbol{u}\right)+\nabla\cdot\left(\rho\boldsymbol{u}\boldsymbol{u}\right)+\nabla\cdot\left(p_\text{eff}\mathbf{I}+\bm{\Pi}_b+\boldsymbol{\sigma}_\text{mix}+\bm{\Pi}_{\text{sp}}\right)&=0,\\
\frac{\partial}{\partial t}\left(\rho E\right)+\nabla\cdot\left(\rho E\boldsymbol{u}\right)+\nabla\cdot\bigg[\left(p_{\text{eff}}\mathbf{I}+\bm{\Pi}_{\text{b,mix}}+\bm{\sigma}_\text{mix}\right)\cdot\boldsymbol{u}+\boldsymbol{q}_\text{mix} \\
+\sum_{s=1}^N h_s \bm{\Phi}_{s}+\bm{\Pi}_{\text{sp}}\cdot\bm{u}+\bm{K}_{\text{diff}} + \sum_s\bm{\sigma}_s\bm{V}_s\bigg]&=-\sum_s e_{f,s}^0\dot{\omega}_s, \\
\frac{\partial \rho Y_s}{\partial t} + \nabla\cdot (\rho Y_s \bm{u}+\bm{\Phi}_s)&=\dot{\omega}_s.
\end{aligned}
\label{eq:equ_NS}
\end{equation}
Here, $Y_s=\rho_s/\rho$ is the mass fraction and $h_s$ is the enthalpy of the species $s$. $\dot{\omega}_s$ is the rate of change of species $s$ due to the chemical reactions. Since we can get the remaining mass fraction $Y_N$ from $Y_N = 1-\sum_{s=1}^{N-1}Y_s$, only $N-1$ independent $Y_s$ values need to be solved. 
$p_\text{eff}$ is the effective pressure obtained based on the effective temperature $T_\text{eff}$:
\begin{equation}
    p_\text{eff}=\rho T_\text{eff} \sum_{s=1}^{N}Y_s R_s,
\end{equation}
where $R_s = k_\text{B}/m_s$ denotes the gas constant for the species $s$. 
Moreover, other parameters, such as the diffusion mass flux $\bm{\Phi}_s$, mixture shear stress $\bm{\sigma}_\text{mix}$, mixture conductive heat flux $\bm{q}_\text{mix}$, and bulk viscous pressure $\bm{\Pi}_b$ (i.e., the isotropic stress correction induced by non-equilibrium relaxation between translational and internal energy modes) in Eq.~\eqref{eq:equ_NS}, will be discussed in the later section.

The specific total energy $E$ is defined according to
\begin{equation}
\label{eq:rhoErhoeTeff05rhou2}
    \rho E=\sum_{s=1}^N\rho_s\left(E_{\text{tra},s}+E_{\text{rot},s}+E_{\text{vib},s}\right)=\rho e(T_\text{eff})+\frac{1}{2}\rho |\bm{u}|^2,
\end{equation}
where the effective internal energy of the gas mixture is expressed as a function of the effective temperature $T_{\text{eff}}$ as,
\begin{equation}
\label{eq:internalenergy}
    \rho e(T_\text{eff})+\sum_{s=1}^N \rho_se_{f,s}^0=\sum_{s=1}^N\left[\rho_s\left(h_{f,s}^0+\int_{T_{\text{ref}}}^Tc_{\text{p},s}\text{d}T\right)\right]-p_\text{eff}.
\end{equation}
Here, $e_{f,s}^0$ denotes the species-dependent zero-point energy offset used to align the internal-energy reference state, which is a constant during the calculation. It accounts for both the heat of formation $h_{f,s}^0$ and the reference contribution associated with the lower integration limit $T_{\mathrm{ref}}$ in Eq.~\eqref{eq:internalenergy}. For numerical stability, the reference energy offset $e_{f,s}^0$ is explicitly separated from the effective internal energy formulation. Consequently, the energy equation contains additional source terms associated with the chemical enthalpy variation. The detailed derivation is provided in \ref{Appendix_meq_derivation}. 

It should be noted that, in previous study~\cite{Luo2026_APDSMCMixtures}, the specific heat at constant pressure $c_{\text{p},s}$ is considered to be constant under the assumption of monatomic gas mixtures.
In the present work, translational, rotational, and vibrational contributions are simultaneously accounted for, and the specific heat $c_{\text{p},s}$ of each species is treated as a function of temperature:
\begin{equation}
    \label{eq:specific_heat}
    c_{\text{p},s}(T_\text{eff})=\underbrace{\frac{3}{2}R_s+\frac{d_{\text{rot},s}}{2}R_s+\left(\frac{\Theta_\text{vib}/T_\text{eff}}{\exp(\Theta_\text{vib}/T_\text{eff})-1}\right)^2\exp\left(\frac{\Theta_\text{vib}}{T_\text{eff}}\right)R_s}_{c_{\text{v},s}(T_\text{eff})} + R_s.
\end{equation}

It should also be noted that several alternative approaches exist for evaluating temperature-dependent specific heats. For example, NASA has developed the NASA-7 and NASA-9 polynomial fits based on experimental data~\cite{NASA_7_1993, NASA_9_2002}, where contributions from electronic energy levels are considered. The present study considers only translational, rotational, and vibrational energy modes, neglecting electronic contributions in order to maintain consistency with the subsequent coupling to DSMC simulations. Therefore, the NASA polynomials are not employed here, although their use could be considered in future extensions of the model.

In the continuum regime, in order to model the chemical source term $\dot{\omega}_{\text{NS},s}$ in the right-hand-side of Eq.~\eqref{eq:equ_NS}, the rate equations for a set of $N_R$ elementary reactions should be considered. 
The chemical source term is given as follows,
\begin{equation}
\label{eq:nssourceterms}
	\begin{aligned}
    \dot{\omega}_{\text{NS},s} = m_s\sum_{l=1}^{N_R}\dot{n}_{ls},
	\end{aligned}
\end{equation}
Note that $\dot{n}_{ls}$ is the rate of change of number density of species $s$ by the $l$-th reaction, which is given by \cite{BLAZEK20157}, 
\begin{equation}
\begin{aligned}
    \dot{n}_{ls} = (\nu_{ls}''-\nu_{ls}')
    \left(k_{\text{f}l}\prod_{m=1}^{N}n_{m}^{\nu_{lm}'}
         -k_{\text{b}l}\prod_{m=1}^{N}n_{m}^{\nu_{lm}''}\right),
\end{aligned}
\end{equation}
where $\nu_{ls}'$ and $\nu_{ls}''$ are the stoichiometric coefficients for species $s$ in the $l$-th forward and backward reaction, respectively. Accordingly, the rate equations involving $N$ species can be written in the general form:
\begin{equation}
    \label{eq:rate_equation}
    \sum_{s=1}^N \nu'_{ls}n_{s} \quad 
    \overset{k_{\text{f}l}}{\underset{k_{\text{b}l}}
    {\rightleftharpoons}}   \quad 
    \sum_{s=1}^N \nu''_{ls}n_{s},\quad\text{for}\quad l=1,2,...,N_R.
\end{equation}
Furthermore, $k_{\text{f}l}$ and $k_{\text{b}l}$ denote the forward and backward reaction rate constants of the $l$-th reaction, respectively. Since only dissociation reactions are considered in the present study, the backward rate constant $k_{\text{b}l}$ is set to zero. The forward rate for the $l$-th reactions is evaluated using the empirical Arrhenius expression:
\begin{equation}
\label{eq:Arrhenius}
k_\text{f}=aT_\text{eff}^b \exp\left(-\frac{E_\text{act}}{k_\text{B}T_\text{eff}}\right),    
\end{equation}
where $E_{\text{act}}$ denotes the activation energy, and $a$ and $b$ are empirical constants. The corresponding values for different reactions of $E_{\text{act}}$, $a$, and $b$ are provided in~\ref{Appendix_QK}. Moreover, due to the rarefaction effects, the higher-order terms should be also introduced into the chemical source terms provided here, which will be discussed in later section.

\subsection{Higher-order terms for stress and heat flux}

The macroscopic equations remain unclosed because the constitutive relations for the shear stress and heat flux are still unknown. In the continuum regime, these properties are obtained from Chapman-Enskog expansion as:
\begin{equation}
\label{eq:consititutiverelation}
\begin{aligned}
    \bm{\sigma}_\text{mix}^\text{NS}= &  -\mu_\text{mix}\left[\nabla\bm{u}+\nabla\bm{u}^{\mathrm{T}}-\frac{2}{3}(\nabla\cdot\bm{u})\mathbf{I}\right], \\
    \bm{q}_\text{mix}^\text{NS}= & -\kappa_\text{mix}\nabla T_\text{eff},\\
    \bm{\Pi}^\text{NS}_{\text{b,mix}} = &-\mu_{\text{b,mix}}\left(\nabla \cdot \bm{u}\right)\mathbf{I},
\end{aligned}
\end{equation}
where $\mu_\text{mix}$, $\kappa_\text{mix}$ and $\mu_\text{b,mix}$ are the shear viscosity, heat conductivity and bulk viscosity of the gas mixture, respectively. 
For the mixture gas, these transport properties are obtained according to the Wilke's mixing rule~\cite{wilke1950viscosity}, i.e.,
\begin{equation}
	\begin{aligned}
    \mu_\text{mix}&=\sum_{s=1}^{N}\frac{X_s \mu_s}{\phi_s},\quad
    \kappa_\text{mix}=\sum_{s=1}^{N}\frac{X_s \kappa_s}{\phi_s},\quad
    \mu_\text{b,mix}=\sum_{s=1}^{N}\frac{X_s \mu_{\text{b},s}}{\phi_s},
    \\
    \phi_s&=\sum_{i=1}^{N}X_i\left[1+\sqrt{\frac{\mu_s}{\mu_i}}\left(\frac{m_i}{m_s}\right)^{1/4}\right]^2
    \left[{8\left(1+\frac{m_s}{m_i}\right)}\right]^{-1/2}.
	\end{aligned}
\end{equation}
In general, the viscosity and thermal conductivity of species $s$ are evaluated using the power-law model and kinetic theory, respectively,
\begin{equation}
    \label{eq:muskappas}
    \mu_s=\mu_{s,\text{ref}}\left(\frac{T}{T_{\text{ref}}}\right)^{\omega_s}, \quad \kappa_s=\frac{15}{4}\mu_s\left(\frac{4}{15}\frac{c_{\text{p},s}}{R_s}+\frac{1}{3}\right),
\end{equation}
where $\mu_{s,\text{ref}}$ is the reference viscosity at the reference temperature $T_{\text{ref}}$, and $\omega_s$ is the viscosity temperature exponent for different species. 

The bulk viscous stress tensor of the gas mixture is also introduced to account for the non-equilibrium relaxation between translational and internal energy modes.  
The bulk viscosity coefficient for each species is evaluated using a relaxation-based formulation involving the internal degrees of freedom and relaxation collision numbers of different internal modes~\cite{Sharma2022}:
\begin{equation}
    \label{eq:mub} \mu_{\text{b},s}=2\mu_s\frac{d_{\text{rot},s}Z_{\text{r},s}+d_{\text{vib},s}Z_{\text{v},s}}{\left(3+d_{\text{rot},s}+d_{\text{vib},s}\right)^2}.
\end{equation}

For the rarefied gas flow, although numerous macroscopic model have been developed, none has yet successfully provided accurate constitutive relations across the entire range of rarefied regimes~\cite{Gu2020AIA}. Consequently, the macroscopic equations are usually closed with the numerical solution of kinetic equations~\cite{Luo2026_APDSMCMixtures}. Here, the shear stress, heat flux and bulk viscous stress tensor are decomposed into two parts: the linear relations in terms of velocities and temperature given by NS equations, and the higher-order terms (HoTs) explicitly extracted from DSMC, i.e.,
\begin{equation}\label{eq:generalrelations}
\begin{aligned}
\bm{\sigma}_\text{mix}= & \bm{\sigma}_\text{mix}^{\text{NS}}+\underbrace{\bm{\sigma}_\text{mix}^{\text{DSMC}}-\bm{\sigma}_\text{mix}^{\text{NS*}}}_{\textbf{HoT}_{\bm{\sigma}}}, \\
    \bm{q}_\text{mix}=& \bm{q}_\text{mix}^{\text{NS}}
    +\underbrace{\bm{q}_\text{mix}^{\text{DSMC}}-\bm{q}_\text{mix}^{\text{NS*}}}_{\textbf{HoT}_{\bm{q}}},\\
\bm{\Pi}_{\text{b,mix}}=&\bm{\Pi}_{\text{b,mix}}^{\text{NS}}+\underbrace{\bm{\Pi}_{\text{b,mix}}^{\text{DSMC}}-\bm{\Pi}_{\text{b,mix}}^{\text{NS*}}}_{\textbf{HoT}_{\bm{\Pi}_\text{b}}},
\end{aligned}
\end{equation}
where $\bm{\sigma}^\text{DSMC}_\text{mix}$ and $\bm{q}_\text{mix}^\text{DSMC}$ denote the time-averaged shear stress and heat flux of the gas mixture obtained from DSMC simulations, respectively, while $\bm{\sigma}_\text{mix}^\text{NS*}$ , $\bm{q}^\text{NS*}_\text{mix}$ and $\bm{\Pi}_{\text{b,mix}}^{\text{NS*}}$ are evaluated using the NS constitutive relations~\eqref{eq:consititutiverelation} based on the macroscopic properties sampled from DSMC. The DSMC bulk viscous stress tensor is constructed from the deviation between the translational pressure and the effective thermodynamic pressure as
\begin{equation}
\label{eq:pidsmc}
    \bm{\Pi}_{\text{b,mix}}^\text{DSMC}=\left[\sum_{s=1}^N p_{\text{t},s}-p(T_\text{eff}^*)\right]\mathbf{I}.
\end{equation}
Here, $T_{\text{eff}}^*$ denotes the effective temperature reconstructed from the macroscopic properties sampled from DSMC through Newton iteration using Eq.~\eqref{eq:internalenergy}. Consequently, all quantities appearing in $\bm{\Pi}_{\text{b,mix}}^\text{DSMC}$ are directly constructed from the DSMC-sampled translational pressure and effective thermodynamic pressure.

It should be noted that variables with superscripts NS and $\text{NS*}$ in Eq.~\eqref{eq:generalrelations} are evaluated at different time steps and thus cannot cancel one another unless a steady state is attained. As demonstrated in subsequent numerical simulations, constructing constitutive relations in this manner yields fast convergence and asymptotic preservation of the NS behavior. Furthermore, it is emphasized that the effective transport coefficients (shear/bulk viscosity and thermal conductivity) may not exactly match those used in DSMC simulations; nevertheless, once steady state is achieved, the stress and heat flux are precisely reproduced by the faithful DSMC solution.

\subsection{Higher-order terms for diffusive terms and chemical source terms}


Since only a single velocity and temperature are considered in the macroscopic synthetic equation,  the diffusion mass flux $\bm{\Phi}_s$ must be explicitly modeled within the NS framework. Following Fick's law, it can be written as
\begin{equation}
\label{eq:massdiffusion}
\bm{\Phi}^{\text{NS}}_s=-\rho D_s \nabla Y_s,\quad D_s=(1-Y_s)/\displaystyle\sum_{i\neq s}^N\frac{X_i}{D_{is}},
\end{equation}
where $D_{is}$ denotes the binary mass diffusion coefficient between species $i$ and $s$, which is evaluated based on the Chapman–Enskog theory~\cite{chapman1990mathematical}.

On the other hand, in the DSMC method, the diffusion flux and chemical source terms are directly obtained from particle sampling. Specifically, they can be evaluated as
\begin{equation}
\begin{aligned}
    \bm{\Phi}_{s}^\text{DSMC}=\frac{m_sN_{\text{eff}}}{V_\text{cell}}\sum_{s=1}^N(\bm{v}_s-\bm{u}),\\
    \dot{\omega}_{s}^\text{DSMC}=\frac{m_sN_{\text{eff}}}{V_\text{cell}\Delta t}\sum_{l=1}^{N_R}(\nu_{ls}''-\nu_{ls}'),
\end{aligned}
\end{equation}
where $\Delta t$ represents the time step used in DSMC method. Similar to the treatment of shear stress and heat flux, the diffusion flux and chemical source terms are decomposed into two parts in order to achieve both rapid convergence and asymptotic-preserving behavior: (i) the model-derived contributions, including the diffusion flux given by Eq.~\eqref{eq:massdiffusion} and the chemical source terms~\eqref{eq:nssourceterms}, which are computed explicitly in terms of macroscopic properties; (ii) the difference between the DSMC-sampled quantities and the model-derived counterparts, which is kept constant during the solution of the macroscopic synthetic equations. Accordingly, they can be written in the following manner:
\begin{equation}\label{eq:PhiOmega}
\begin{aligned}
    \bm{\Phi}_s=\bm{\Phi}^{\text{NS}}_s+\underbrace{\bm{\Phi}^{\text{DSMC}}_s-\bm{\Phi}^{\text{NS*}}_s}_{{\textbf{HoT}}_{\bm{\Phi}_s}},\\ \dot{\omega}_{s}=\dot{\omega}^{\text{NS}}_s+\underbrace{\dot{\omega}^{\text{DSMC}}_s-\dot{\omega}^{\text{NS*}}_s}_{{\text{HoT}}_{\dot{\omega}_s}}.
\end{aligned}
\end{equation}

Moreover, physically, the diffusive enthalpy transport remains species-dependent due to the different internal energies and formation enthalpies of each species. Therefore, under the framework of single-temperature formation, the total diffusive enthalpy flux cannot be fully represented using a unified mixture enthalpy evaluated from the effective temperature. Accordingly, the diffusive enthalpy flux is decomposed into a Fick-law contribution evaluated from the effective temperature and an additional higher-order correction term extracted from DSMC, i.e.,
\begin{equation}
    \sum_s h_s\bm{\Phi}_s=\sum_sh_s(T_\text{eff})\bm{\Phi}_s^{\text{NS}}+\underbrace{\left[\sum_sh_s^{\text{DSMC}}(T_{\text{tra},s},T_{\text{rot},s},T_{\text{vib},s})\bm{\Phi}_s^\text{DSMC}-\sum_sh_s(T_\text{eff}^*)\bm{\Phi}_s^{\text{NS*}}\right]}_{{\textbf{HoT}_{\bm{h\Phi}}}}.
\end{equation}
However, other diffusion-induced terms, including the diffusive shear stress tensor $\bm{\Pi}_{\text{sp}}$, the diffusion kinetic energy term $K_{\text{diff}}$, and the stress-diffusion work term $\sum_s \bm{\sigma}_s \cdot \bm{V}_s$, are directly extracted from DSMC simulations and remain constant during the solution of the macroscopic synthetic equations. Such a treatment enables the synthetic equations to retain the species-diffusion effects associated with diffusion velocities in rarefied regimes. In the continuum limit, where the diffusion velocities $\bm{V}_s$ gradually vanish, these additional terms naturally disappear, and the macroscopic equations recover the conventional single-temperature NS equations.

Again, given that NS‑type constitutive terms are added and subtracted in Eq.~\eqref{eq:PhiOmega}, the solution to the macroscopic synthetic equation coincides with that from DSMC once steady state is attained. Consequently, the precise formulation of the mass diffusion coefficient and chemical reaction rates is of minor importance within our DIG framework.

\section{The DIG method for chemical reacting flows} \label{sec:4}

In this section, the procedure for incorporating the updated macroscopic properties of individual species into the DSMC framework is described in detail.
First, a time-averaged sampling strategy is briefly introduced to reduce the statistical fluctuations inherent in DSMC. 
Next, a set of redistribution coefficients is constructed to recover the species-resolved macroscopic properties from the updated mixture properties obtained from the synthetic equations.
Finally, a particle modification procedure within each computational cell is briefly presented, which follows a similar treatment to our previous work~\cite{Luo2026AMS}.
The detailed formulations and implementation of these procedures are provided in the following subsections.

\subsection{Time-averaged sampling method}

In chemically reacting flows, continuous species production and consumption significantly alter the local particle population. This may induce pronounced statistical fluctuations in DSMC, especially for species with low number densities. Thus, the macroscopic properties determined in Eq.~\eqref{equ_2} cannot be directly applied when solving macroscopic synthetic equations. A time-averaged sampling method is required to reduce the thermal fluctuations~\cite{Jenny2010JCP}, i.e., the summation of the macroscopic variables
\begin{equation*}
   \bm{\Xi}_s=\left\{n_s,\rho_s,(\rho\bm{u})_s,(\rho E_\text{tra})_s,(\rho E_\text{rot})_s,(\rho E_\text{vib})_s,\bm{\sigma}_{s},\bm{q}_{\text{tra},s},\bm{q}_{\text{rot},s},\bm{q}_{\text{vib},s},\right\}, 
\end{equation*} for $s$ species can be obtained as follows:
\begin{equation}
\label{eq:exponentialweighted}
\begin{aligned}
    &\bm{\Xi}_s(t)=\frac{n_\text{a}-1}{n_\text{a}}\bm{\Xi}_s(t-\Delta t)+\frac{1}{n_\text{a}}\frac{m_sN_{\text{eff}}}{V_{\text{cell}}}\sum_{p=1}^{N_{s,p}}\bm{\zeta}_s(t),
\end{aligned}
\end{equation}
where 
\begin{equation}
    \bm{\zeta}_s=\left\{\frac{1}{m_s},1,\bm{v}_s,\frac{1}{2}v_s^2,\frac{I_{\text{r},s}}{m_s},\frac{I_{\text{v},s}}{m_s},\left(\bm{c}_s\bm{c}_s-\frac{c_s^2}{3}\text{\textbf{I}}\right),\frac{1}{2}\bm{c}_sc_s^2,\bm{c}_sI_{\text{r},s},\bm{c}_sI_{\text{v},s}\right\},  
\end{equation}
and $n_\text{a}$ is the number of sampling steps applied in the time-averaging process. Meanwhile, the higher-order terms $\text{HoT}$ of diffusion flux, shear stress and heat flux in Eqs~\eqref{eq:consititutiverelation} and~\eqref{eq:PhiOmega} are determined according to the time-averaged macroscopic properties determined above. 

Moreover, since the higher-order term associated with the chemical source term, $\text{HoT}_{\dot{\omega}_s}$, is evaluated based on the DSMC time step $\Delta t$. When $\Delta t$ becomes excessively small, the resulting sampling process is prone to amplified statistical fluctuations, which may in turn induce numerical instabilities when solving the macroscopic synthetic equations. To alleviate this issue, the higher-order chemical source term is time-averaged from the beginning of the simulation.


\subsection{Redistribution of momentum and energy}

Upon solving the synthesized equations, the updated macroscopic variables $\bm{W}^{n+1}$ are employed to drive subsequent particle evolution in the DSMC simulation until steady state is reached. Nevertheless, the macroscopic solutions only yield a single velocity and temperature for the gas mixture. Consequently, species‑resolved momentum and specific energy remain unknown after solving the macroscopic synthesized equations, with the sole exception of the species mass density $\rho_s^{n+1}$. To retrieve these missing quantities, the updated momentum and energy of each species within a computational cell are reconstructed via the pre‑defined scaling and partitioning factors.

After performing a prescribed number of DSMC steps, the time-averaged conservative variables $\bm{W}^{n+\frac{1}{2}}$ are evaluated as
\begin{equation}
\label{eq:conservativevariables}
    \begin{aligned}
\rho^{n+\frac{1}{2}}&=\sum_{s=1}^N\rho_s^{n+\frac{1}{2}},\quad
(\rho Y_s)^{n+\frac{1}{2}}=\rho_s^{n+\frac{1}{2}},\quad
(\rho\bm{u})^{n+\frac{1}{2}} = \sum_{s=1}^N (\rho_s\bm{u}_s)^{n+\frac{1}{2}},\\
        (\rho E)^{n+\frac{1}{2}}& =\sum_{s=1}^N (\rho_sE_s)^{n+\frac{1}{2}}\\
        &= \sum_{s=1}^N \left[\underbrace{\frac{3}{2}R_s(\rho_sT_{\text{tra},s})^{n+\frac{1}{2}}+\frac{1}{2}(\rho_s|\bm{u}_s|^2)^{n+\frac{1}{2}}}_{(\rho_sE_{\text{tra},s})^{n+\frac{1}{2}}}+(\rho_sE_{\text{rot},s})^{n+\frac{1}{2}}+(\rho_sE_{\text{vib},s})^{n+\frac{1}{2}}\right],
    \end{aligned}
\end{equation}
where the translational, rotational, and vibrational energy components are explicitly considered for different species. Moreover, based on these quantities, the scaling factors for the momentum $\bm{K}_s^{\bm{u}*}$ and total energy $K_s^E$ of each species are defined as 
\begin{equation}
    \label{eq:scalingfactor}
    \begin{aligned}
    \bm{K}_{s}^{\bm{u}*}& = \begin{pmatrix}
\frac{(\rho_su_{s,x})^{n+\frac{1}{2}}}{\sum_{s=1}^N (\rho_su_{s,x})^{n+\frac{1}{2}}} & 0  \\
0 & \frac{(\rho_su_{s,y})^{n+\frac{1}{2}}}{\sum_{s=1}^N (\rho_su_{s,y})^{n+\frac{1}{2}}} \\
\end{pmatrix}
,\\
K_s^{E*} & = \frac{(\rho_sE_s)^{n+\frac{1}{2}}}{\sum_{s=1}^N (\rho_sE_s)^{n+\frac{1}{2}}}.
\end{aligned}
\end{equation}
Since the translational, rotational, and vibrational energy modes are treated separately for each species, the corresponding energy partitioning factors are introduced as
\begin{equation}
    \label{eq:energyfactor}
    \bm{A}^*=\left[A_\text{tra}^*, 
    A_\text{rot}^*,  A_\text{vib}^*\right]=\left[
    \frac{(\rho_sE_{\text{tra},s})^{n+\frac{1}{2}}}{(\rho_sE_s)^{n+\frac{1}{2}}}, ~
    \frac{(\rho_sE_{\text{rot},s})^{n+\frac{1}{2}}}{(\rho_sE_s)^{n+\frac{1}{2}}}, ~
    \frac{(\rho_sE_{\text{vib},s})^{n+\frac{1}{2}}}{(\rho_sE_s)^{n+\frac{1}{2}}}\right].
\end{equation}

Then, the species momentum and the translational, rotational, and vibrational energy components are evaluated as
\begin{equation}
\label{eq:momentumenergyafter}
\begin{aligned}
    (\rho_s\bm{u}_s)^{n+1}&=\bm{K}_s^{\bm{u}*}(\rho\bm{u})^{n+1}, \\
    (\rho_sE_{\text{tra},s})^{n+1}&= K_s^{E*} \cdot A_\text{tra}^*\cdot(\rho E)^{n+1}, \\
    (\rho_sE_{\text{rot},s})^{n+1}&= K_s^{E*} \cdot A_\text{rot}^*\cdot(\rho E)^{n+1}, \\
    (\rho_sE_{\text{vib},s})^{n+1}&= K_s^{E*} \cdot A_\text{vib}^*\cdot(\rho E)^{n+1}. \\
\end{aligned}
\end{equation}
Consequently, the species-resolved macroscopic quantities $\bm{M}^{n+1}_s$ can be determined and utilized in the subsequent step to modify the particle information in the DSMC simulation.




\subsection{Modification of particle distribution}

After obtaining the macroscopic properties $\bm{M}^{n+1}$ for each species $s$, these properties are fed back into the DSMC framework to bypass unnecessary intermediate particle evolutions before the steady state. This feedback procedure is implemented in two steps. Firstly, the target number of simulated particles for species $s$ within each computational cell is determined based on the updated number density $\rho_s^{n+1}$ as
\begin{equation}
    N_{s,p}^{n+1}=\mathrm{Iround}\left( \frac{\rho_s^{n+1}V_{\text{cell}}}{N_\text{eff}}\right),\quad
    \mathrm{Iround}(x)=\begin{cases}
        \left \lfloor x \right \rfloor+1,  & \text{ with probability } x-\left \lfloor x \right \rfloor, \\
        \left \lfloor x \right \rfloor,  & \text{ with probability } 1-x+\left \lfloor x \right \rfloor,
        \end{cases}
\end{equation}
where $\left \lfloor x \right \rfloor$ is the integer part of $x$. If $N_{s,p}^{n+1} > N_{s,p}^{n}$, a total of $N_{s,p}^{n+1} - N_{s,p}^{n}$ new particles are introduced into the cell. Their velocities and internal energies are assigned by randomly replicating those of existing particles of the same species within the cell, while their spatial positions are sampled uniformly over the cell volume. Conversely, if $N_{s,p}^{n+1} < N_{s,p}^{n}$, a total of $N_{s,p}^{n} - N_{s,p}^{n+1}$ particles are randomly selected and deleted from the current cell.

Secondly, after adjusting the number of simulated particles to match the updated number density $\rho_s^{n+1}$, the temporary macroscopic quantities within each computational cell, including the velocity $\bm{u}_s^{*}$, translational temperature $T_{\text{tra},s}^{*}$, rotational temperature $T_{\text{rot},s}^{*}$, and vibrational temperature $T_{\text{vib},s}^{*}$, are evaluated by sampling the current particle distribution according to Eq.~\eqref{equ:sampleMar}. Subsequently, the properties of the simulated particles of species $s$ are scaled through a linear transformation, which can be expressed as,
\begin{equation}
\label{eq:linearmodification}
\begin{aligned}
    \bm{v}_s^{n+1} & =\sqrt{\frac{T_{\text{tra},s}^{n+1}}{T_{\text{tra},s}^*}}(\bm{v}^{n}_s-\bm{u}_s^*)+\bm{u}_s^{n+1},\\
    I_{\text{r},s}^{n+1} & =\frac{T_{\text{rot},s}^{n+1}}{T_{\text{rot},s}^*}I_{\text{r},s}^{n},\\
    I_{\text{v},s}^{n+1} &=\frac{\exp(\Theta_v/T_{\text{vib},s}^*)-1}{\exp(\Theta_v/T_{\text{vib},s}^{n+1})-1}I_{\text{v},s}^{n}.
\end{aligned}
\end{equation}
Overall, the particle replication and deletion procedures adjust the number of simulated particles to ensure consistency with the updated number density predicted by the macroscopic synthetic equations. Subsequently, the scaling procedure modifies the particle velocities and internal energies so that the mean flow velocity and the translational, rotational, and vibrational temperatures are consistent with the updated macroscopic properties. Since the macroscopic synthetic equations are formulated to preserve mass, momentum, and energy, these conservation laws are maintained after the particle information is fully adjusted.

\begin{figure}[t]
	\centering
    \includegraphics[scale=0.6]{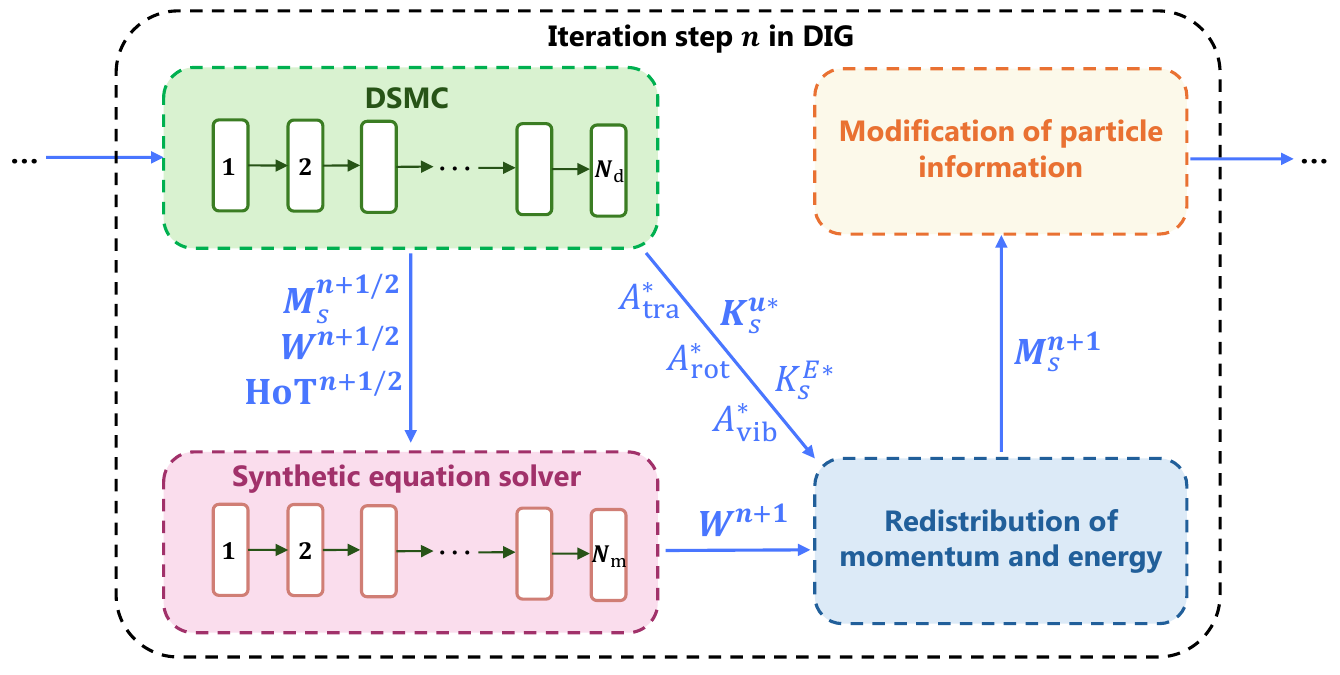}
	\caption{Flowchart of the DIG algorithm for chemically reacting flows. In each iteration, the DSMC method is first executed for $N_\text{d}=100$ steps. Subsequently, the single-velocity and single-temperature macroscopic synthetic equations with chemical reactions are solved for $N_\text{m}=200 \sim 1000$ steps, or until the maximum relative change of macroscopic properties falls below $10^{-5}$.}
	\label{fig:3}
\end{figure}

\subsection{General algorithm}

The general flowchart of the DIG method for simulating chemically reacting flows is illustrated in Fig.~\ref{fig:3}. Within the unit circle, the conventional DSMC method coupled with the QK chemical reaction model is performed for a prescribed number of steps, denoted by $N_d$, during which the time-averaged macroscopic properties ($\bm{M}_s^{n+\frac{1}{2}}=[\rho_s^{n+\frac{1}{2}},\,\bm{u}_s^{n+\frac{1}{2}},\,T_s^{n+\frac{1}{2}},\,Y_s^{n+\frac{1}{2}}]$), conservative macroscopic variables $\bm{W}^{n+\frac{1}{2}}=[\rho^{n+\frac{1}{2}},\,(\rho\bm{u})^{n+\frac{1}{2}},\,(\rho E)^{n+\frac{1}{2}},\,(\rho Y_s)^{n+\frac{1}{2}}]$ and higher-order terms (HoT) are sampled. Based on the sampled macroscopic properties $\bm{M}_s$, the redistribution coefficients are subsequently evaluated. The synthetic equations are subsequently solved using a time-implicit finite-volume scheme, which is detailed in~\ref{Finite_volume_synthetic}, with boundary condition imposed according to the Grad-13 moment equation~\cite{liu2024further}.
After the solution procedure, the updated conservative variables $\bm{W}^{n+1}$ are obtained. Based on these variables and the redistribution coefficients, the momentum and energy for each species are redistributed, and the updated macroscopic properties $\bm{M}_s^{n+1}$ can be subsequently obtained. 
These macroscopic properties are then used to modify the particle information for each species, thereby adjusting the particle distribution and accelerating its convergence toward the steady state. The whole DIG method for chemically reacting flows is summarized in Algorithm~\ref{algo:DIG_chem}.

\begin{algorithm}[!t]
    \caption{Overall algorithm of DIG for chemically reacting flows} 
    \label{algo:DIG_chem}
    \begin{algorithmic}[1]
        \Require
            Initial distribution of macroscopic properties $\bm{M}_s$ for species $s$;
        \Ensure
            Time-averaged macroscopic properties $\bm{M}_s$ after steady state;
        \State Initialize $N_{s,p}$ simulation particles for species $s$
        within each computational cell based on the initial macroscopic properties and Maxwellian distribution;
        \State Run standard DSMC for $n_\text{a}$ steps to obtain sufficient samples;
        \State Set iteration step $n = 1$;
        \While {$n \le \text{MaxSteps}$}
            \State Run standard DSMC for $N_d$ time steps;
            \State Calculate the time-averaged macroscopic properties $\bm{M}_s^{n+\frac{1}{2}}$ based on Eq.~\eqref{eq:exponentialweighted};
            \State Determine the conservative variables $\bm{W}^{n+\frac{1}{2}}$ based on Eq.~\eqref{eq:conservativevariables};
            \State Obtain the high-order terms $\text{HoT}^{n+\frac{1}{2}}$ according to Eqs.~\eqref{eq:consititutiverelation} and~\eqref{eq:PhiOmega};
            \State Compute redistribution coefficients $\bm{K}^{\bm{u}*}_s$, $K_s^{E*}$ and $\bm{A}^*$ based on Eqs.~\eqref{eq:scalingfactor} and \eqref{eq:energyfactor};
            \State Solve Eqs.~\eqref{eq:equ_NS} by $N_\text{m}$ steps to obtain $\bm{W}^{n+1}$;
            \State Redistribute momentum and energy for each species to obtain $\bm{M}_s^{n+1}$ based on~\eqref{eq:momentumenergyafter};
            \State Replicating and discarding particles to match species mass density $\rho_s^{n+1}$;
            \State Particle velocity scaling to match $\bm{u}_s^{n+1}$, $T_{\text{tra},s}^{n+1}$, $T_{\text{rot},s}^{n+1}$, $T_{\text{vib},s}^{n+1}$ based on Eq.~\eqref{eq:linearmodification} ;
            \State $n ++$;
        \EndWhile
    \end{algorithmic}
\end{algorithm}

\section{Numerical results}\label{sec:Num}
As discussed in the previous section, DIG requires a sufficient number of statistical samples to suppress the stochastic fluctuations when solving the macroscopic synthetic equations. In the present study, $n_a=500\sim1000$ samples are used, and the macroscopic synthetic equations are solved every $N_d=100$ steps. The benchmark DSMC results are obtained using the Stochastic PArallel Rarefied-gas Time-accurate Analyzer (SPARTA, \url{https://sparta.sandia.gov/}). We have modified the QK chemical reaction model in SPARTA to make it more consistent with analytic results. A comparison of the results before and after the modification is given in \ref{Appendix_sparta}. All simulations are performed on an Intel(R) Xeon(R) Gold 6148 CPU @ 2.40GHz processor.

To assess the performance of the proposed DIG method for chemically reacting flows, hypersonic dissociating flow over a cylinder is simulated. The incoming gas flow is nitrogen and modeled by the VHS collision model ($\alpha=1.0$), with the associated reaction mechanisms listed in Table~\ref{tab:Arrhenius}. Moreover, the nitrogen gas is taken as the reference species, with the reference molecular mass $m_0=4.65\times10^{-26}\,\text{kg}$ and the reference collision diameter $d_0=4.17\times10^{-10}\,\text{m}$. The reference temperature for defining the collision diameter is $T_0=273.15\,\text{K}$. Accordingly, the global Knudsen number is defined as
\begin{equation}
	\begin{aligned}
    \text{Kn}=\frac{\mu_{0}}{p_{0}L_\text{ref}}\sqrt{\frac{\pi k_BT_{0}}{2m_{0}}},
    \end{aligned}
    \label{equ:Kn_equ}
\end{equation}
where the reference viscosity $\mu_0$ is obtained at the reference temperature $T_0$ for nitrogen gas, and $L_{\text{ref}}$ denotes the characteristic length for different systems. The reference pressure $p_0$ is computed based on reference number density $n_0=2.685\times10^{25}\,\text{m}^{-3}$ and $T_0$ for nitrogen gas. It should be noted that all macroscopic properties in this section are non-dimensionlized by reference number density $n_0$, reference temperature $T_0$, reference velocity $c_0=\sqrt{2k_\text{B}T_0/m_0}$ and reference length $L_{\text{ref}}$. 
The rotational and vibrational collision numbers are $Z_\text{r}=3.24$ and $Z_\text{v}=32.4$, respectively.

\begin{figure}[!t]
\centering
\includegraphics[width=0.45\textwidth,trim=10pt 10pt 40pt 40pt,clip]{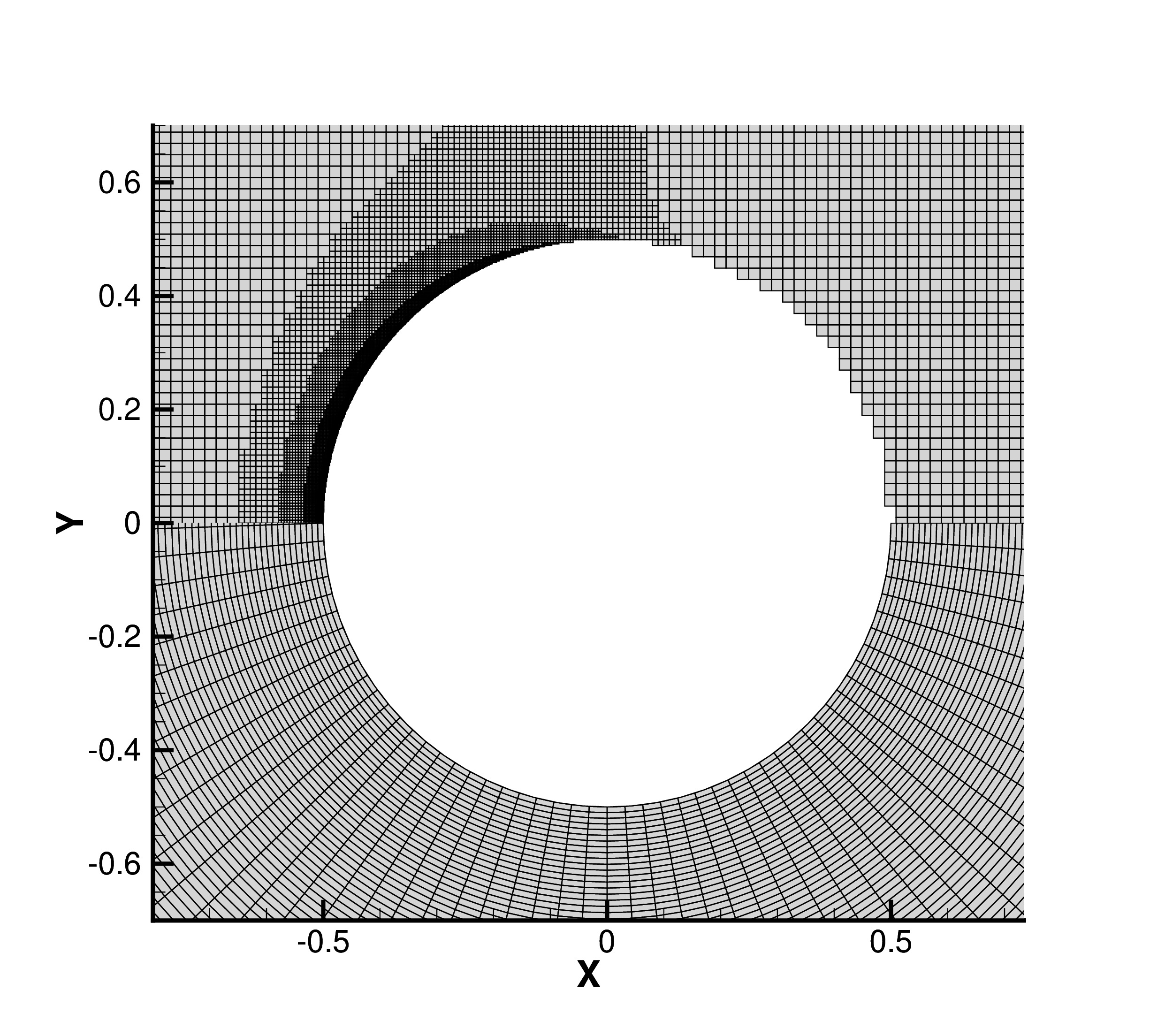}
\includegraphics[width=0.45\textwidth,trim=10pt 10pt 40pt 40pt,clip]{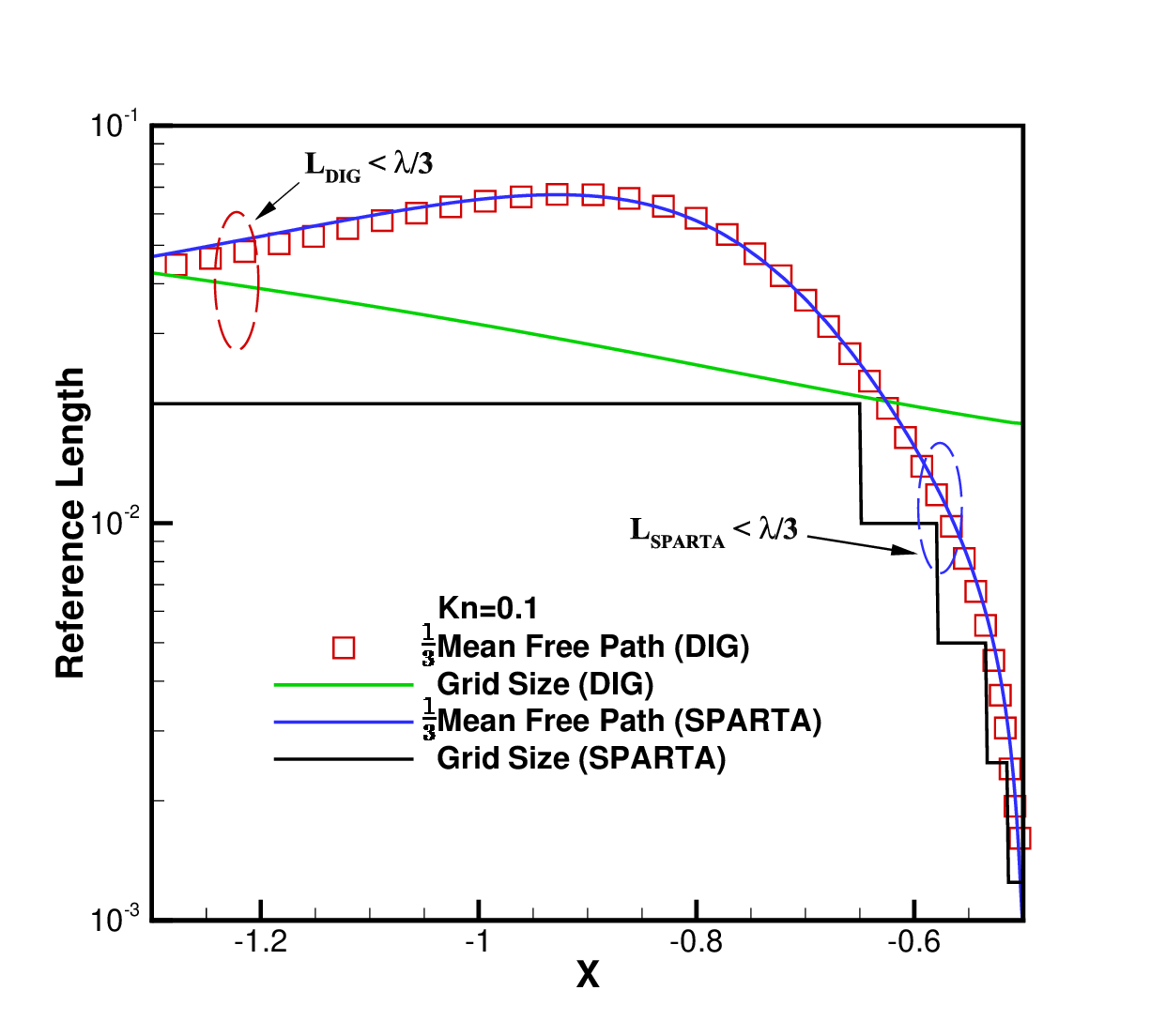}
\\
\vspace{-1.5mm}
\includegraphics[width=0.45\textwidth,trim=10pt 10pt 40pt 40pt,clip]{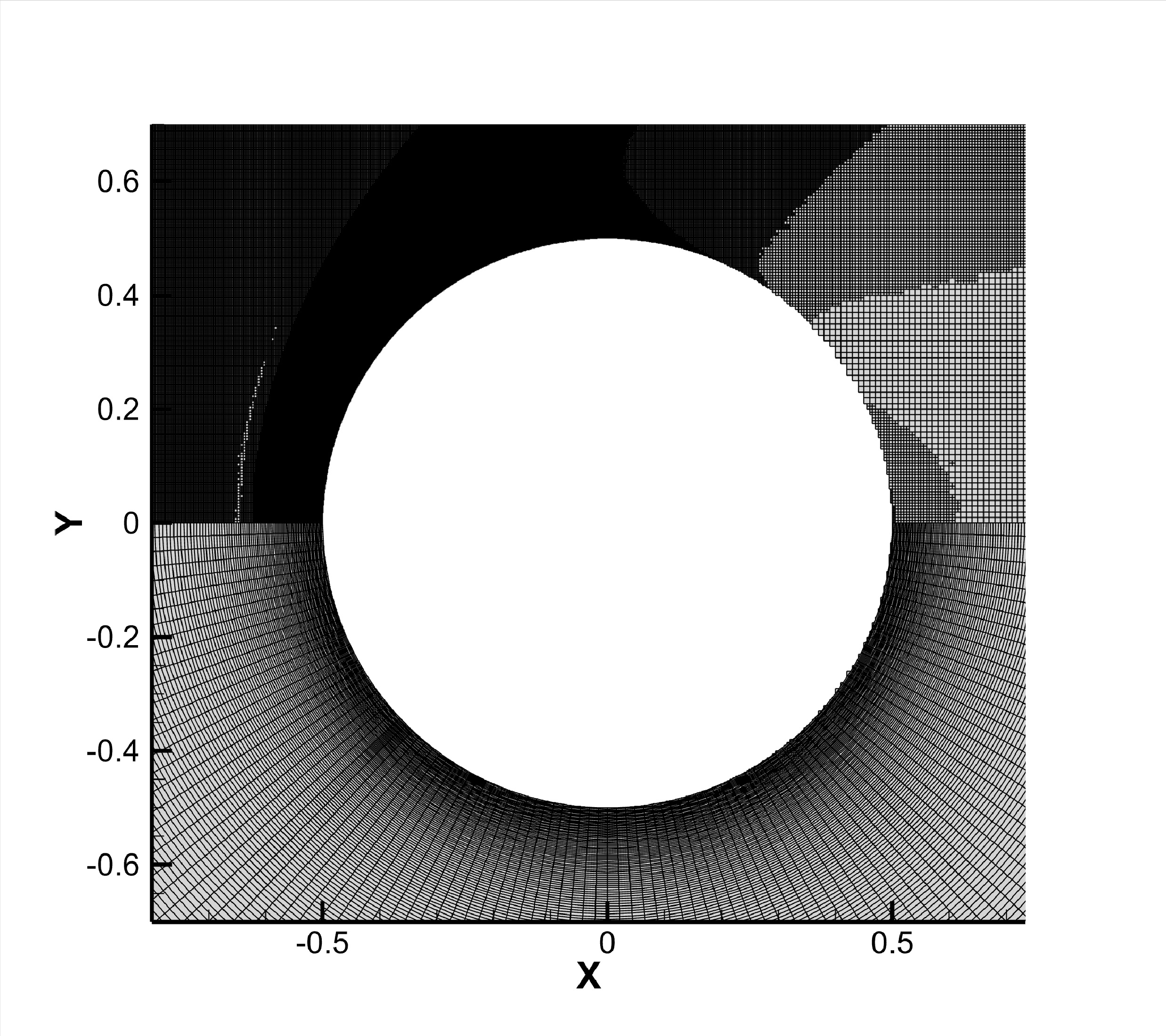}
\includegraphics[width=0.45\textwidth,trim=10pt 10pt 40pt 40pt,clip]{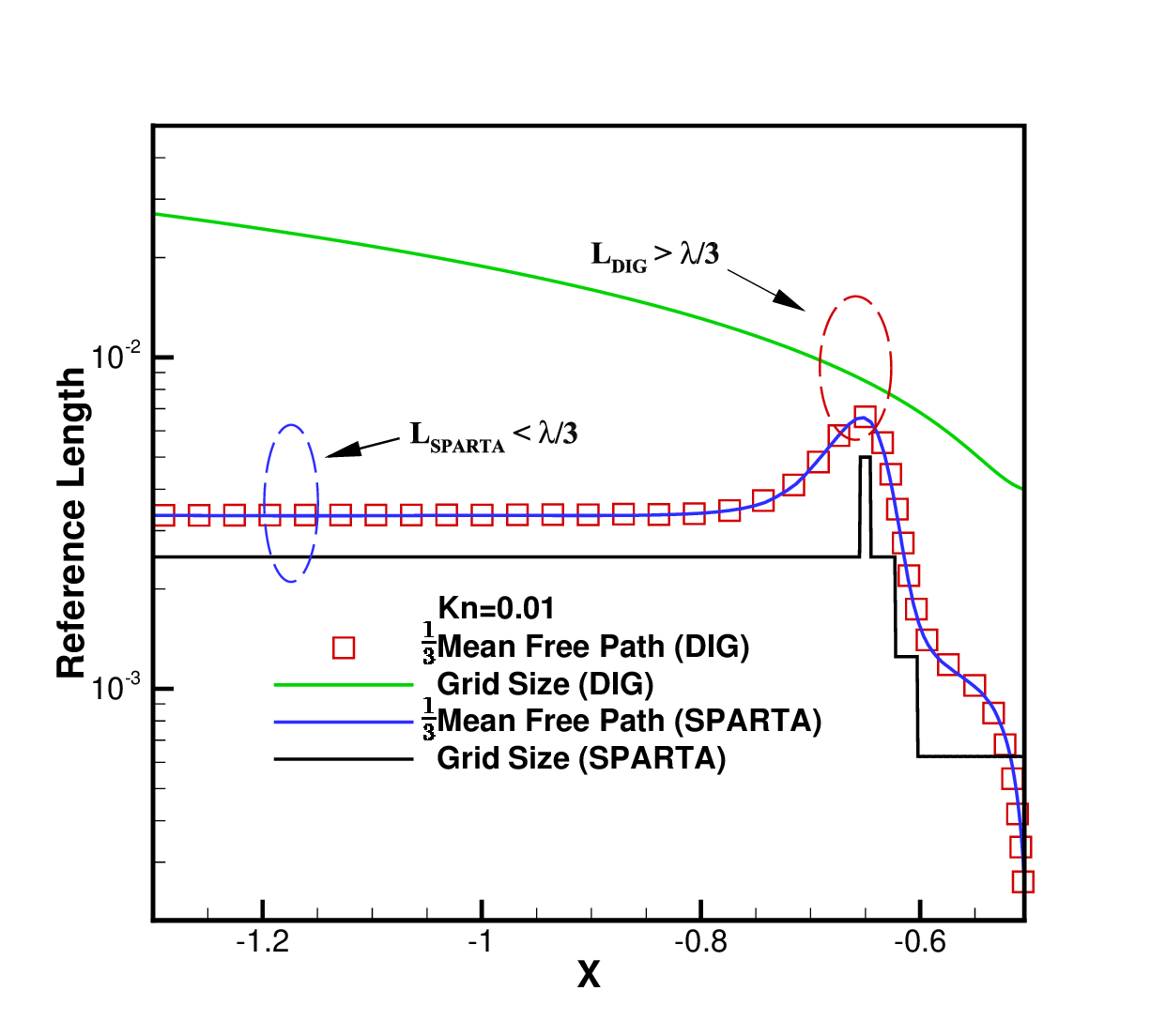}
\caption{Left column: the computational grids applied in SPARTA (upper half) and DIG (lower half) in the steady state. Right column: the comparison of the local mean free path with cell size in DIG and DSMC. The grid size is defined as the square root of the cell volume. The global Knudsen numbers are 0.1 and 0.01 for the top and bottom rows, respectively. Note that both the local mean free path and grid size are normalized by the characteristic length $L_0$. In SPARTA, the grid is adaptively refined such that the cell size remains smaller than one-third of the local mean free path of $\text{N}_2$. In addition, the time step $\Delta t$ applied in SPARTA satisfies the constraint that the product of $\Delta t$ and the most probable molecular speed is smaller than the smallest grid size. }
\label{fig:Ma20mesh}
\end{figure}

We evaluate the performance of DIG in two-dimensional hypersonic dissociating nitrogen gas flow passing over a cylinder. The global Knudsen numbers, defined based on nitrogen properties, are specified as 0.1 and 0.01. 
The freestream and wall temperatures are specified as $T_\infty=273.15\,\text{K}$ and $T_\text{w}=1000\,\text{K}$, respectively, thereby establishing a strong dissociating reaction in the shock wave region.
The freestream is assumed to consist of pure nitrogen with the number density of $n_0$, corresponding to a mass fraction $Y_1=1$.
Moreover, the freestream Mach number is 20, with the speed of sound defined as $a_\infty = \sqrt{\gamma k_B T_\infty / m_0}$, where $\gamma=1.4$ is the specific heat ratio of nitrogen gas. Initially, the number of simulated particles in each cell is set to $N_{s,p}=100$ to achieve a balance between numerical stability and computational efficiency.

\begin{figure}[!t]
\centering
\vspace{-1.5mm}
\hspace{-11mm}
\includegraphics[width=0.38\textwidth,trim=10pt 0pt 10pt 0pt,clip]{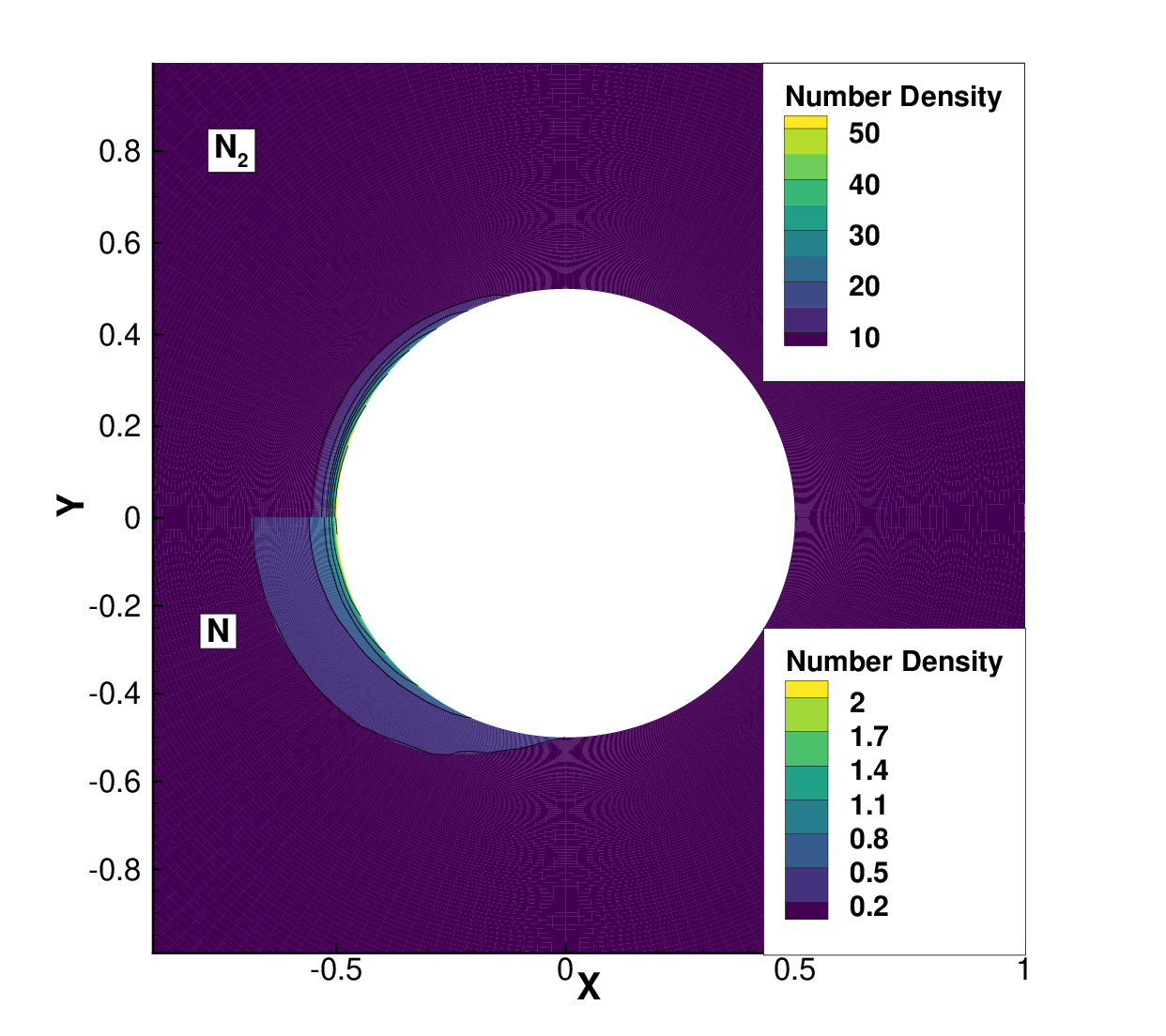}
\hspace{-9.5mm}
\includegraphics[width=0.38\textwidth,trim=10pt 0pt 10pt 0pt,clip]{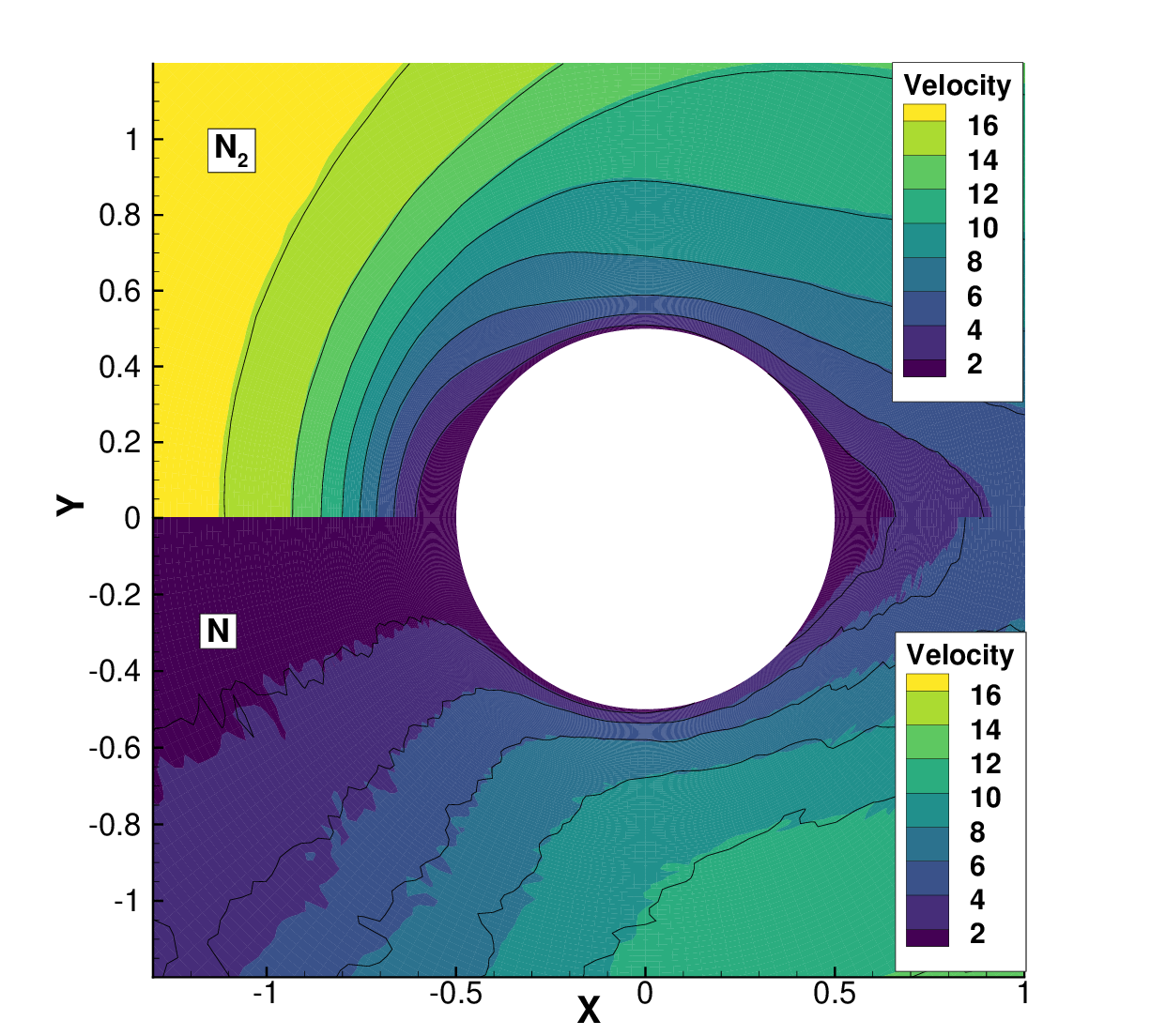}
\hspace{-9.5mm}
\includegraphics[width=0.38\textwidth,trim=10pt 0pt 10pt 0pt,clip]{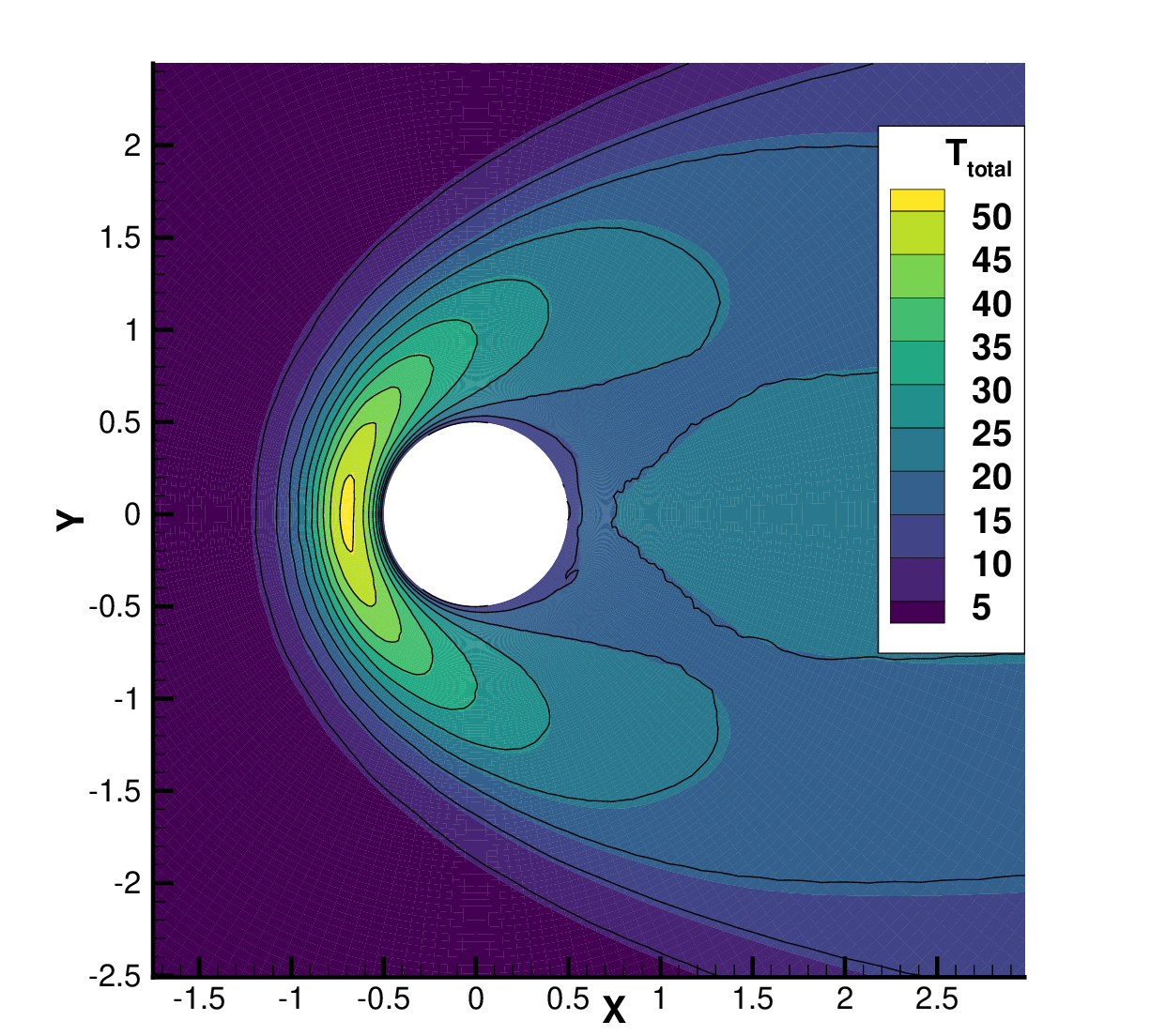}
\\
\hspace{-11mm}
\includegraphics[width=0.38\textwidth,trim=10pt 20pt 10pt 0pt,clip]{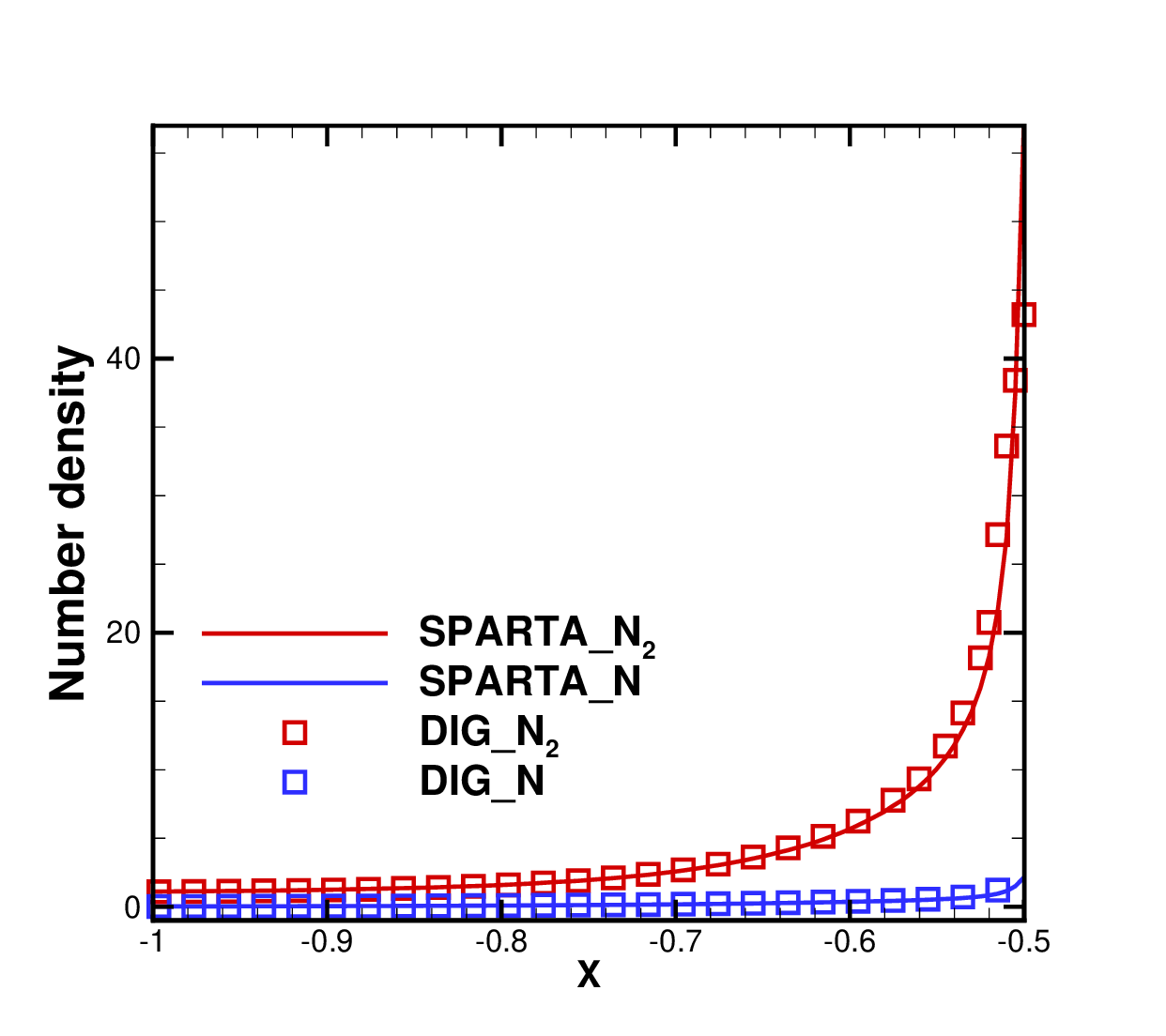}
\hspace{-9.5mm}
\includegraphics[width=0.38\textwidth,trim=10pt 20pt 10pt 0pt,clip]{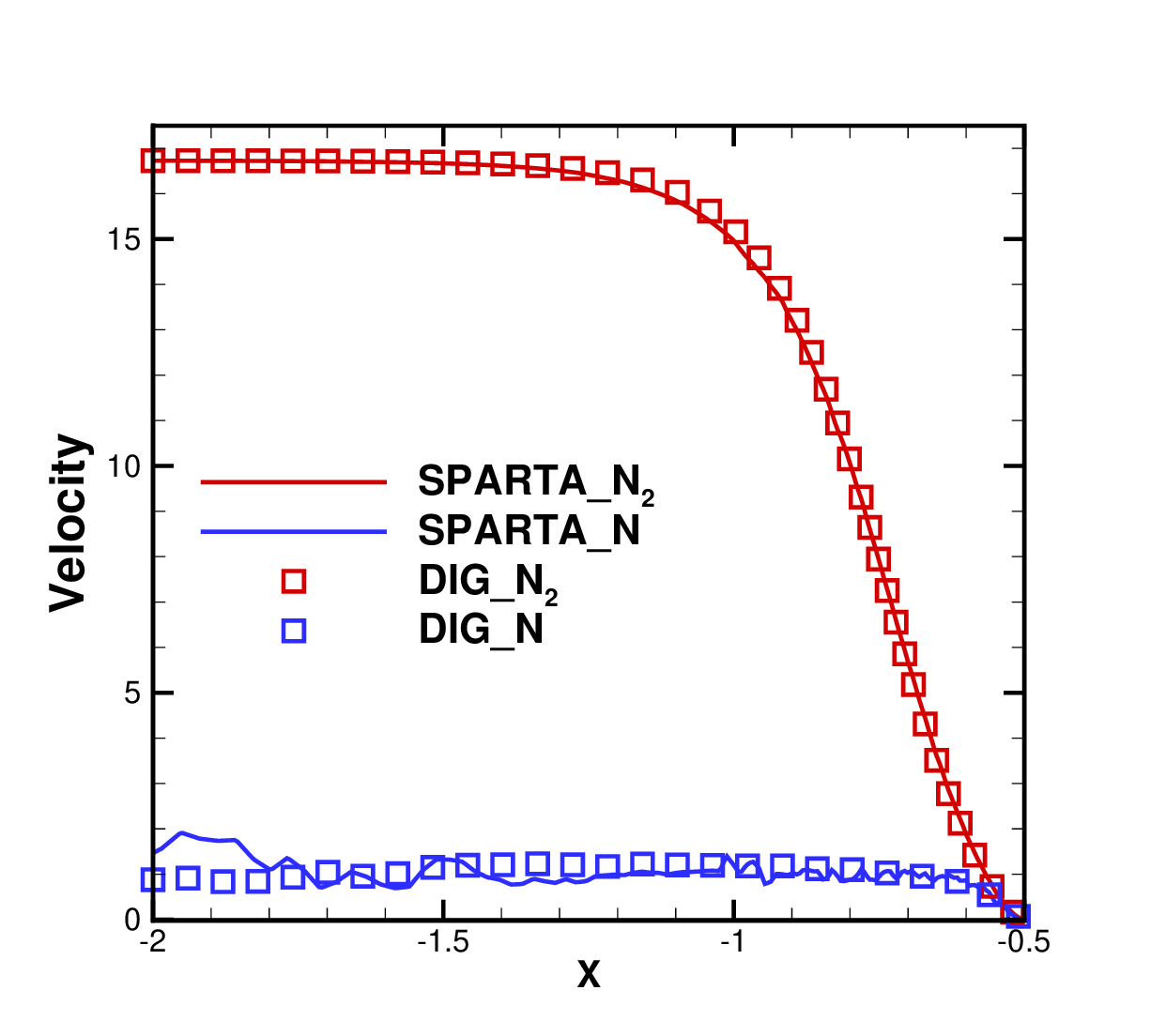}
\hspace{-9.5mm}
\includegraphics[width=0.38\textwidth,trim=10pt 20pt 10pt 0pt,clip]{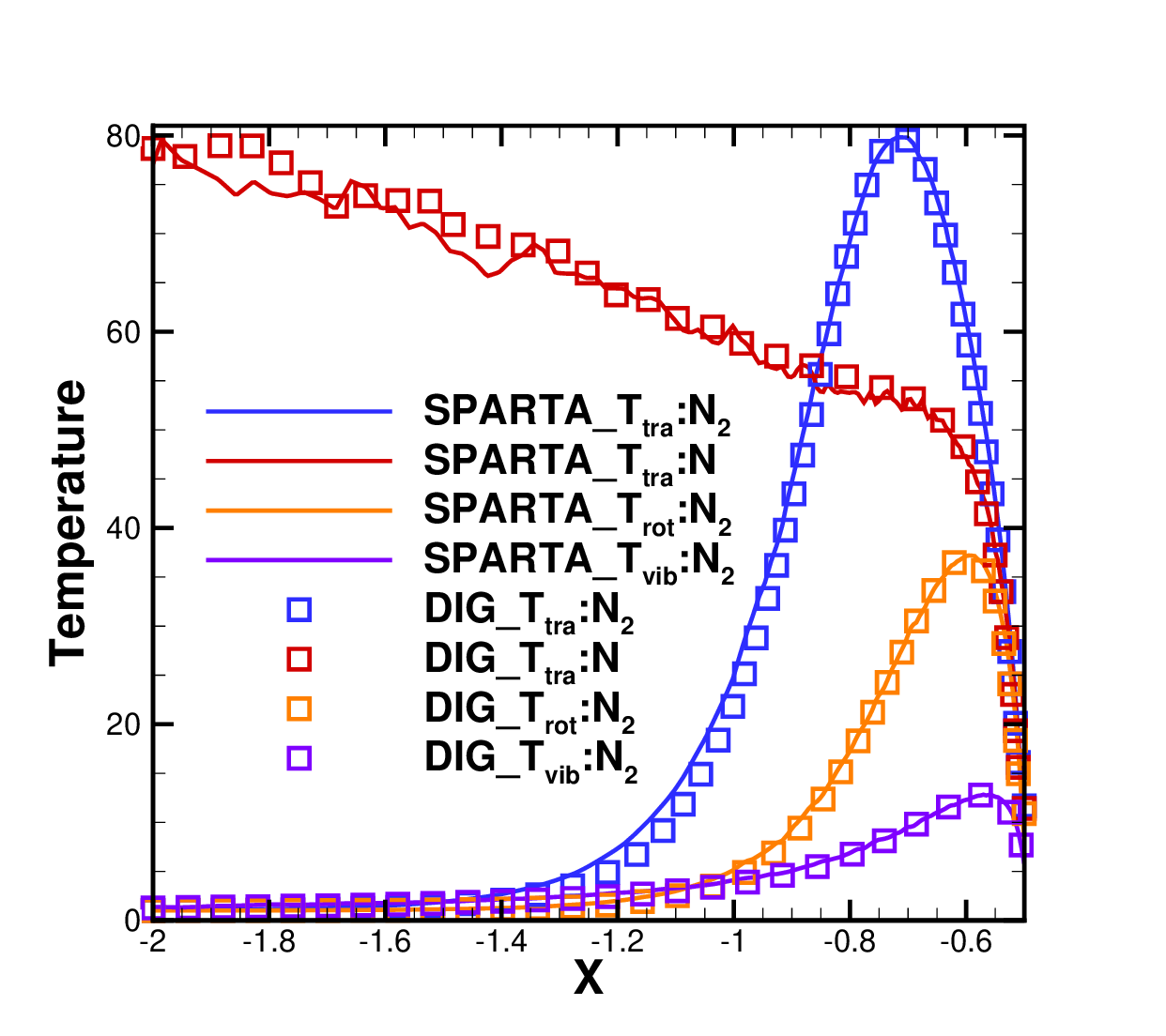}
\caption{Comparisons of macroscopic properties predicted by DIG (lines) and SPARTA (contours) for the incoming Mach number of 20 with global Knudsen numbers of 0.1. The bottom row represents the macroscopic properties along the stagnation stream line in the windward side of the cylinder.}
\label{fig:Contour_Kn01}
\end{figure}

The computational setup is defined with the cylinder diameter taken as the reference length $L_{\text{ref}}$ for both DIG and SPARTA. In DIG, the computational domain is constructed as an annular region, with the inner boundary corresponding to the cylinder surface and the outer boundary representing the far field. For $\text{Kn}=0.1$, the outer radius is set to $5.5L_{\text{ref}}$, while for $\text{Kn}=0.01$, it is reduced to $2.5L_{\text{ref}}$ due to the thinner shock layer. In SPARTA, the computational domain is defined as a square region $[-L,L]\times[-L,L]$, where $L/L_{\text{ref}}=5.5$ for $\text{Kn}=0.1$ and $L/L_{\text{ref}}=3$ for $\text{Kn}=0.01$.

Figure~\ref{fig:Ma20mesh} illustrates the computational domains and grid employed in both DIG and SPARTA. 
In the open-source DSMC code SPARTA, Cartesian grid cells are adaptively refined to ensure that the cell size remains smaller than one-third of the local mean free path. As a result, approximately 291,066 cells are generated for $\text{Kn}=0.1$, while the number increases to about $10^7$ for $\text{Kn}=0.01$. In contrast, the cell size adopted in DIG is significantly larger than the mean free path, particularly in the near-continuum regime. For instance, when $\text{Kn}=0.01$, the cell size near the cylinder surface is nearly 20 times larger than the local mean free path. Consequently, the total number of cells in DIG method is substantially reduced, with only 12,800 cells for $\text{Kn}=0.1$ and 40,000 cells for $\text{Kn}=0.01$.

\begin{figure}[!t]
\centering
\vspace{-1.5mm}
\hspace{-11mm}
\includegraphics[width=0.38\textwidth,trim=10pt 0pt 10pt 0pt,clip]{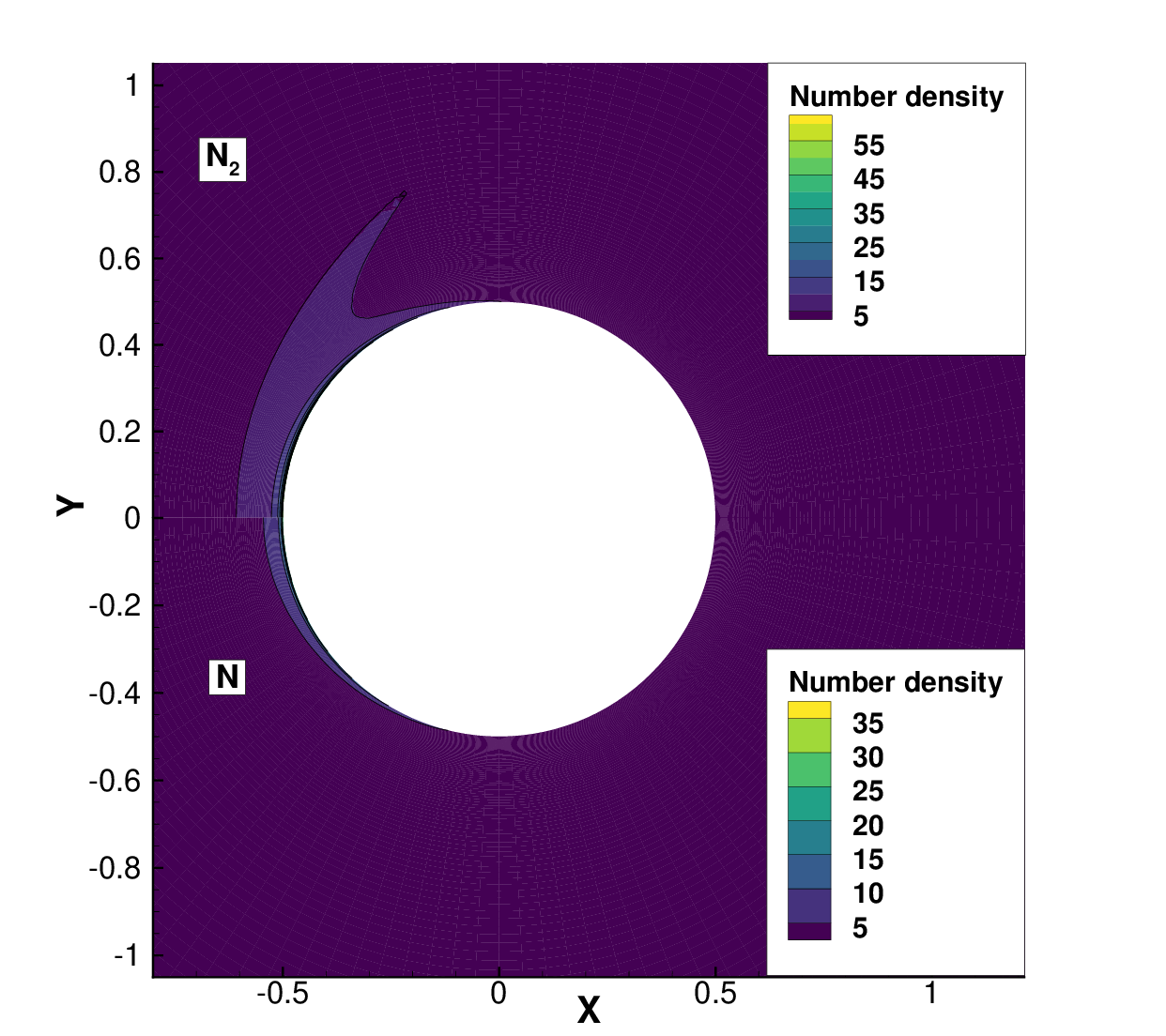}
\hspace{-9.5mm}
\includegraphics[width=0.38\textwidth,trim=10pt 0pt 10pt 0pt,clip]{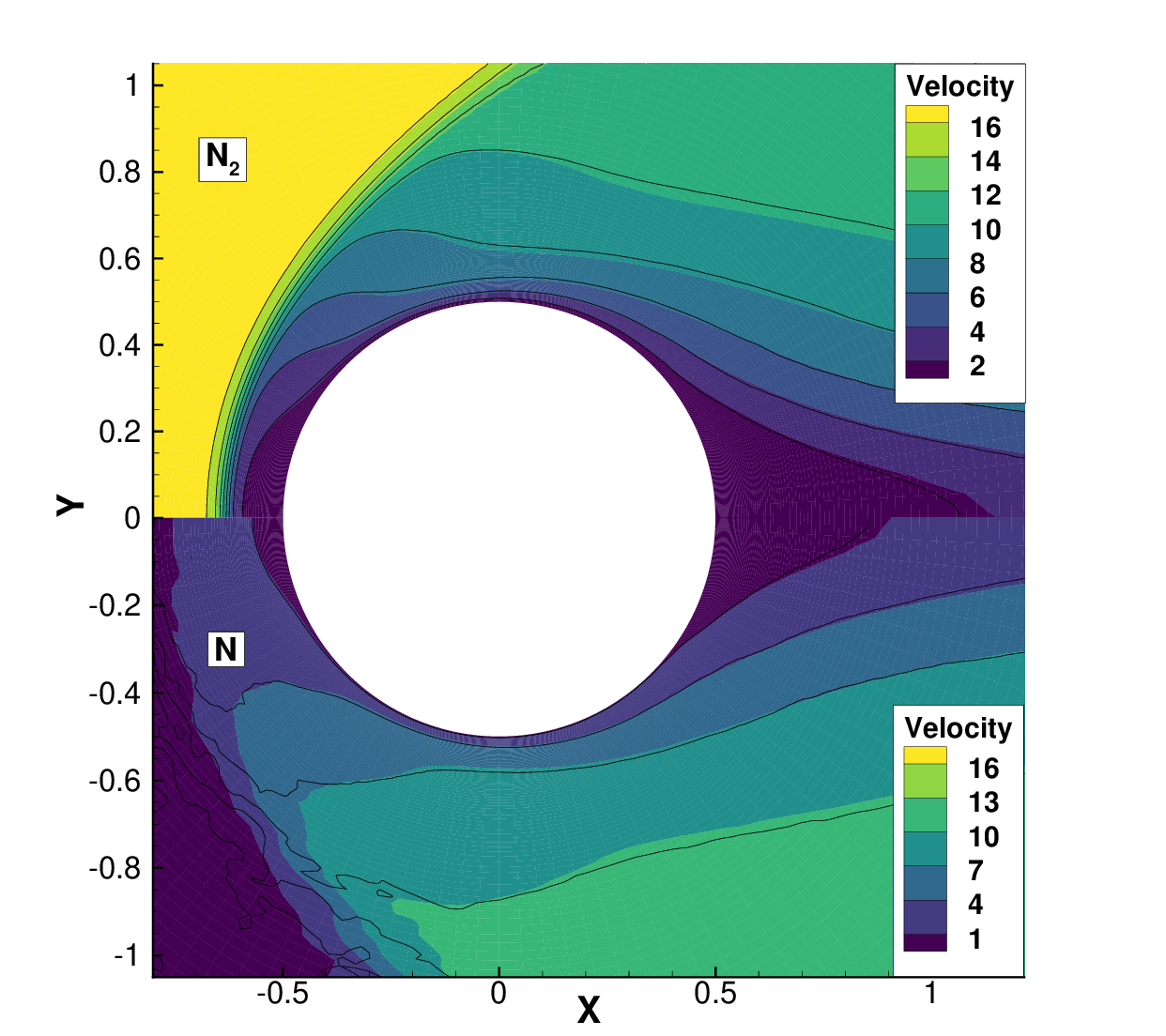}
\hspace{-9.5mm}
\includegraphics[width=0.38\textwidth,trim=10pt 0pt 10pt 0pt,clip]{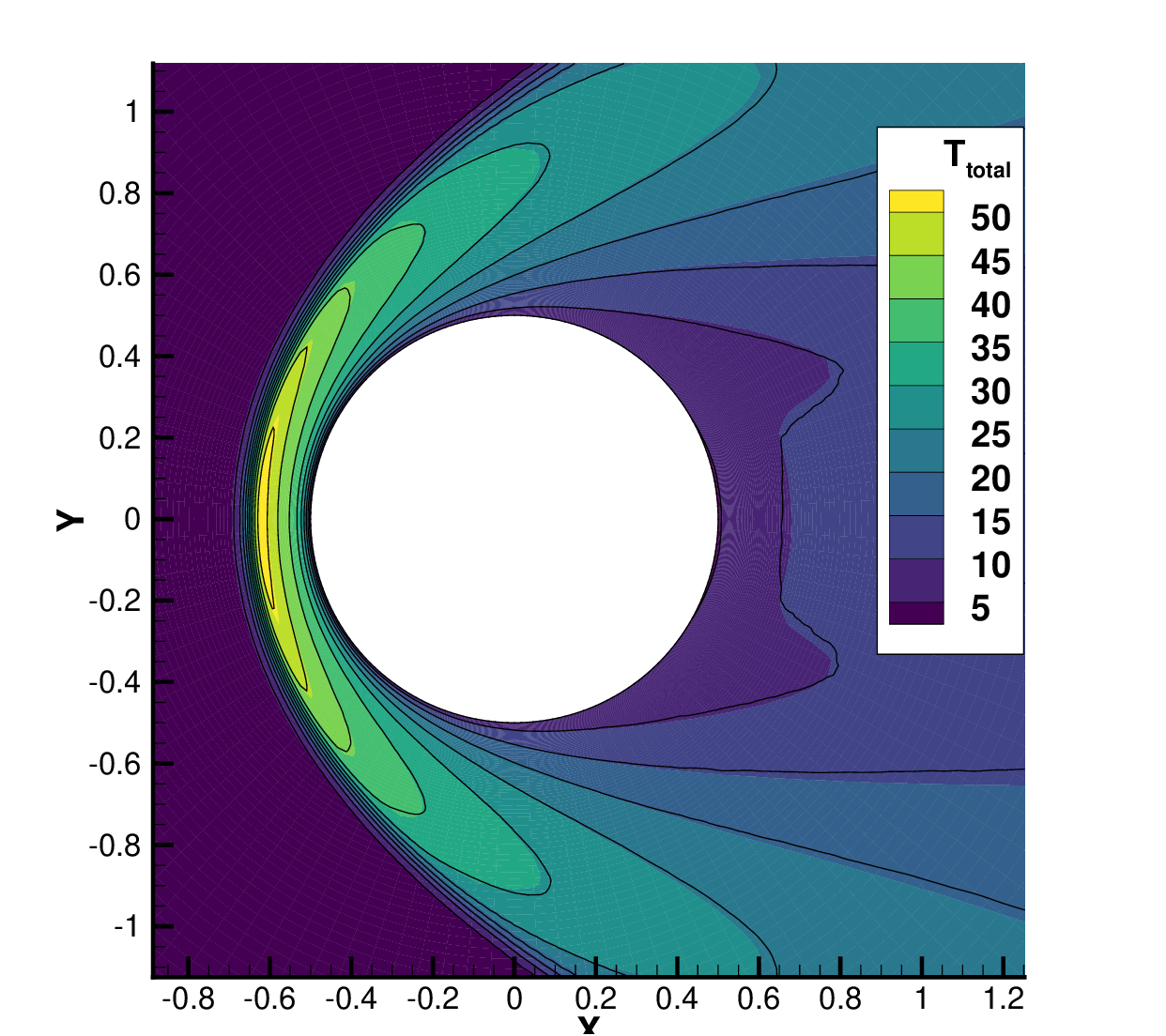}
\\
\hspace{-11mm}
\includegraphics[width=0.38\textwidth,trim=10pt 20pt 10pt 0pt,clip]{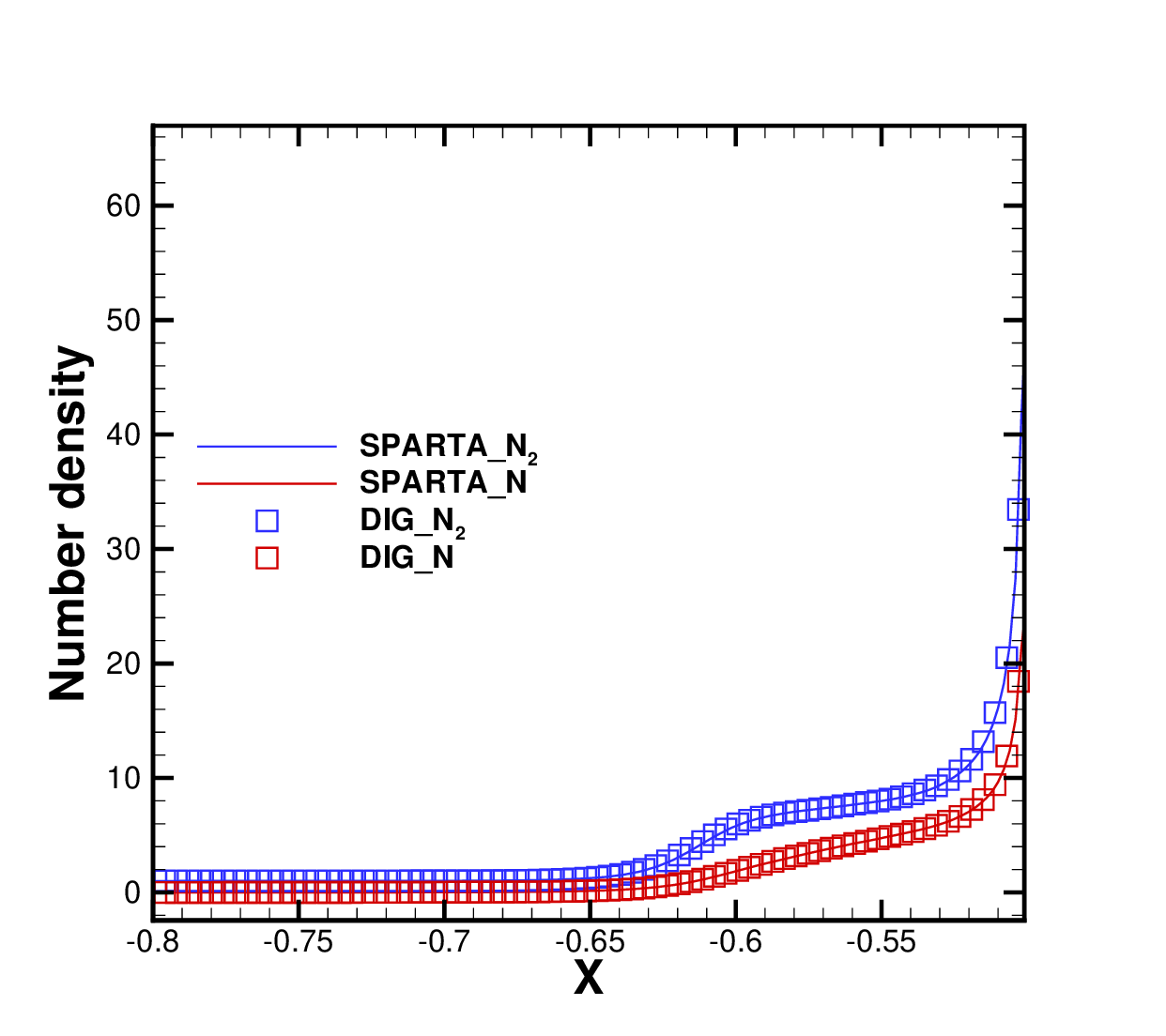}
\hspace{-9.5mm}
\includegraphics[width=0.38\textwidth,trim=10pt 20pt 10pt 0pt,clip]{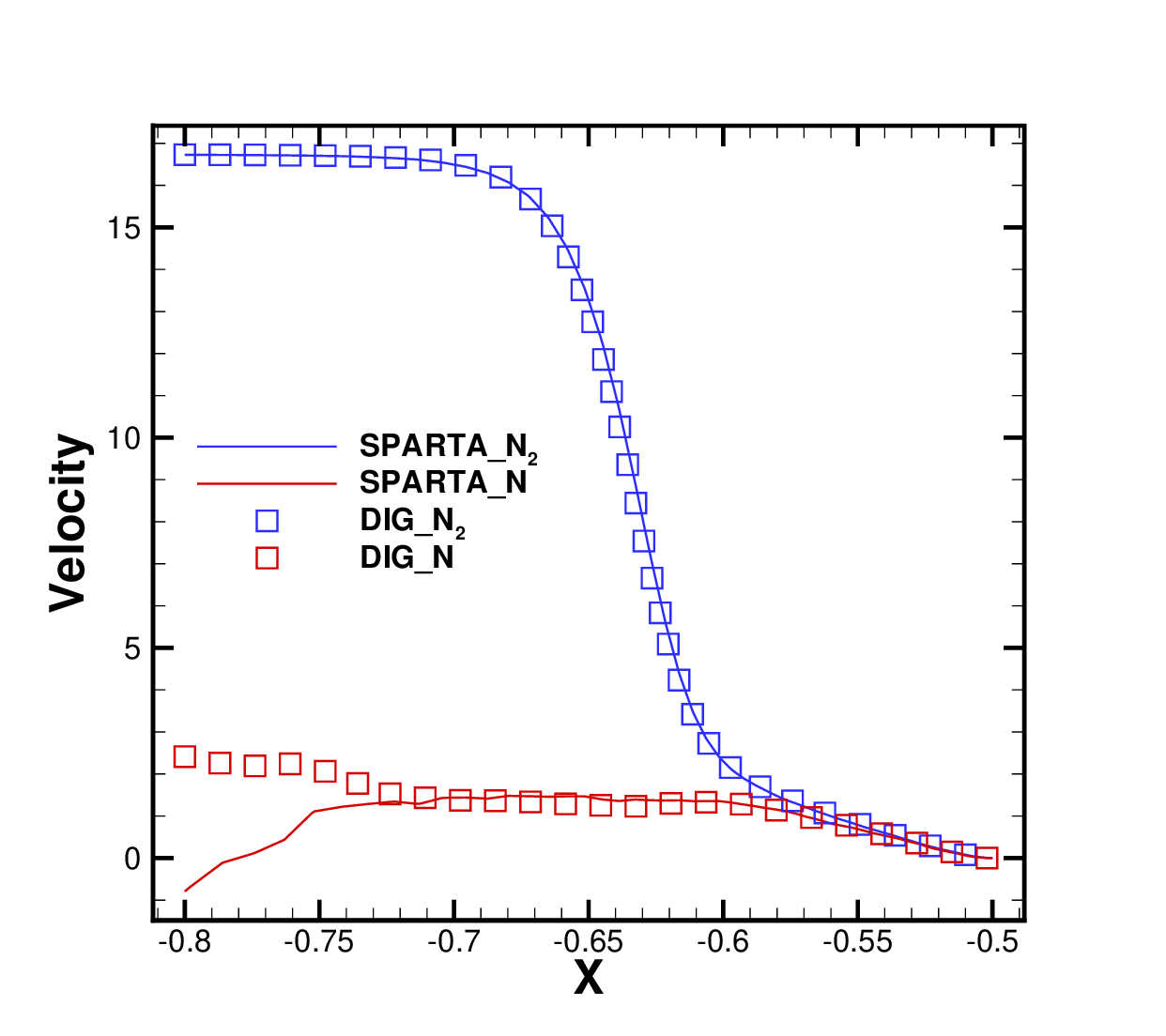}
\hspace{-9.5mm}
\includegraphics[width=0.38\textwidth,trim=10pt 20pt 10pt 0pt,clip]{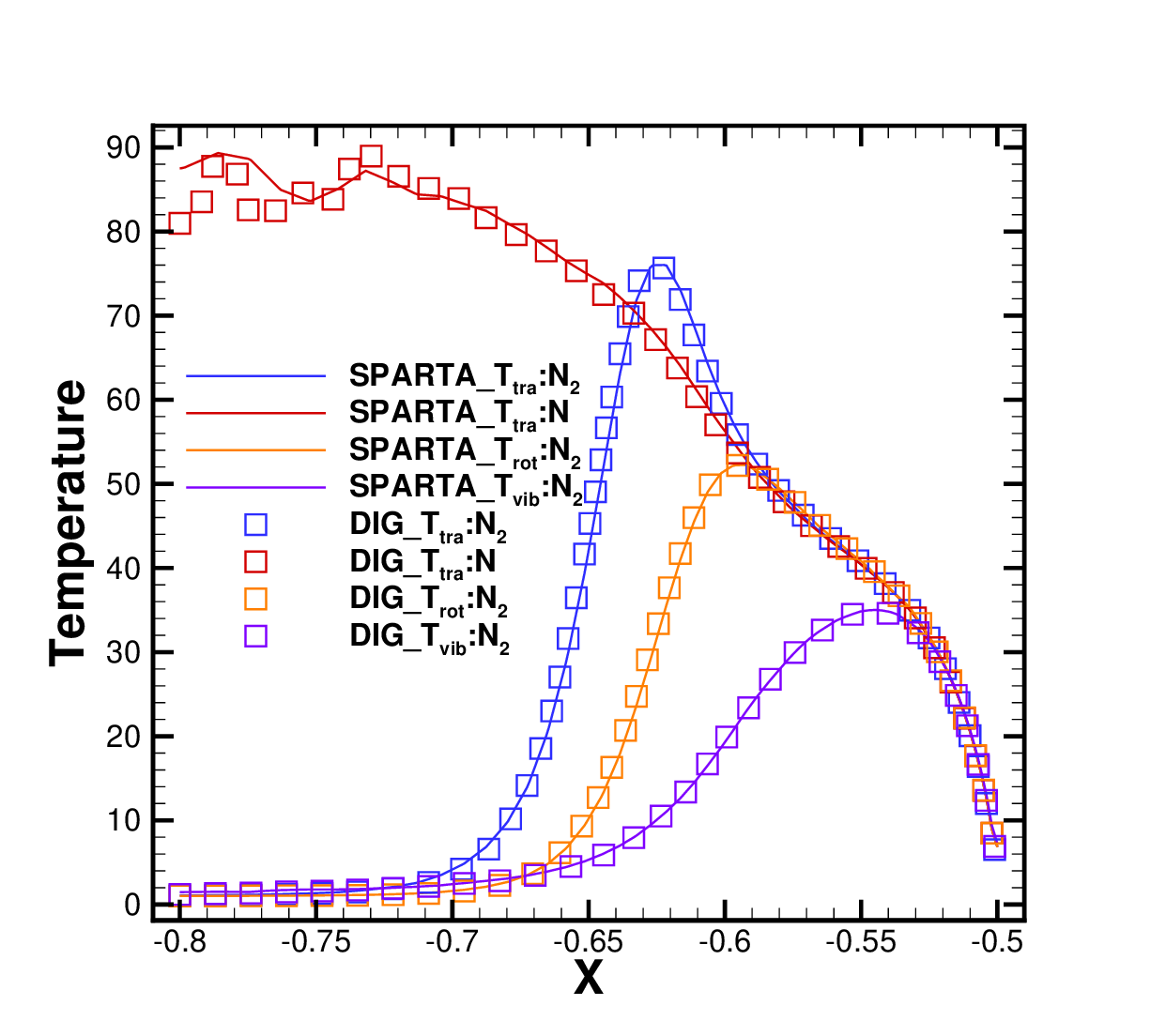}
\caption{Comparisons of macroscopic properties predicted by DIG (lines) and SPARTA (contours) for the incoming Mach number of 20 with global Knudsen numbers of 0.01. The bottom row represents the macroscopic properties along the stagnation stream line in the windward side of the cylinder.}
\label{fig:Contour_Kn001}
\end{figure}

Figure~\ref{fig:Contour_Kn01} presents a comparison of macroscopic flow properties predicted by DIG and SPARTA at $\text{Kn}=0.1$. Despite the temperature reaching approximately $53T_0$ in the shock region, the level of chemical reaction remains weak under rarefied conditions, leading to only a small amount of nitrogen dissociation. The low number density of nitrogen atoms upstream results in pronounced statistical fluctuations, which give rise to discrepancies in their velocity and temperature between DIG and SPARTA. These differences primarily stem from the limited in-situ production of nitrogen atoms near the stagnation region, rather than inflow transport. Overall, good agreement between DIG and SPARTA is observed for the major flow features, including density, velocity, and temperature distributions.

Figure~\ref{fig:Contour_Kn001} compares the macroscopic flow properties predicted by DIG and SPARTA at $\text{Kn}=0.01$. Compared with the case of $\text{Kn}=0.1$, the shock layer becomes significantly thinner and the dissociation of nitrogen molecules is substantially enhanced, leading to a much higher number density of nitrogen atoms. Due to the relatively low number density of nitrogen atoms in regions away from the shock, noticeable statistical fluctuations persist, which result in discrepancies in the predicted velocity and temperature between DIG and SPARTA in these regions. However, within the shock layer, where the number density is sufficiently high, the macroscopic properties predicted by the two methods exhibit good agreement.

\begin{figure}[t]
	\centering
    \includegraphics[width=0.48\linewidth,trim={10 20 30 50},clip]{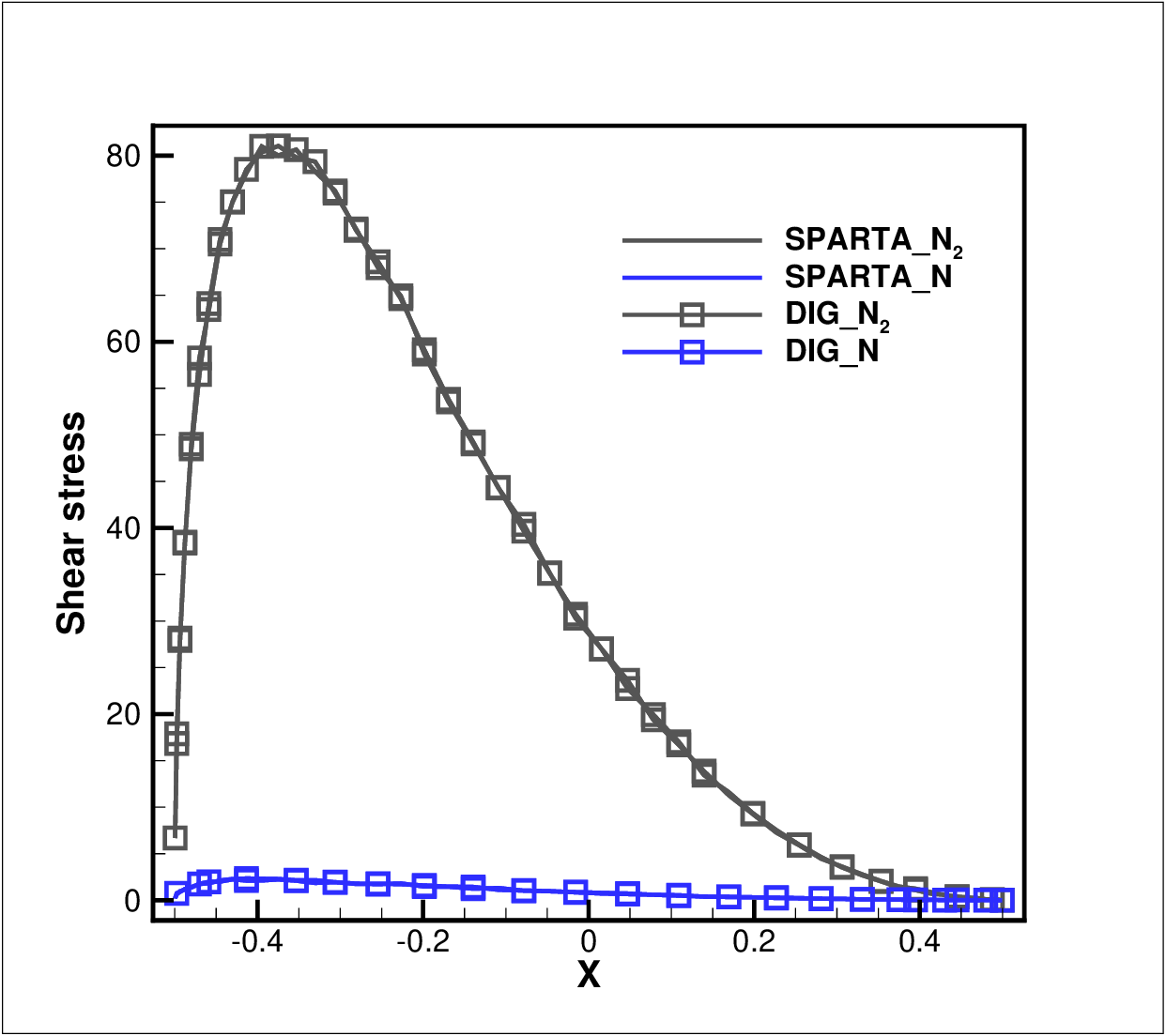}
    \hspace{0.1cm}
	\includegraphics[width=0.48\linewidth,trim={10 20 30 50},clip]{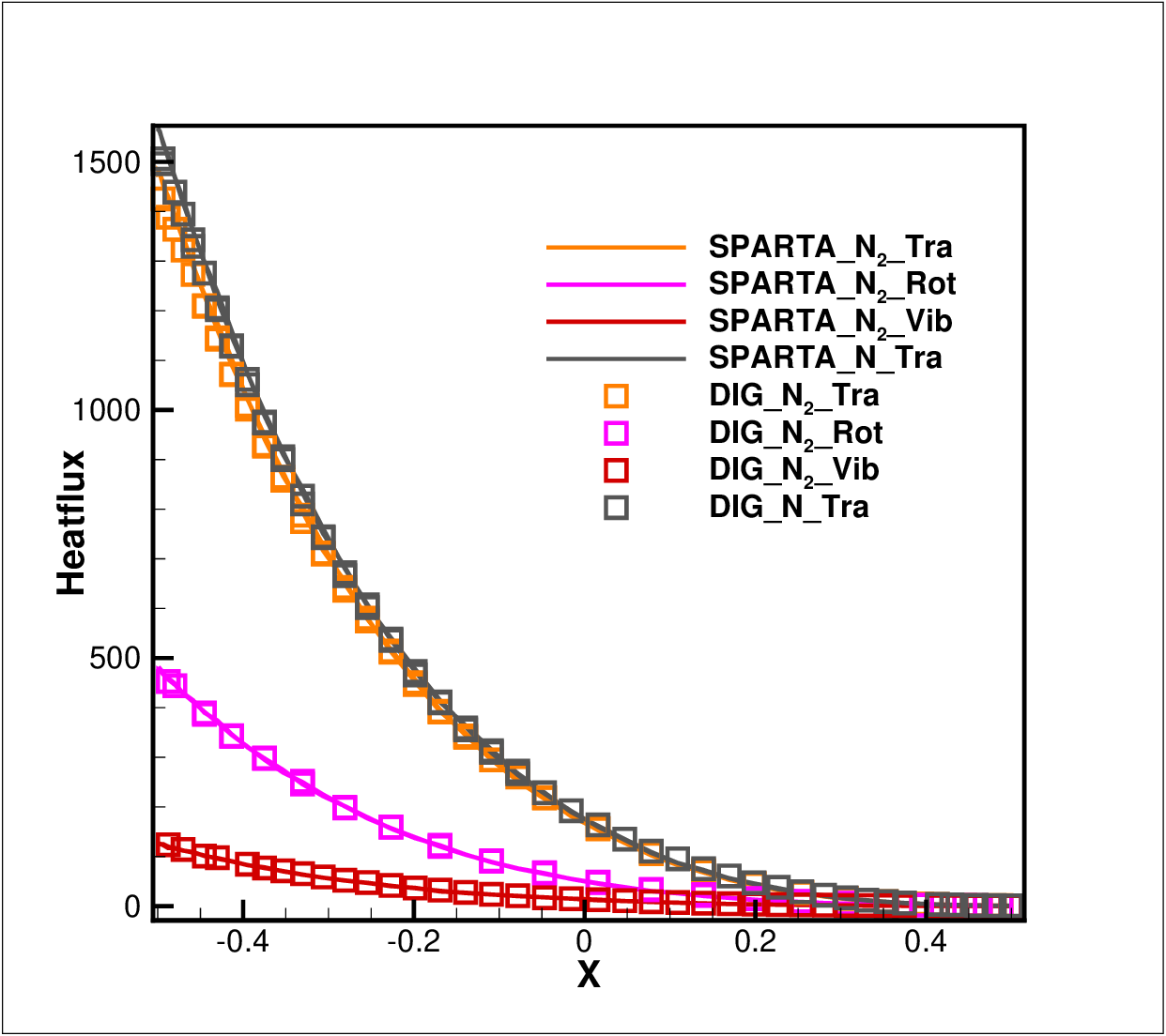}\\
	\includegraphics[width=0.48\linewidth,trim={10 20 30 50},clip]{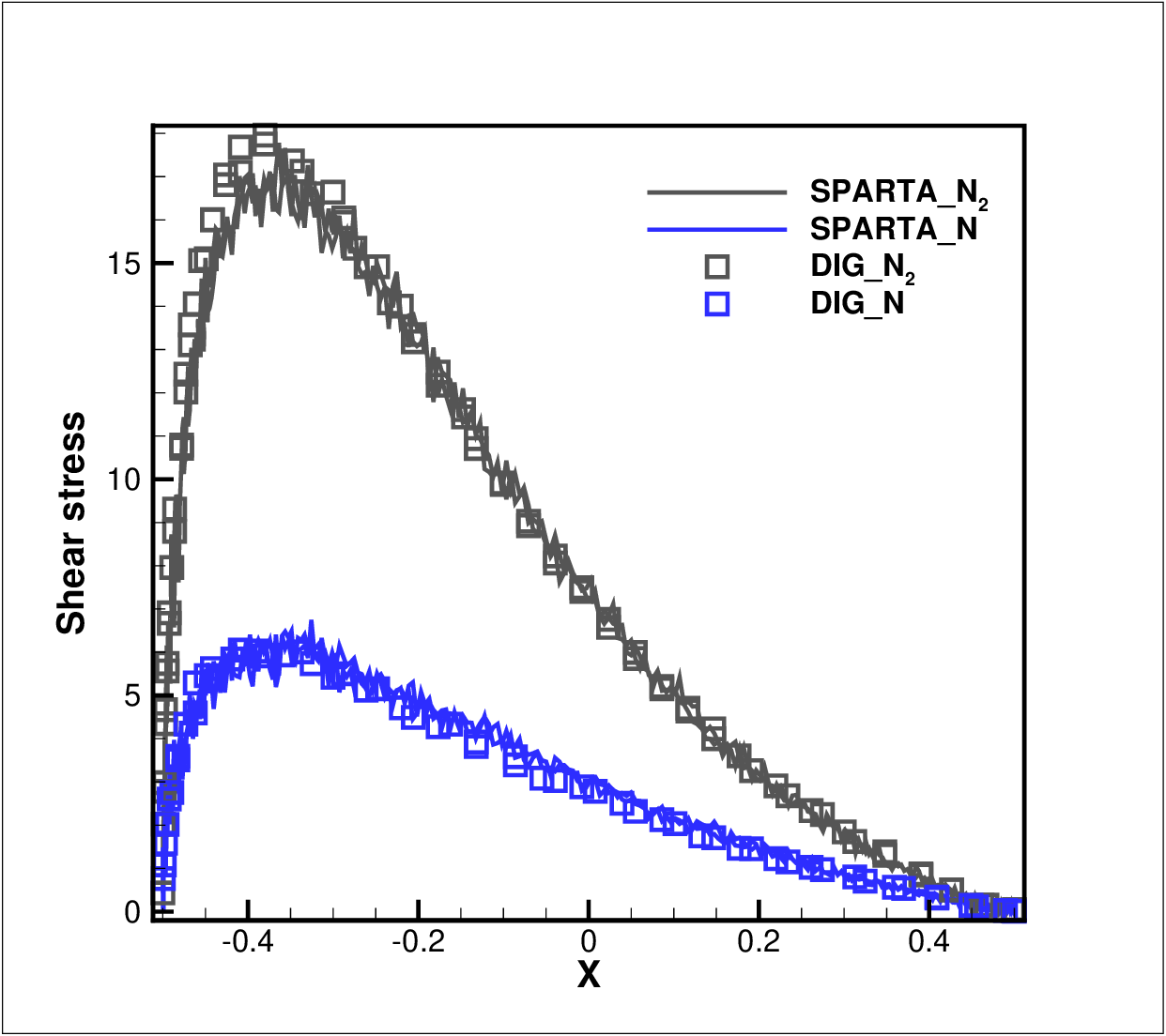}
    \hspace{0.1cm}
	\includegraphics[width=0.48\linewidth,trim={10 20 30 50},clip]{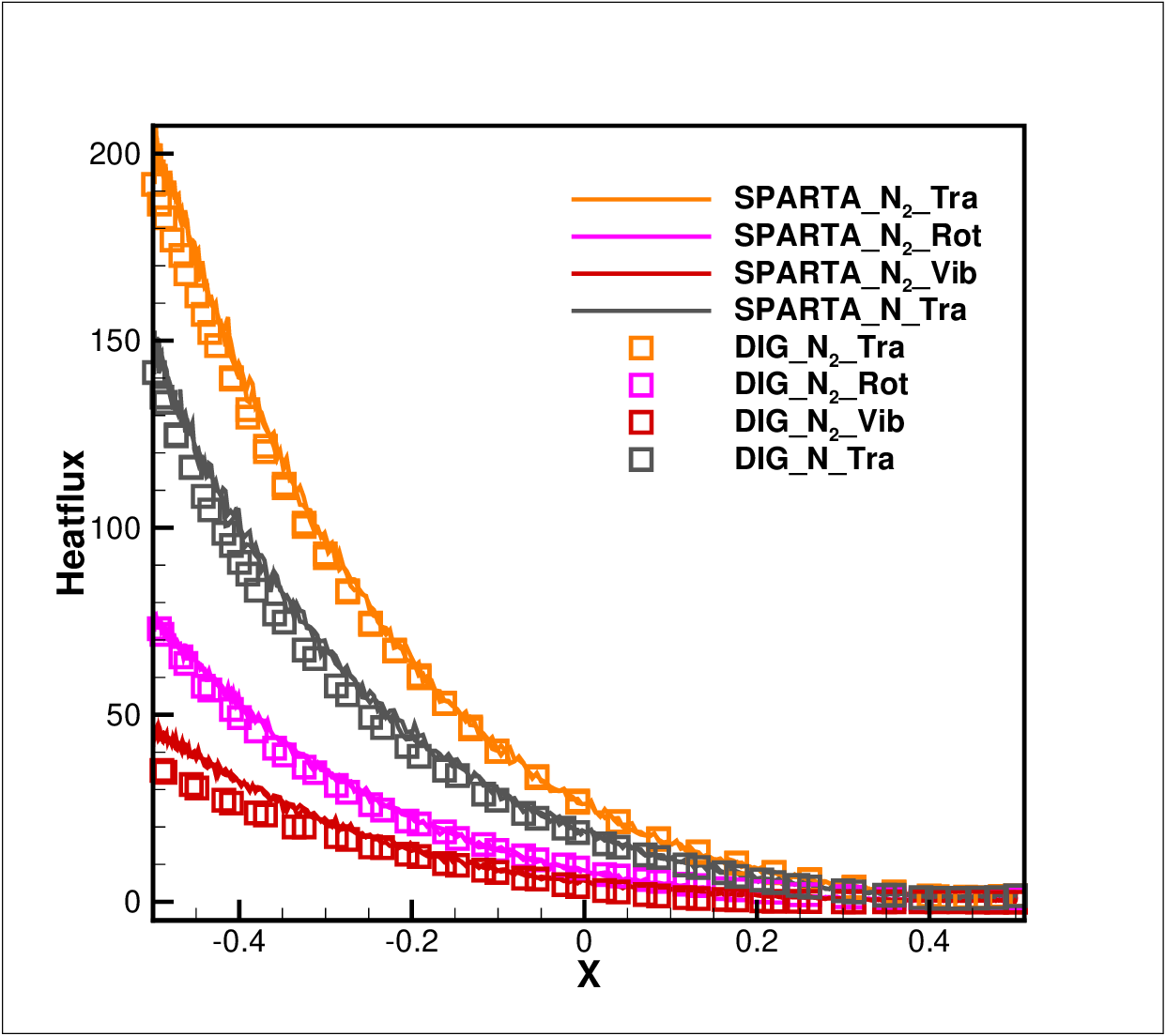}
	\caption{Comparison of the surface shear stress and heat flux of different internal modes at Kn = 0.1 (upper row) and 0.01 (lower row).}
	\label{fig:dsmcdig_HeatfluxShear}
\end{figure}
Figure~\ref{fig:dsmcdig_HeatfluxShear} compares the wall distributions of shear stress and heat flux obtained by SPARTA and DIG. For both $\mathrm{Kn}=0.1$ and $\mathrm{Kn}=0.01$, the wall shear stress and heat flux predicted by DIG agree well with the SPARTA results, demonstrating that DIG is capable of maintaining satisfactory accuracy even on substantially coarser meshes.

\begin{figure}[t]
	\centering
    \includegraphics[width=0.48\linewidth,trim={10 20 30 50},clip]{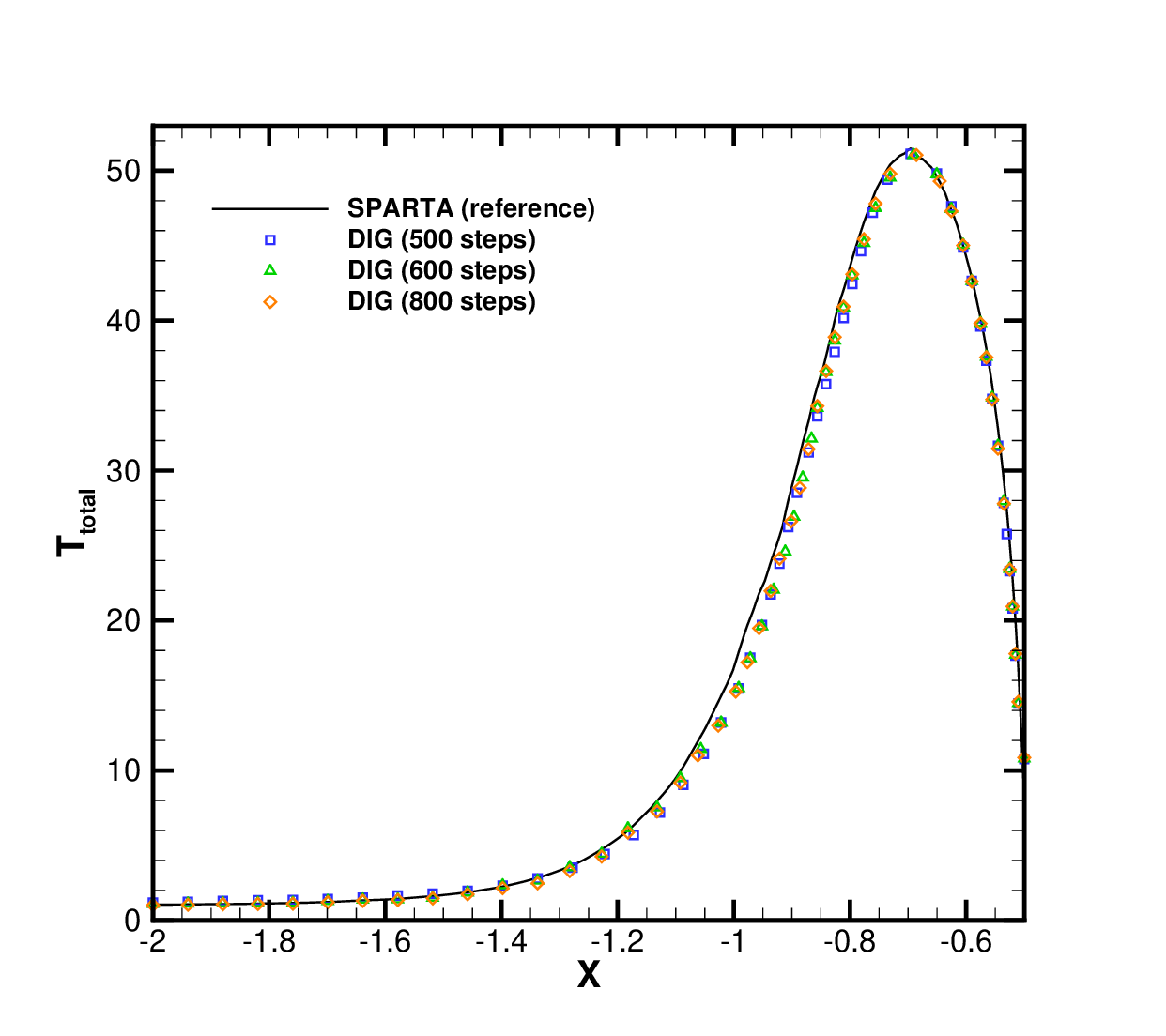}
    \hspace{0.2cm}
	\includegraphics[width=0.48\linewidth,trim={10 20 30 50},clip]{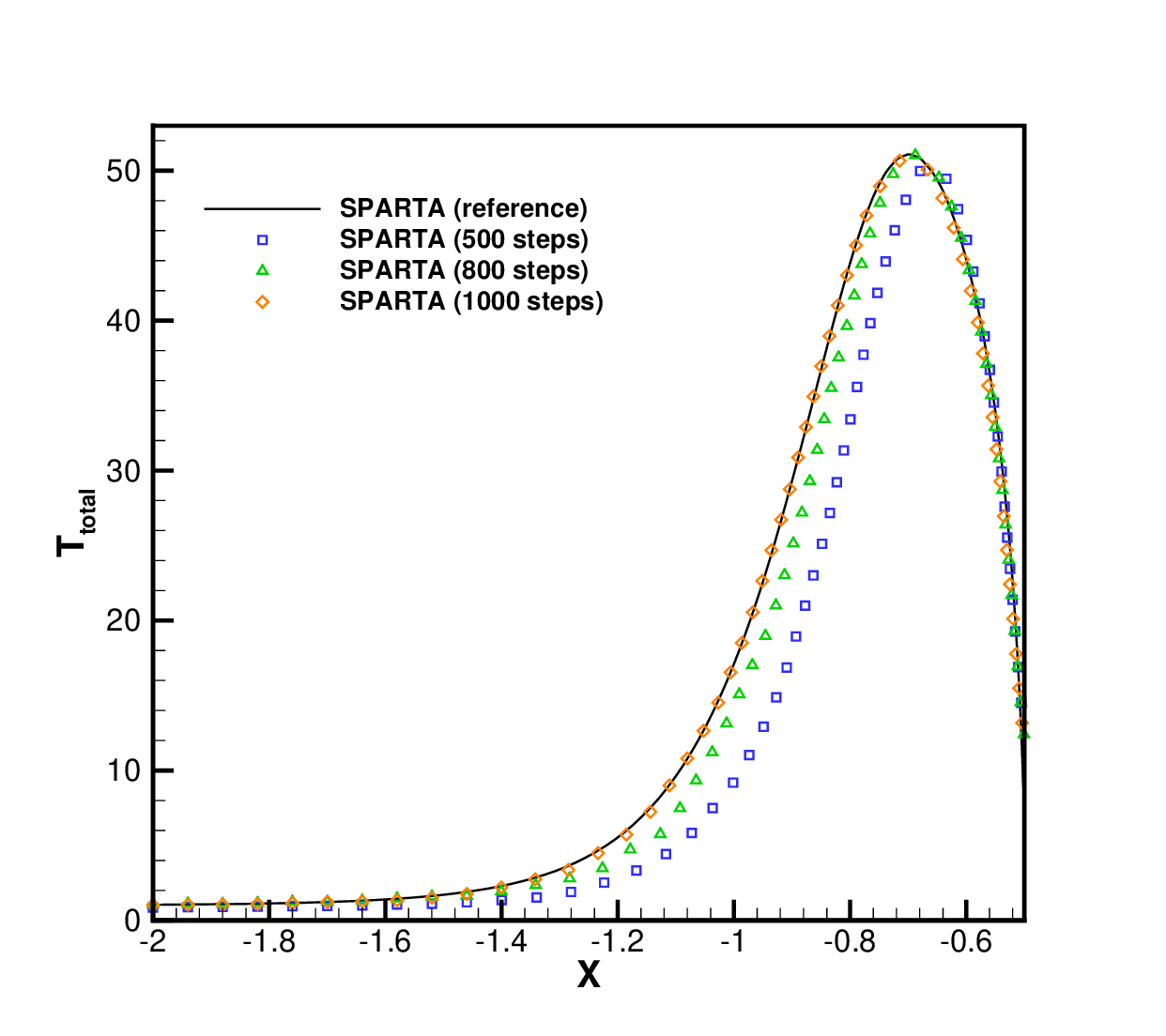}\\
    \vspace{-0.1cm}
	\includegraphics[width=0.48\linewidth,trim={10 20 30 50},clip]{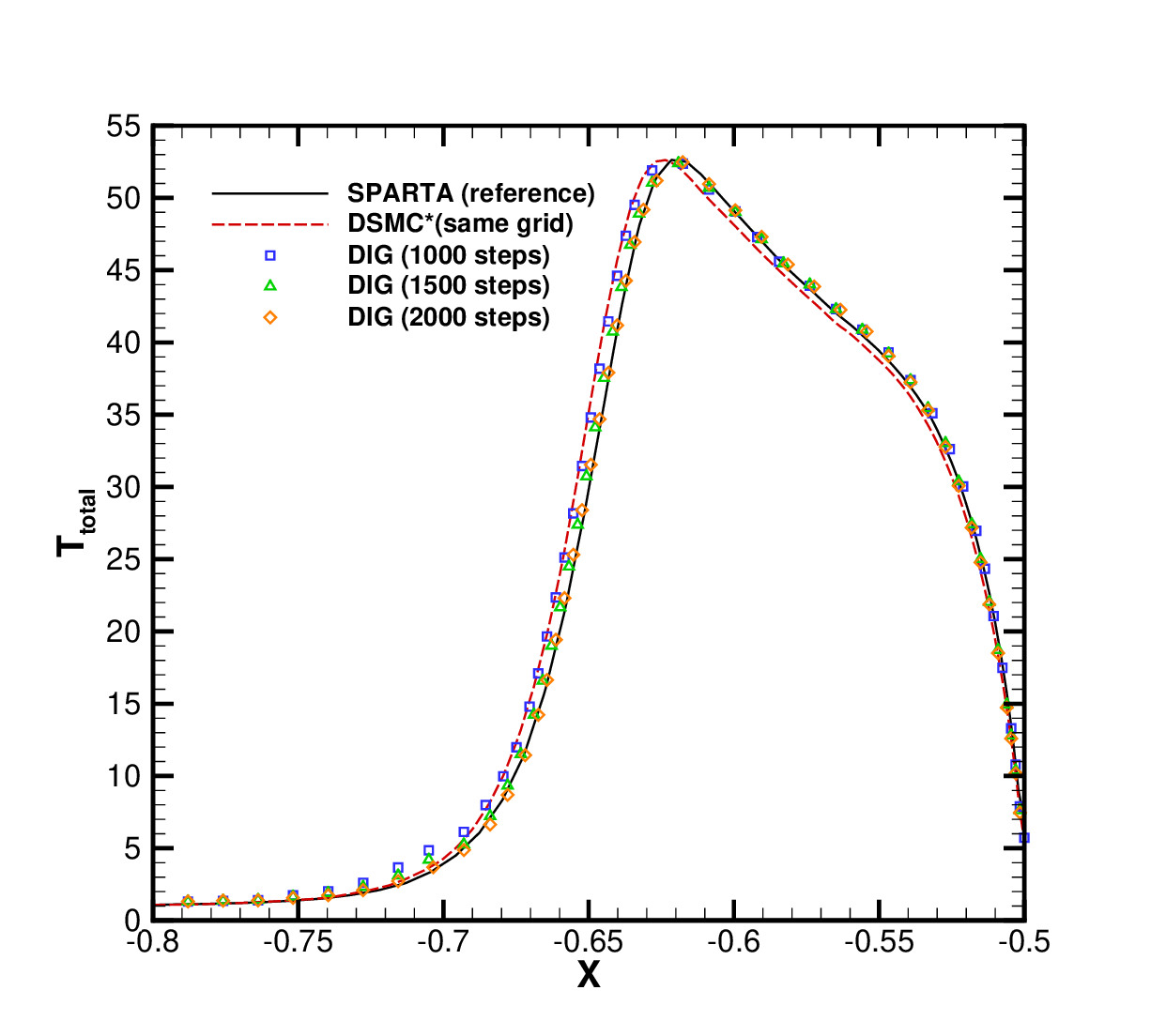}
    \hspace{0.2cm}
	\includegraphics[width=0.48\linewidth,trim={10 20 30 50},clip]{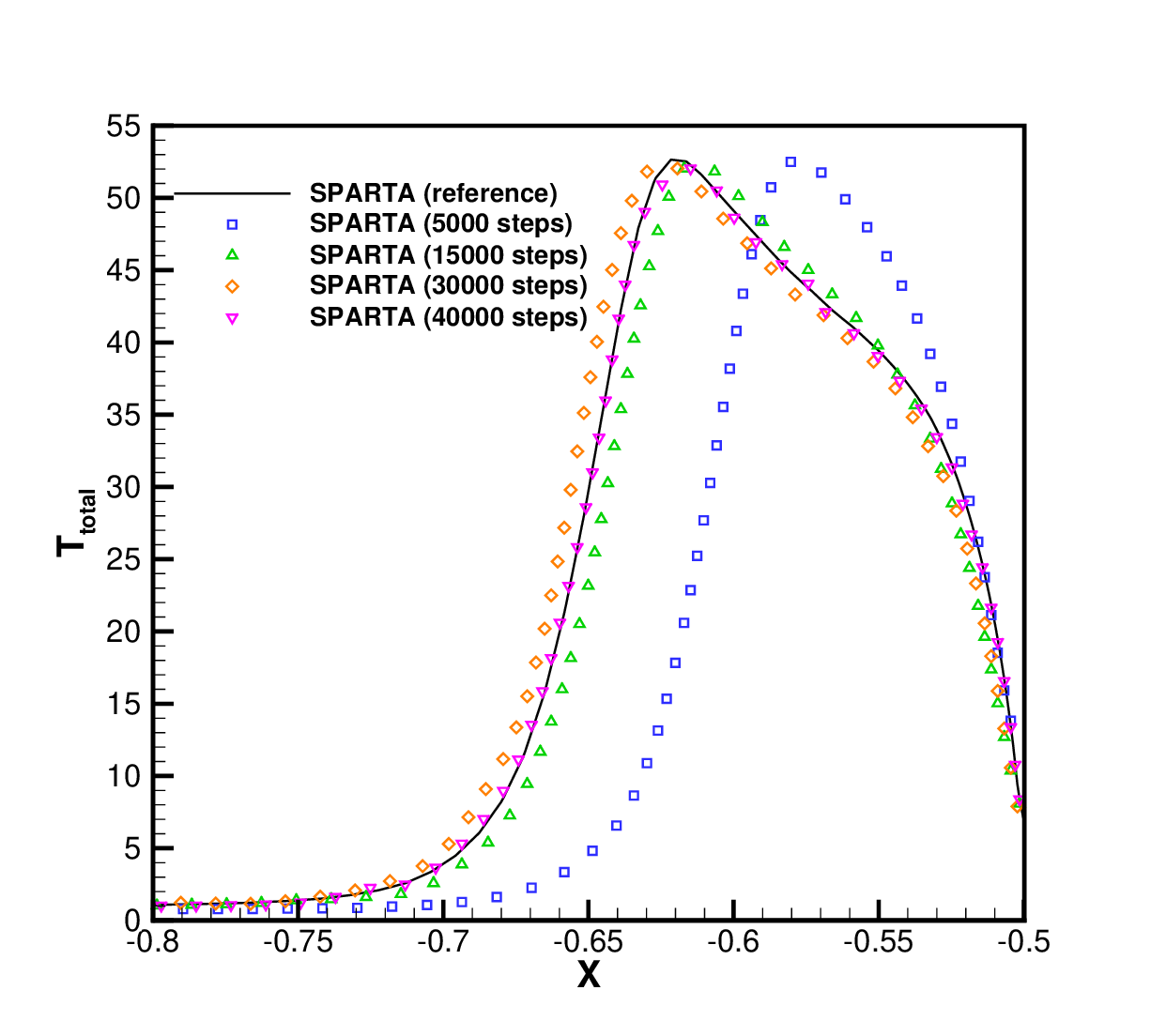}
	\caption{ The evolution of total temperature along the stagnation line on the windward side of the cylinder obtained by DIG and SPARTA at Kn = 0.1 (upper row) and 0.01 (lower row). Note that the superscript $*$ represents the DSMC results obtained based on the same grid of DIG method.}
	\label{fig:dsmcdig_step}
\end{figure}

Figure~\ref{fig:dsmcdig_step} compares the temporal evolution of the total temperature during the transition state for DIG and SPARTA. The same time step is employed in both methods, and the particle velocities and internal energies, including rotational and vibrational modes, are initialized by sampling from Maxwellian distributions corresponding to the freestream conditions. As shown in the figure, when $\text{Kn}=0.1$, SPARTA requires approximately 1000 time steps to reach steady state, whereas DIG converges in about 800 steps. However, the acceleration effect becomes significantly more pronounced at $\text{Kn}=0.01$, where SPARTA requires approximately 40,000 steps, while DIG converges within only about 2000 steps. This corresponds to more than an order-of-magnitude reduction in the number of time steps required to reach steady state. 
It should be noted that, in SPARTA, the grid is adaptively refined according to the local mean free path. As a result, the solution initially evolves on relatively coarse grids, leading to deviations from the converged solution, and gradually approaches the correct steady state as the grid resolution increases. As a result, in the near-continuum regime, the number of grid cells employed in DIG is approximately two orders of magnitude smaller than that in SPARTA. It can be witnessed from the figure that, if the same coarse grid used in DIG is directly applied in our in-house DSMC solver, the resulting solution exhibits significant deviations from the SPARTA reference solution. This observation highlights the asymptotic-preserving property of the proposed DIG method, which enables accurate predictions even on meshes that are much coarser than those required in DSMC.

\begin{table}[!t]
 \centering
 \caption{\label{tab:tab1}Computational overheads of the DSMC and DIG for hypersonic nitrogen gas past a cylinder. The DSMC simulations are performed using the SPARTA code with 160 cores and adaptive mesh refinement, whereas our in-house DIG uses 40 cores. For each Knudsen number, both methods employ the same time step and are initialized from uniform freestream conditions. The simulation time is given in core$\times$hours. Note that without DIG, our in-house  DSMC code is slower than SPARTA by one order of magnitude, due to the use of structured mesh (we will soon enhance the efficiency of particle transport to match the level of SPARTA). 
}
\begin{threeparttable} 
  \begin{tabular}{c c c c c c c c}\toprule
\multirow{2}{*}{Ma}  & \multirow{2}{*}{Kn} & \multirow{2}{*}{method} & \multirow{2}{*}{$N_{\text{cell}}$}  & \multicolumn{2}{c}{Transition state}  &  \multicolumn{2}{c}{Steady state} \\ \cmidrule(r){5-8}
  ~ & ~ & &  & steps&time&steps   &    time\\ \hline
 \multirow{4}{*}{20} & \multirow{2}{*}{0.1} &SPARTA & 291,066  &    1000   & 1.77 &  5000 &      12.8    \\
 ~  & ~ &      DIG       & $100\times 128$    &  800   & 0.53 &  1000 &  0.67     \\ \cmidrule(r){2-8}
 ~  & \multirow{2}{*}{0.01} &SPARTA & 10,675,858   &   40,000  & 504 &  10,000 &      365    \\ 
~  & ~ &            DIG & $200\times 200$    &  2000   & 4.01 &  5000 &      10.63     \\
\bottomrule
\end{tabular}
 \end{threeparttable}
\end{table}

Table~\ref{tab:tab1} presents the comparison of CPU time between SPARTA and DIG. It should be noted that the parallel efficiency of our in-house DSMC code is approximately one order of magnitude lower than that of SPARTA, primarily due to the Cartesian grid structure and the more advanced parallelization strategy employed in SPARTA. 
For $\mathrm{Kn}=0.1$ and $\mathrm{Kn}=0.01$, the total computational time of SPARTA is approximately 15 and 53 times larger than that of DIG, respectively. In particular, for $\mathrm{Kn}=0.01$, the asymptotic-preserving property of DIG allows the use of substantially fewer computational cells in near-continuum regimes, thereby leading to a significant reduction in computational cost compared with the adaptive-mesh SPARTA simulation.

\section{Conclusions and outlook}\label{sec:conclusion}

In summary, the DIG method is developed for efficient simulation of nonequilibrium chemical reactions. Numerical simulations verify that, relative to SPARTA‑DSMC, results from the DIG method agree well with those of the conventional DSMC approach. Thanks to its fast-converging property and noise reduction property from the guidance of macroscopic synthetic equation, even with identical time‑step sizes, the DIG method achieves a nearly one‑order‑of‑magnitude reduction in the number of time steps needed to reach steady state in the near‑continuum regime. Due to its asymptotic-preserving property, the cell size can be much larger than the molecular mean free path. When these properties are combined, significantly reduction of computational memory and simulation time are achieved.

Owing to the flexibility of the single-velocity macroscopic framework, the DIG method can be readily extended to incorporate more complex physicochemical processes, including electromagnetic effects and multi-component chemically reacting flows.

\section*{Declaration of competing interest}
The authors declare that they have no known competing financial interests or personal relationships that could have appeared to influence the work reported in this paper.

\section*{Acknowledgments}
This work was supported by the National Natural Science Foundation of China (Grant No.~12450002). Special thanks are given to the Center for Computational Science and Engineering at the Southern University of Science and Technology.

\appendix
\section{SPARTA result after modification}\label{Appendix_sparta}

\begin{algorithm}[!t]
    \caption{Overall algorithm of origin QK model in SPARTA} 
    \label{algo:Origin_SPARTAQK}
    \begin{algorithmic}[1]
        \Require Information of two particles;
        \Ensure Reaction products and total energy after reaction $post\_etotal$;
        
        \State Input two particles information; 
        \State Compute the sum of translational, rotational and vibrational energy of two particles $pre\_etotal$;
        \For{Loop all possible reactions for these two particles}
        \State Compute the sum of translational and rotational energy $ecc$;
        \If{$ecc < E_{\text{act}}(\text{activation energy})$}
            \State Break the current step and continue to judge next reaction;
        \EndIf

        \If{Reaction type is dissociation}
            \State Compute the sum of translational and vibrational energy $e_{\text{tv}}$;
            \State Compute the maximum vibrational level $i_{\text{max}}=e_{\text{tv}}/(k_\text{B}\Theta_{\text{vib}})$;
            \If{$i_{\text{max}} > \Theta_{\text{dis}}/\Theta_{\text{vib}}$}
                \State $post\_etotal$ = $pre\_etotal$ - $E_{\text{act}}$;
                \State Reaction takes place; break loop;
            \EndIf
        \EndIf
        \If{Reaction type is others}
            \If{Reaction takes place}
                \State Break loop;
            \EndIf
        \EndIf
        \EndFor

        \If{Reaction takes place}
            \State Updating the particles species as reaction products and $post\_etotal$;
            \State Return 1;
        \EndIf
        \State Return 0;
    \end{algorithmic}
\end{algorithm}

In SPARTA, the implementation of the QK chemical reaction model is primarily encapsulated within the \texttt{react\_qk.cpp} file; the corresponding workflow is summarized in Algorithm \ref{algo:Origin_SPARTAQK}.
It can be found that there is an extraneous constraint on translational and rotational energy in Line 15 of the algorithm, which deviates from the QK model mentioned by Bird \cite{bird2013dsmc} that the condition for dissociation of the molecule is $i_{\text{max}} > \Theta_{\text{d}}/\Theta_{\text{v}}$.
The rotational energy can be effectively coupled and contribute through translational energy, due to the rapid thermal equilibration of the rotational and translational energy modes \cite{gallis2010jtht}.
Imposing this unnecessary constraint artificially reduces the probability of chemical reaction. Therefore, we modified the \texttt{react\_qk.cpp} file in SPARTA by removing the judgment statement in Line 15 of Algorithm \ref{algo:Origin_SPARTAQK}.

\begin{figure}[t]
	\centering
	\includegraphics[width=0.48\linewidth,trim={10 20 30 30},clip]{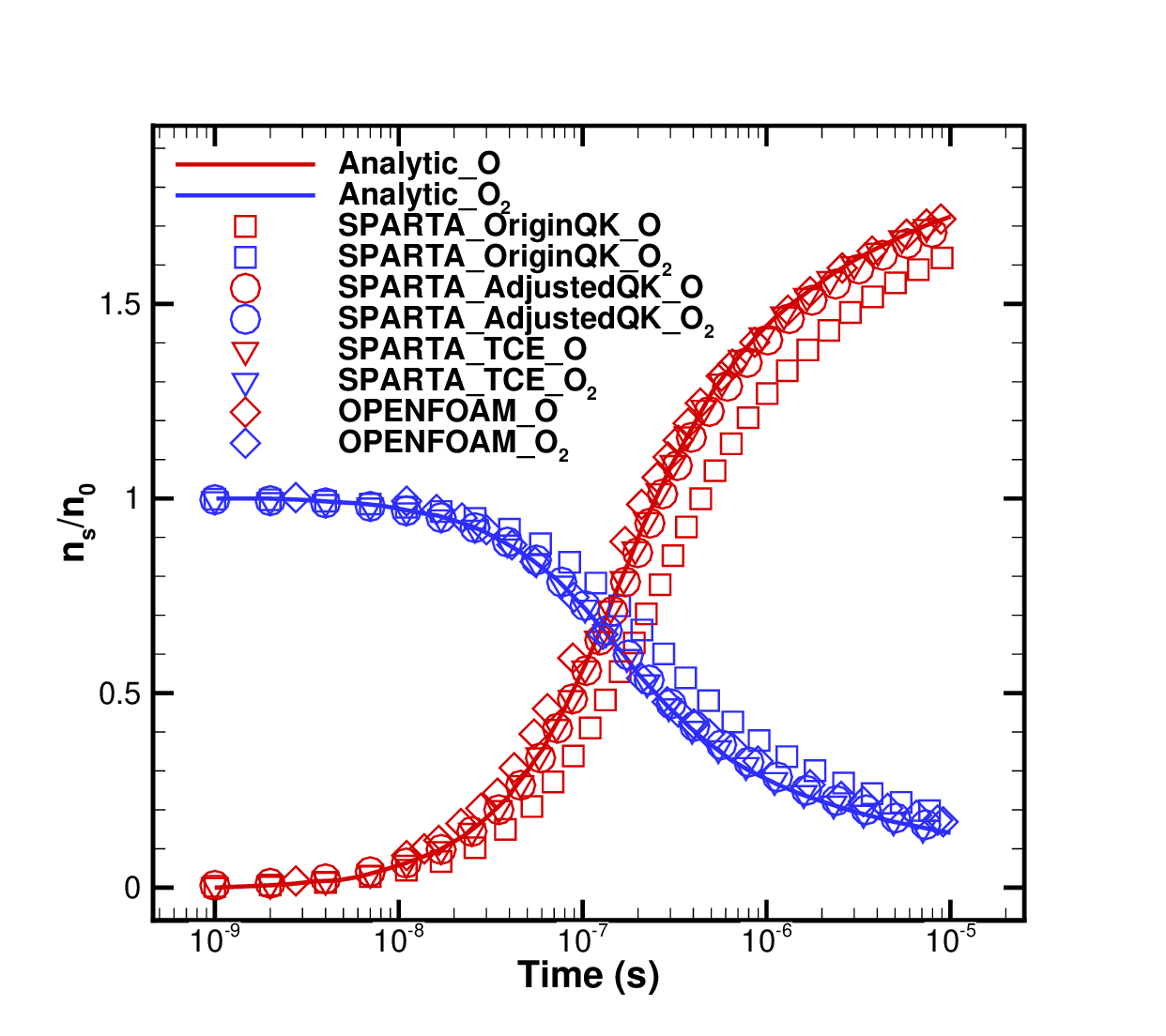}
    \hspace{0.2cm}
	\includegraphics[width=0.48\linewidth,trim={10 20 30 30},clip]{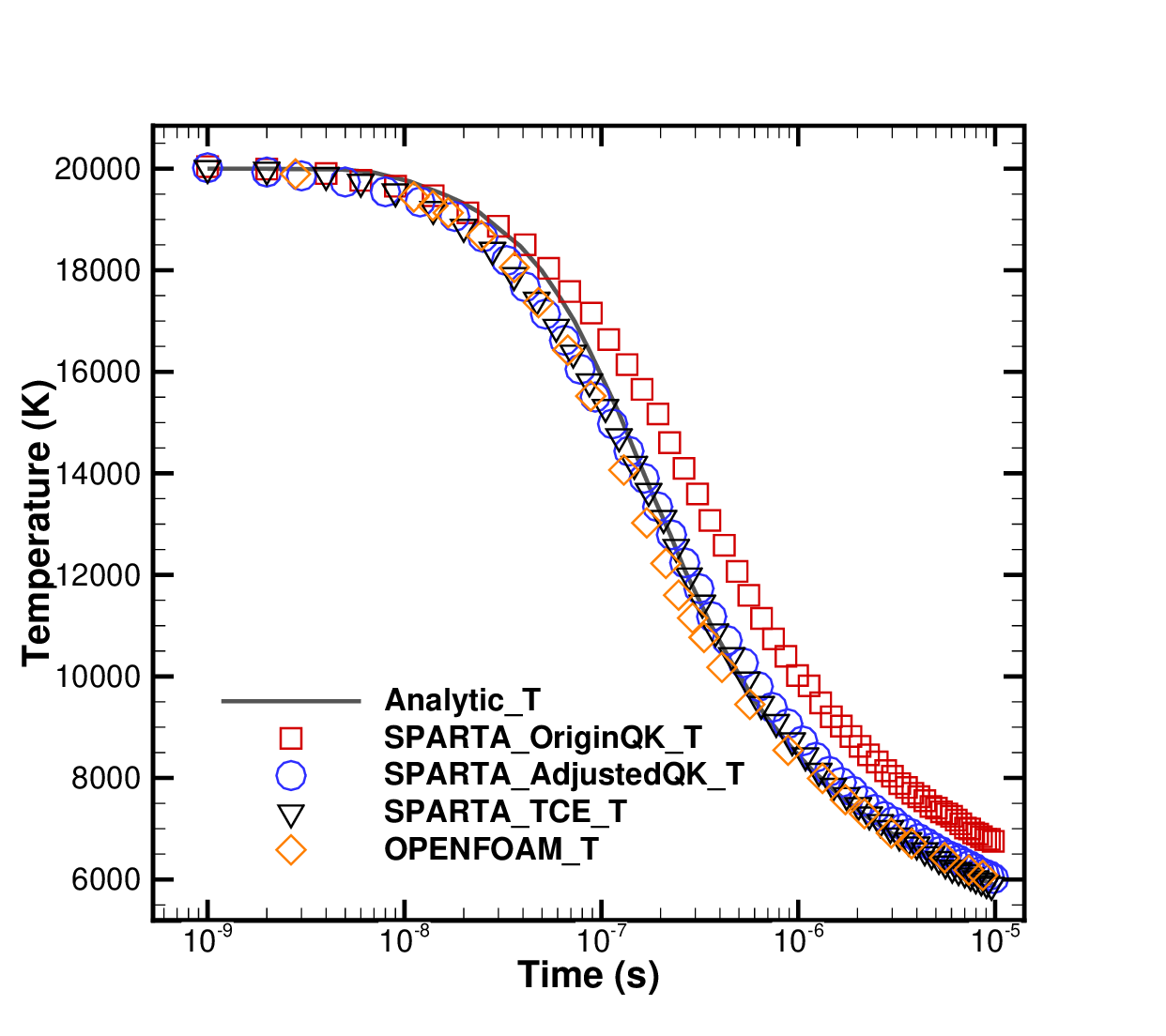}
	\caption{ Non-equilibrium dissociation of O$_2$ (undergoing reaction 1 and 2 in Table \ref{tab:Arrhenius}) simulated by different methods. The temporal evolutions of species concentrations and species temperatures are shown in left and right, respectively.}
	\label{fig:sparta}
\end{figure}

As shown in Figure \ref{fig:sparta}, a comparative analysis is conducted for the non-equilibrium dissociation of oxygen using different methods. The theoretical data and the OpenFOAM results obtained by \cite{Rodrigo2014phd} are taken as reference analytical and numerical solutions, respectively, to assess the accuracy of the different methods. In addition, the parameters used in the total collision energy model are listed in Table \ref{tab:Arrhenius}. The corresponding equilibrium reaction rates are identical to those of the QK model, as verified using the \texttt{QKrates.exe} utility developed by Bird~\cite{bird2013dsmc}. 
It is observed that, without any modification to \texttt{react\_qk.cpp}, the original SPARTA implementation yields an oxygen atom concentration of approximately 1.6 and a temperature of about 6700 K at $1\times10^{-5}$ s. 
These values deviate noticeably from the theoretical prediction and other computational results, which give corresponding values of approximately 1.7 and 6000 K. By contrast, the results obtained using the modified SPARTA show good agreement with both the theoretical and the other computational results.



\section{Equilibrium and non-equilibrium cases}\label{Appendix_QK}

In the QK model, the equilibrium chemical reaction rate of the dissociation reaction in the VHS model is given by
\begin{equation}
	\begin{aligned}
		k_\text{f}=[R_\text{coll}\gamma(i_\text{max})]^{\text{AB,C}},
        \label{equ:eq_rate}
	\end{aligned}
\end{equation}
where $R_\text{coll}$ is the collision rate parameter, which can be computed as the collision rate between species AB and C divided the number density, i.e.,
\begin{equation}
	\begin{aligned}
		R_{\text{coll}}=\frac{2\pi^{1/2}}{\epsilon}
        \left( r_{\text{ref}}^{\text{AB}}+r_{\text{ref}}^\text{C} \right)^2
		\left(\frac{T}{T_\text{ref}}\right)^{1-\omega^{AB,C}}\left(\frac{2k_\text{B}T_\text{ref}}{m_\text{r}^{\text{AB,C}}}\right)^{1/2}.
	\end{aligned}
\end{equation}
Here, $r_{\text{ref}}$ is the molecular reference radius at the temperature $T_\text{ref}$. The quantities $T$, $m_\text{r}$, and $\epsilon$ represent the equilibrium temperature, the reduced mass of the collision pair, and the symmetry parameter, respectively. It should be noted that $\epsilon=1$ for collisions between different species and $\epsilon=2$ for collisions between identical species.
Moreover, in Eq.~\eqref{equ:eq_rate}, $\gamma(i_\text{max})$ is the fraction of collision that satisfies the dissociation reaction. For the VHS gas,
\begin{equation}
	\begin{aligned}
		\gamma(i_\text{max})=\frac{\sum_{i=0}^{i_\text{max}-1} \left \langle Q\left \{ (\frac{5}{2}-\omega^{\text{AB,C}}),\left [(i_\text{max}-i)\frac{\Theta_\text{vib}}{T} \right ]\right \}  
		\exp\left(-i\frac{\Theta_\text{vib}}{T}\right)\right \rangle }{z^{\text{AB}}_\text{vib}(T)}.
	\end{aligned}
\end{equation}
where $Q(a,x)=\Gamma(a,x)/\Gamma(a)$ is a form of incomplete Gamma function and 
\begin{equation}
    z^{\text{AB}}_\text{v}(T)=\left[1-\exp\left(-\frac{\Theta_\text{vib}}{T}\right)\right]^{-1}
\end{equation} 
is the vibrational partition function in the harmonic oscillator model.

To validate the QK chemical reaction model in our in-house code, simulations of equilibrium and non-equilibrium chemical reactions are performed. Both test cases are carried out in a cubic reactor with a side length of 10 micrometers, where all surfaces are defined as specular reflection boundaries. A total of 50,000 simulation particles are initialized; these particles move and collide within the cube, with the internal energy of colliding particles being redistributed. The probabilities of inelastic collision for rotational and vibrational modes are set to 1, and the time step is configured as 1 nanosecond.

\begin{figure}[t]
	\centering
	\includegraphics[width=0.45\linewidth]{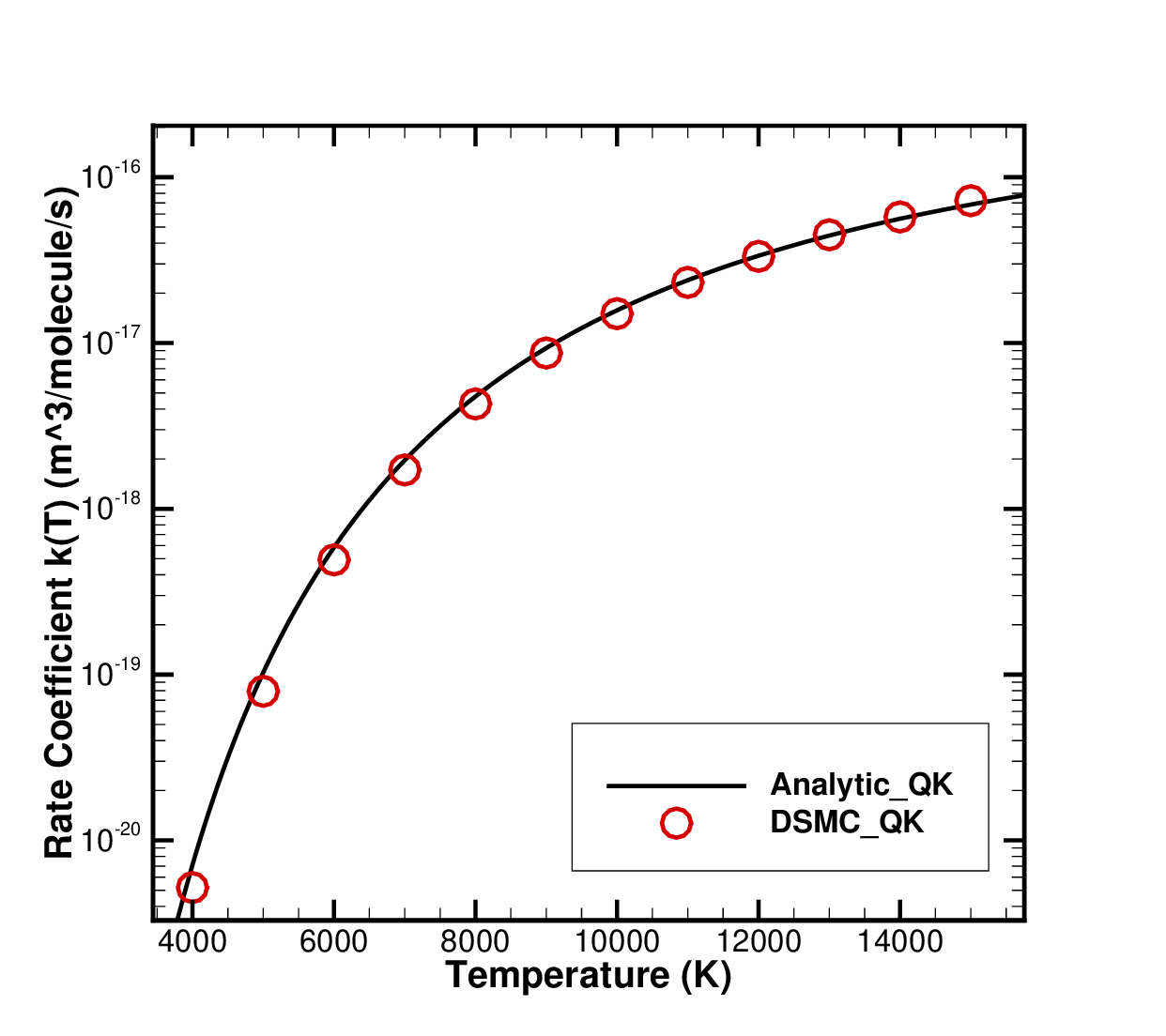}
    \hspace{0.3cm}
	\includegraphics[width=0.45\linewidth]{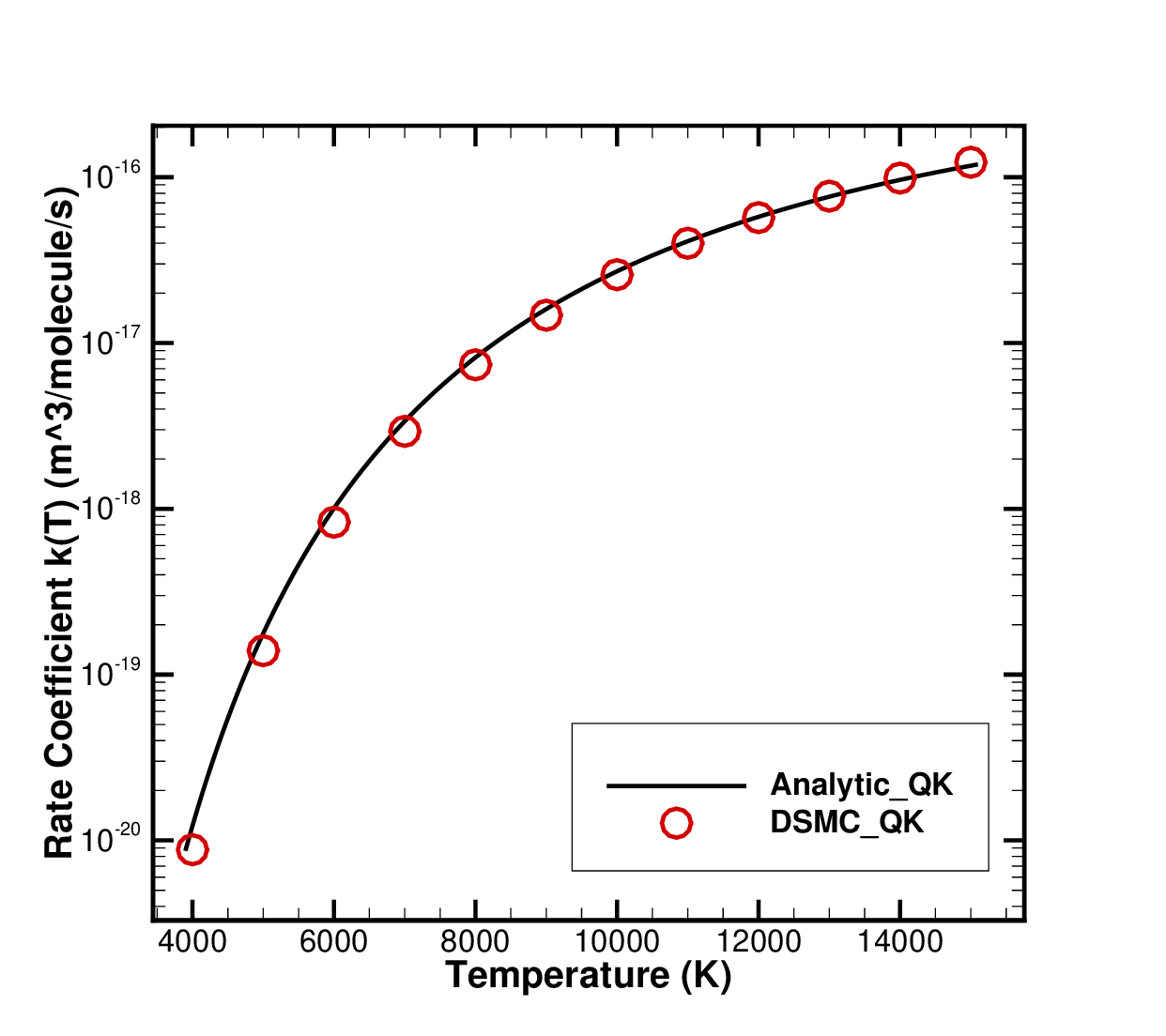}
	\caption{ Equilibrium dissociation rate coefficient of type 1 reaction (left) and type 2 reaction (right), which is represented as a function of temperature. }
	\label{fig:1}
\end{figure}

For the equilibrium chemical reaction simulations, whenever a colliding particle pair satisfies the reaction criterion, the reaction counter is increased by one, but no actual chemical reaction is performed. Instead, the pair undergoes internal energy redistribution followed by an elastic collision. As a result, the total energy of the collision pair is conserved, while the macroscopic temperature remains unchanged. Figure \ref{fig:1} presents the equilibrium reaction rates for reactions 1 and 2 in Table \ref{tab:Arrhenius}. The analytical equilibrium reaction rates are determined from Eq.~\ref{equ:eq_rate}, which can also be expressed in the Arrhenius form given by Eq.~\eqref{eq:Arrhenius}, with the corresponding constants taken from \cite{Scanlon2015AIAA}. In general, the figure shows that, for equilibrium chemical reactions, the results obtained from our DSMC code are in excellent agreement with the theoretical values.

\begin{table}[t]

    \centering
\caption{Chemical reaction list and Arrhenius rate coefficients}
\label{tab:Arrhenius}
\begin{tabular}{c c c c c}
\toprule
 Number &Reaction & $a$ & $b$ & Activation energy (J) \\ 
 \hline
 1&$\text{O}_2+\text{O}_2\to \text{O}+\text{O}+\text{O}_2$ & $5.33\times10^{-11}$ & -1 & 8.197$\times 10^{-19}$ \\
 2&$\text{O}_2+\text{O}\to \text{O}+\text{O}+\text{O}$ & $1.5\times10^{-10}$ & -1.05 & 8.197$\times 10^{-19}$ \\
 3&$\text{N}_2+\text{N}_2\to \text{N}+\text{N}+\text{N}_2$ & $4.1\times10^{-12}$ & -0.62 & 15.67$\times 10^{-19}$ \\
 4&$\text{N}_2+\text{N}\to \text{N}+\text{N}+\text{N}$ & $1\times10^{-11}$ & -0.68 & 15.67$\times 10^{-19}$ \\
\bottomrule
\end{tabular}
\end{table}

\begin{figure}[t!]
	\centering
	\subfigure[]{
    \label{fig:2_a}
    \includegraphics[width=0.45\linewidth,trim={10 20 30 40},clip]{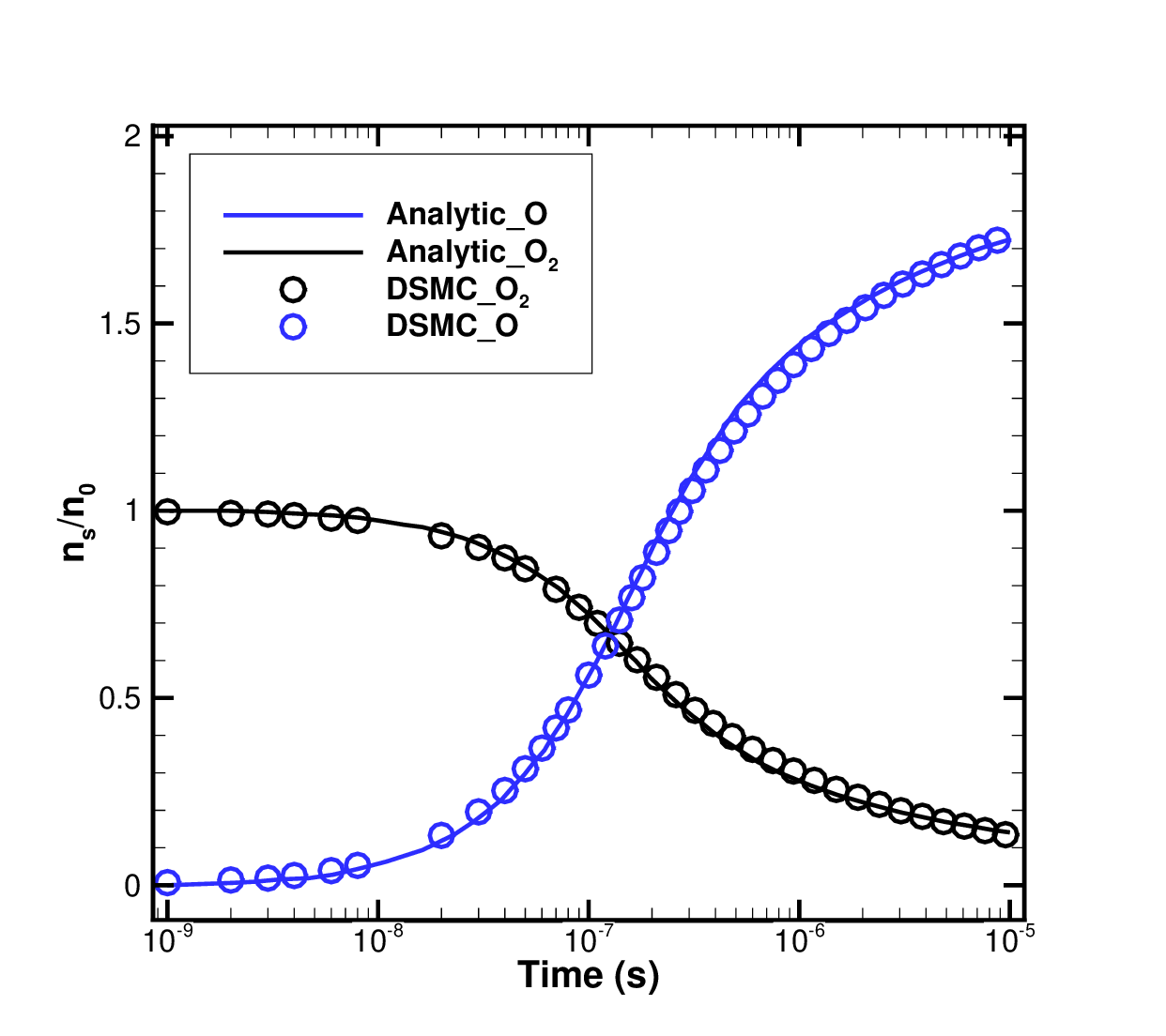}}
	\hspace{0.2cm}
	\subfigure[]{
    \label{fig:2_b}
    \includegraphics[width=0.45\linewidth,trim={10 20 30 40},clip]{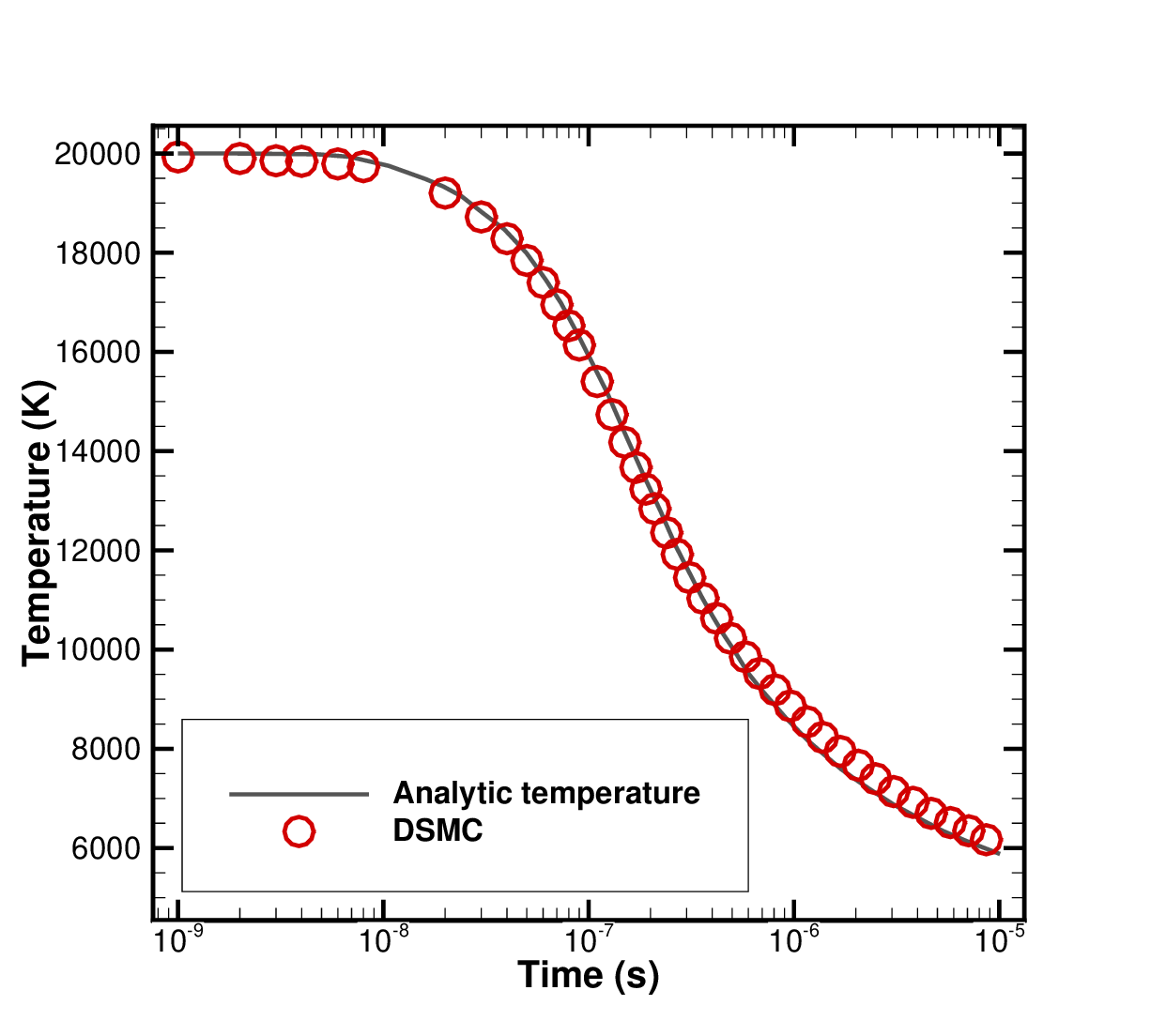}}\\
    \subfigure[]{ 
    \label{fig:2_c}
    \includegraphics[width=0.45\linewidth,trim={10 20 30 40},clip]{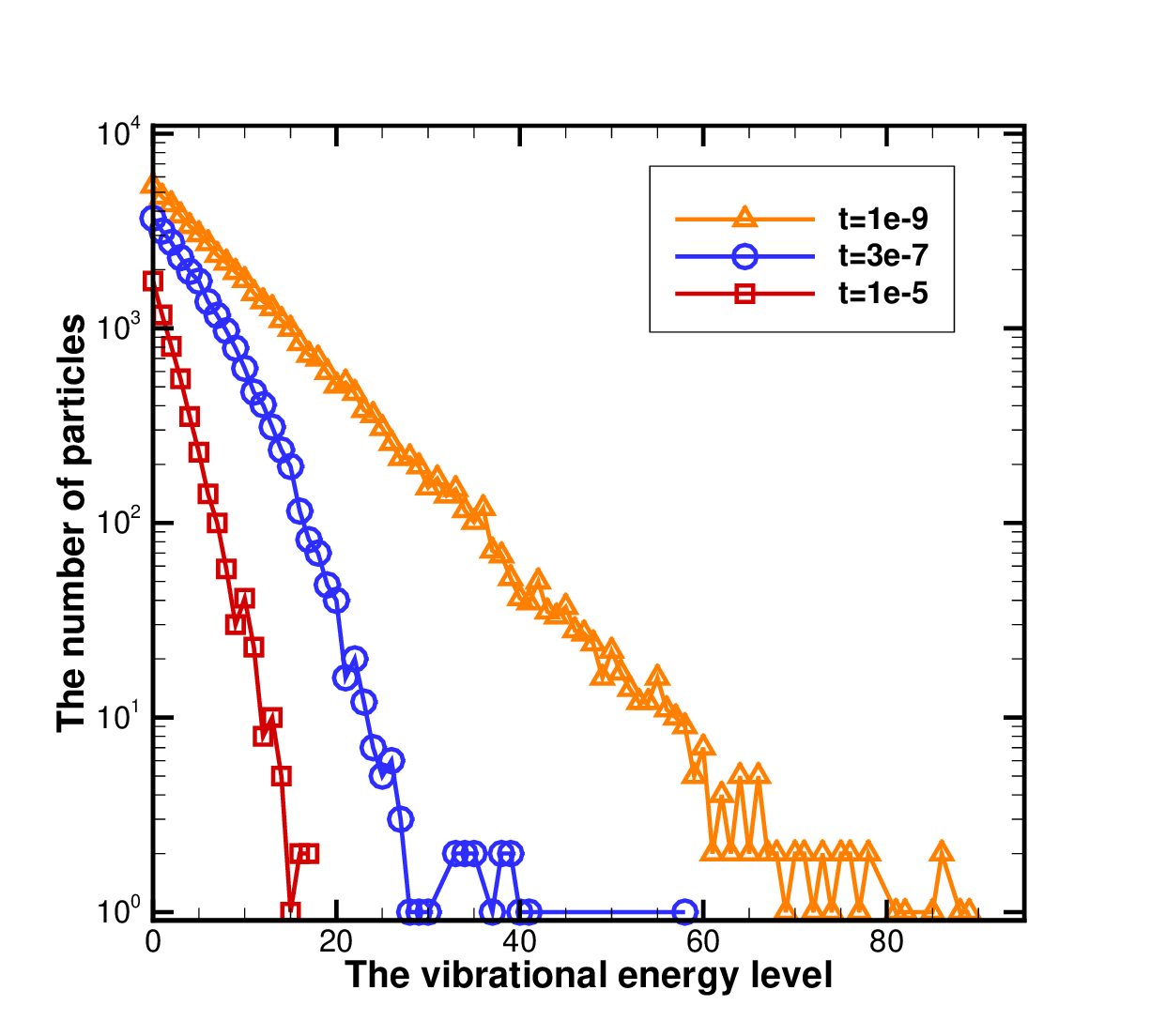}}
	\caption{Non-equilibrium dissociation of O$_2$ (undergoing reaction 1 and 2 in Table \ref{tab:Arrhenius}) at $T_0=20,000$~K and $p_0=0.063$~atm. Variation of (a) species concentrations and (b) temperature with respect to the time. (c) Particle number distribution across vibrational energy levels at different time points during non-equilibrium reactions.}
	\label{fig:2}
\end{figure}

For non-equilibrium chemical reaction simulations, an actual chemical transformation is performed once a colliding particle pair satisfies the reaction criterion. For a general reaction of the form $\text{A}+\text{B}\to \text{C}+\text{D}$, the concentration $[X_\text{A}]$ of species A evolves according to
\begin{equation}
	\begin{aligned}
        \frac{\text{d}[X_\text{A}]}{\text{d}t}=-k_\text{f}[X_\text{A}][X_\text{B}]+k_\text{b}[X_\text{C}][X_\text{D}],
	\end{aligned}
\end{equation}
where $k_\text{f}$ and $k_\text{b}$ are the forward and backward reaction rate coefficients, respectively. The analytical results are obtained using the method proposed by Hass~\cite{Hass1993PoF}.
When the dissociation reaction occurs, the molecule AB dissociates into two distinct atoms, A and B. During the subsequent internal energy redistribution, a portion of the total energy is assigned to the relative translational energy of the newly formed A and B particles.
Figure \ref{fig:2} presents the results of oxygen dissociation at an initial temperature of $T_0=20{,}000$ K and an initial pressure of $p_0=0.063$ atm, with only the forward reactions considered.
Figure \ref{fig:2}\subref{fig:2_a} shows the temporal evolution of the species number density normalized by its initial value, i.e., $n_s/n_0$, while Fig.~\ref{fig:2}\subref{fig:2_b} presents the corresponding temperature variation with time. Figure \ref{fig:2}\subref{fig:2_c} further illustrates the evolution of the particle number distribution over energy levels at different time instants, revealing the transition of the reaction process from an initial equilibrium state to a non-equilibrium state, followed by a gradual relaxation back toward equilibrium.

\section{Derivation of single-temperature governing equations}\label{Appendix_meq_derivation}

The multi-fluid macroscopic equations for species $s$ can be obtained by taking moments of Eq.~\eqref{eq:wcu-type-reaction}. Firstly, the mixture continuity equation can be obtained by summing the species continuity equations over all species as, 
\begin{equation}
    \label{eq:continuityequation_sum}
    \sum_s \frac{\partial \rho_s}{\partial t}+\sum_s\nabla\cdot(\rho_s\bm{u}_s)=\sum_s\dot{\omega}_s,
\end{equation}
where $\sum_s\rho_s=\rho$ and $\sum_s\rho_s\bm{u}_s=\rho\bm{u}$ are the mixture density and momentum, respectively. Due to the conservation of mass, $\sum_s\dot{\omega}_s=0$. Moreover, the species continuity equation can be obtained by introducing the mass diffusivity term $\bm{\Phi}_s$, which can be obtained according to,
\begin{equation}
    \rho_s\bm{u}_s= \rho_s\left(\bm{u}+\bm{V}_s\right)=\rho\bm{u}+\bm{\Phi}_s.
\end{equation}
Secondly, the summation of species momentum equations over all species yields the total momentum equation:
\begin{equation}
    \label{eq:momentumequation_sum}
    \sum_s\frac{\partial (\rho_s\bm{u}_s)}{\partial t}+\nabla\cdot\left(\sum_s\rho_s\bm{u}_s\bm{u}_s+\sum_sp_{t,s}\bm{I}+\sum_s\bm{\sigma}_s\right)=\sum_s\bm{Q}_{s}^\text{M},
\end{equation}
where $\sum_s\rho_s\bm{u}_s\bm{u}_s=\rho\bm{u}\bm{u}+\bm{\Pi}_\text{sp}$ by using the relations $\sum_s\rho_s\bm{V}_s=0$. According to the conservation of momentum, $\sum_s\bm{Q}_{s}^\text{M}=0$. Based on Eqs.~\eqref{eq:generalrelations} and \eqref{eq:pidsmc}, the summation of the pressure term can be given by,
\begin{equation}
    \sum_sp_{t,s}\mathbf{I}= p(T_\text{eff}) + \bm{\Pi}_\text{b,mix}.
\end{equation}
Finally, for the total energy equation, the summation of the species energy equations gives
\begin{equation} 
\label{eq:energyquation_sum}
\begin{aligned} \sum_s\frac{\partial (\rho_s E_s)}{\partial t}+\nabla\cdot&\left[\sum_s\rho_s E_s\bm{u}_s+\sum_sp_{t,s}\bm{u}_s+\sum_s\bm{\sigma}_s\cdot\bm{u}_s \right.\\ &\left. +\sum_s\left(\bm{q}_{\text{tra},s}+\bm{q}_{\text{tra},s}+\bm{q}_{\text{tra},s}\right)\right]+Q_{e^0_{f,s}} =\sum_sQ_{s}^\text{E}. 
\end{aligned} 
\end{equation}
Owing to total energy conservation during inter-species collisions and chemical reactions, one has $\sum_s Q_s^E=0$. Furthermore, according to Eqs.~\eqref{eq:rhoErhoeTeff05rhou2} and~\eqref{eq:internalenergy}, together with the following relation:
\begin{equation}
    \sum_s \frac{1}{2}\rho_s |\bm{u}_s|^2 \bm{u}_s=\frac{1}{2}\rho |\bm{u}|^2 \bm{u}
+
\sum_s \rho_s (\bm{u}\cdot\bm{V}_s)\bm{V}_s
+
\frac{1}{2}\bm{u}\sum_s \rho_s |\bm{V}_s|^2
+
\frac{1}{2}\sum_s \rho_s |\bm{V}_s|^2 \bm{V}_s,
\end{equation}
the total energy equation in Eq.~\eqref{eq:equ_NS} can be obtained. Finally, $Q_{e^0_{f,s}}$ represents the additional term considering the zero-point energy offset, which can be written as,
\begin{equation}
\label{eq:zeropoint_energy}
    Q_{e^0_{f,s}}=\frac{\partial}{\partial t}\sum_s\rho_se_{f,s}^0+\nabla\cdot \left(\sum_s\rho_se_{f,s}^0\bm{u}+\sum_se_{f,s}^0\bm{\Phi}_s\right).
\end{equation}
Substituting the species continuity equation in Eq.~\eqref{eq:equ_NS} into \eqref{eq:zeropoint_energy}, we have,
\begin{equation}
    \frac{\partial}{\partial t}\sum_s\rho_se_{f,s}^0=\sum_se_{f,s}^0\left[\dot{\omega}_s-\nabla\cdot(\rho_s\bm{u}+\bm{\Phi}_s)\right].
\end{equation}
Thus,
\begin{equation}
    Q_{e^0_{f,s}}=\sum_se_{f,s}^0\dot{\omega}_s.
\end{equation}

\section{Finite volume method for macroscopic synthetic equations}\label{Finite_volume_synthetic}


The macroscopic synthetic equations for chemically reacting flows may be regarded as the conventional NS equations supplemented by the HoTs in Eqs.~\eqref{eq:generalrelations} and~\eqref{eq:PhiOmega}. They can therefore be solved efficiently using standard computational fluid dynamics methods. In this study, the cell-centered finite volume method is employed, and the synthetic equations are discretized as follows:
\begin{equation}
\frac{\partial \bm{W}_i}{\partial t} + \frac{1}{V_i} \sum_{j \in N(i)} (\bm{F}_{ij}+\bm{F}_{ij}^{\text{HoT}}) \bm{S}_{ij} = \bm{Q}_i.
\label{eq:discreteform}
\end{equation}
Here, $\boldsymbol{W} =[\rho, \rho \bm{u}, \rho E, \rho Y]^\top$ is the vector of macroscopic variables for the discrete cell $i$, and $N(i)$ represents the set of neighboring cells adjacent to cell $i$, with $j$ indicating a specific neighboring cell.
The volume of cell $i$ is $V_i$, while $S_{ij}$ denotes the area of the interface $ij$, with its outward normal vector directed from cell $i$ to cell $j$. $F_{ij}$ is the interface flux $\bm{F}_{ij}=\bm{F}_{c,ij}+\bm{F}_{{v},ij}$ including both convection and viscosity flux. Moreover, the viscous flux term depends on the HoTs in the mass diffusion, shear stress and heat flux $\bm{F}_{v,ij}^\text{HoT}=\bm{F}_{v}(\bm{\Phi}_s^{\text{DSMC}},\bm{\sigma}_s^{\text{DSMC}},\bm{q}_s^{\text{DSMC}})-\bm{F}_{v}(\bm{\Phi}_s^{\text{NS}*},\bm{\sigma}_s^{\text{NS}*},\bm{q}_s^{\text{NS}*})$. The source term $\bm{Q}$ is determined based on the macroscopic variables $\bm{W}$. In two-dimensional cases for two species gas mixture, the conservative variables, together with the convective flux vector $\bm{F}_{\bm{c}}$ and the viscous flux vector $\bm{F}_{\bm{v}}$ in Eq.~\eqref{eq:discreteform} can be written as, 
\begin{equation}
\label{eq:macroscopicvariables}
\bm{W}=\left[\begin{matrix}
 \rho \\ \rho u_x \\ \rho u_y \\ \rho E \\ \rho Y_1 
\end{matrix}\right],\,\,
\boldsymbol{F}_{c} = \begin{bmatrix}
\rho u_{n} \\
\rho u_{x} u_{n} + n_{x} p \\
\rho u_{y} u_{n} + n_{y} p \\
u_{n}(\rho E + p) \\ 
\rho Y_1u_n
\end{bmatrix},\,\,
\bm{F}_{\bm{v}}=\left[\begin{matrix}
0 \\ n_x\sigma_{xx}+n_y\sigma_{xy} \\ n_x\sigma_{yx}+n_y\sigma_{yy} \\ n_x\Theta_x+n_y\Theta_y \\ n_x\Phi_{x,1}+n_y\Phi_{y,1}
\end{matrix}\right],\,\, 
\bm{Q}=\left[\begin{matrix}
 0 \\ 0 \\ 0 \\ 0 \\ {\dot{\omega}_1+{\text{HoT}}_{\dot{\omega}_1}} 
\end{matrix}\right],
\end{equation}
where
\begin{equation}
\begin{aligned}
    \Theta_x& =u_x\sigma_{xx}+u_y\sigma_{xy}+q_x+\sum_{s}{h_s\Phi_{x,s}},\\
    \Theta_y & =u_x\sigma_{yx}+u_y\sigma_{yy}+q_y+\sum_{s}{h_s\Phi_{y,s}}, 
\end{aligned}
\end{equation}
and $u_{n}=u_{x}n_x+u_{y}n_y$ is  the scalar product of velocity vector and the unit normal vector. 

Equation~\eqref{eq:discreteform} can be further discretized by the implicit backward Euler scheme:
\begin{equation}
\frac{\bm{W}_{i}^{k+1} - \bm{W}_{i}^{k}}{\Delta t_{i}} + \frac{1}{V_{i}} \sum_{j \in N(i)} (\bm{F}_{ij}^{k+1}+\bm{F}_{v,ij}^\text{HoT}) \bm{S}_{ij} = \bm{Q}_{i}^{k+1},
\label{eq:discrete_conservation}
\end{equation}
where $\Delta t=t^{k+1}-t^k$ is the numerical time step given in the implicit process. By introducing the incremental variables $\Delta\bm{W}^{k}_i=\bm{W}^{k+1}_i-\bm{W}^k_i$ and $\bm{F}_{ij}^{k+1}=\bm{F}_{ij}^k+\Delta\bm{F}_{ij}^k$, the delta-form governing equation of Eq.~\eqref{eq:discrete_conservation} for implicit iterative algorithm can be written as,
\begin{equation}
\scalebox{1.0}{ $
\displaystyle 
\left[\frac{1}{\Delta t_{i}} - \left(\frac{\partial \bm{Q}_{i}}{\partial \bm{W}_{i}}\right)\right]
\Delta \bm{W}_{i}^{k} + 
\frac{1}{V_{i}}\sum_{j\in N(i)}\Delta \bm{F}_{ij}^{k}\bm{S}_{ij} = 
\underbrace{
-\frac{1}{V_{i}}\sum_{j\in N(i)}(\bm{F}_{ij}^{k}+\bm{F}_{v,ij}^\text{HoT})\bm{S}_{ij} + \bm{Q}_{i}^{k}}_{\bm{R}_{i}^{k}},
$}
\label{eq:deltaform_discretization}
\end{equation}
where $\bm{R}_{i}^{k}$ is the macroscopic residuals in the $k$-th step. 

In general, the macroscopic implicit fluxes in left-hand-side of Eq.~\eqref{eq:deltaform_discretization} is approximated by the first-order flux in the Euler equation:
\begin{equation}
\Delta\bm{F}_{ij}^k = \frac{1}{2}\left[\Delta\bm{F}_i^k+\Delta\bm{F}_j^k+\Gamma_{ij}\left(\Delta \bm{W}_i^k-\Delta \bm{W}_j^k\right)\right],\,\, \bm{F}_{ij}=\bm{F}(\bm{W}_L,\bm{W}_R,S_{ij}),
\label{eq:Eulerfluxes}
\end{equation}
where $\Gamma_{ij}=|u_{n}|+c+2\mu/\rho|\bm{n}_{ij}\cdot(\bm{x}_{j}-\bm{x}_i)|$ is the approximate spectral radius for each species, and $c$ is the speed of sound for the mixture gas. The reconstructed macroscopic variables of the left and right sides of the interface can be obtained as $\bm{W}_{L/R}=\bm{W}_{i/j}+\phi\nabla(\bm{W}_{i/j}\cdot\bm{x})$, where $\phi$ is calculated using the Venkatakrishnan limiter. In this paper, we apply the Rusanov scheme~\cite{sod-1978} for reconstruction to enhance the numerical stability. Moreover, since the control volume satisfies $\sum_{j\in N(i)}\bm{F}_i\bm{S}_{ij}=0$, the flux can be directly represented by the convection flux, and thus the incremental flux from $j$-th cell can be expressed as $\Delta\bm{F}_j^k=\bm{F}_c(\bm{W}_j^k+\Delta \bm{W}_j^k)-\bm{F}_c(\bm{W}_j^k)$. Thus, the general implicit governing equations for macroscopic properties in Eq.~\eqref{eq:deltaform_discretization} can be expressed as: 
\begin{equation}
\scalebox{1.0}{ $
\displaystyle 
\left[\frac{1}{\Delta t_{i}} + \frac{1}{2 V_{i}}\sum_{j \in N(i)} \Gamma_{i j}S_{ij} - \left(\frac{\partial \bm{Q}_i}{\partial \bm{W}_i}\right)^{k}\right] \Delta \bm{W}_{i}^{k} + \frac{1}{2 V_{i}} \sum_{j \in N(i)} \left( \Delta \bm{F}_{j}^{k} - \Gamma_{ij} \Delta \bm{W}_{j}^{k} \right) \bm{S}_{ij} = \bm{R}_{i}^{k},$ }
\end{equation}
which can be efficiently solved using the classical Lower-Upper Symmetric Gauss-Seidel iteration technique. Furthermore, the Jacobian $\frac{\partial \bm{Q}_i}{\partial \bm{W}_i}$ can be explicitly formulated in terms of macroscopic properties. 
In order to increase the stability of macroscopic solver, the source term Jacobi matrix should be considered in the implicit scheme, which can be written as,
\begin{equation}
\label{eq:Jacobian}
\begin{aligned}
    \frac{\partial \bm{Q}}{\partial \bm{W}} =
    \left[\begin{matrix}
 0 & 0 & 0 & 0 & 0 \\ 0 & 0 & 0 & 0 & 0 \\ 0 & 0 & 0 & 0 & 0 \\ 0 & 0 & 0 & 0 & 0 \\ 
 \frac{\partial Q(5)}{\partial W(1)} & \frac{\partial Q(5)}{\partial W(2)} & \frac{\partial Q(5)}{\partial W(3)} & 
 \frac{\partial Q(5)}{\partial W(4)} & 
 \frac{\partial Q(5)}{\partial W(5)}
\end{matrix}\right] 
   \approx 
    \left[\begin{matrix}
 0 & 0 & 0 & 0 & 0 \\ 0 & 0 & 0 & 0 & 0 \\ 0 & 0 & 0 & 0 & 0 \\ 0 & 0 & 0 & 0 & 0 \\ 
 \frac{\partial\dot{\omega}_1}{\partial \rho} & 0 & 0 & 
 \frac{\partial\dot{\omega}_1}{\partial \rho E} & 
 \frac{\partial\dot{\omega}_1}{\partial \rho Y_1}
\end{matrix}\right],
\end{aligned}
\end{equation}
where
\begin{equation}
\scalebox{0.9}{ $
\begin{aligned}
    \frac{\partial \dot{\omega}_1}{\partial \rho }&=m_1\sum_{l=1}^{N_R}\left[ \nu_{1\text{b}}^{(l)} - \nu_{1\text{f}}^{(r)} \right] \left[ k_{\text{f}l} \, n_{1}^{\nu_{1\text{f}}^{(l)}} \, n_{2}^{\nu_{2\text{f}}^{(l)}}  \frac{\nu_{2\text{f}}^{(l)}}{m_{2} n_{2}} - k_{\text{b}l} \, n_{1}^{\nu_{1\text{b}}^{(l)}} \, n_{2}^{\nu_{2\text{b}}^{(l)}}  \frac{\nu_{2\text{b}}^{(l)}}{m_{2} n_{2}}\right],
    \\
    \frac{\partial \dot{\omega}_1}{\partial \rho E }&=m_1\sum_{l=1}^{N_R}\left[ \nu_{1\text{b}}^{(l)} - \nu_{1\text{f}}^{(r)} \right]\left[\frac{\text{d}k_{\text{f}l}}{\text{d}T}  \, n_{1}^{\nu_{1\text{f}}^{(l)}} \, n_{2}^{\nu_{2\text{f}}^{(l)}} -\frac{\text{d}k_{\text{b}l}}{\text{d}T}  \, n_{1}^{\nu_{1\text{b}}^{(l)}} \, n_{2}^{\nu_{2\text{b}}^{(l)}}    \right]\frac{1}{\rho c_{v,\text{mix}}},
    \\
    \frac{\partial \dot{\omega}_1}{\partial \rho Y_1}&=m_1\sum_{l=1}^{N_R}\left[ \nu_{1\text{b}}^{(l)} - \nu_{1\text{f}}^{(l)} \right] \left[ k_{\text{f}l} \, n_{1}^{\nu_{1\text{f}}^{(l)}} \, n_{2}^{\nu_{2\text{f}}^{(l)}} \left( \frac{\nu_{1\text{f}}^{(l)}}{m_{1} n_{1}} - \frac{\nu_{2\text{f}}^{(l)}}{m_{2} n_{2}}  \right) - k_{\text{b}l} \, n_{1}^{\nu_{1\text{b}}^{(l)}} \, n_{2}^{\nu_{2\text{b}}^{(l)}} \left( \frac{\nu_{1\text{b}}^{(l)}}{m_{1} n_{1}} - \frac{\nu_{2\text{b}}^{(l)}}{m_{2} n_{2}}  \right)\right].
\end{aligned}
$}
\end{equation}
Note that in Eq.~\eqref{eq:macroscopicvariables}, the higher-order source term $\text{HoT}_{\dot{\omega}_1}$ remains constant during the iteration of the macroscopic equations. Consequently, its derivative with respect to variables in $\bm{W}$ vanishes. 

\bibliographystyle{elsarticle-num}
\bibliography{ref}
\end{document}